\documentclass[12pt]{article}
\pdfoutput=1
\usepackage{amssymb}
\usepackage{amsmath}
\usepackage{cases}
\usepackage{latexsym}
\usepackage[usenames]{color}
\usepackage{fancybox}
\usepackage{simplewick}
\usepackage{comment}
\usepackage{cite}
\usepackage{subcaption}
\definecolor{shadecolor}{rgb}{0.9,0.9,0.95}
\usepackage{setspace}
\usepackage{mathtools}
\usepackage[usenames,dvipsnames]{xcolor}
\definecolor{darkgreen}{rgb}{0,0.5,0}
\definecolor{darkblue}{cmyk}{0.9,0.9,0,0}
\definecolor{darkred}{rgb}{0.6,0,0.3}

\usepackage{graphicx} 
\usepackage[setpagesize=false,pagebackref=false, linktocpage, bookmarksopen=true, colorlinks=true, linkcolor=RoyalBlue,citecolor=Maroon,urlcolor=Maroon]{hyperref}
\usepackage{hyperref}

\renewcommand{\thefootnote}{\arabic{footnote}}

\def\del{\partial}

\def\fn#1{\footnote{#1}}
\def\nn{\nonumber}
\def\eqref#1{(\ref{#1})}
\def\comma{\,,}
\def\period{\,.}

\def\bDelta{\bar{\Delta}}
\def\beq{\begin{equation}}
\def\eeq{\end{equation}}

\def\O{\mathcal{O}}
\def\tO{\tilde{\mathcal{O}}}
\newcommand{\be}{\begin{equation}}
\newcommand{\ee}{\end{equation}}
\newcommand{\bea}[1]{\begin{align*}#1\end{align*}}
\newcommand\eqn{\addtocounter{equation}{1}\tag{\theequation}}
\def\m{\mu}
\def\n{\nu}
\def\dt{\delta \theta}
\def\dx{\delta x}

\def\tk{\theta_k}

\def\p{\partial}
\def\D{\Delta}

\def\t{\tau}
\def\C{\mathcal{C}}
\def\Tf{T_{D5}}
\def\Tt{T_{D3}}
\def\t{\tau}
\def\du{\delta u}
\def\dy{\delta y}
\def\vu{\mathbf{u}}
\DeclareMathOperator{\csch}{csch}

\def\NO#1{:\! #1 \!:\,}

\def\red#1{#1}

\numberwithin{equation}{section}

\begin{document}
\thispagestyle{empty}

\renewcommand{\thefootnote}{\fnsymbol{footnote}}
\setcounter{page}{1}
\setcounter{footnote}{0}
\setcounter{figure}{0}
\begin{flushright}
\end{flushright}
\vspace{0.7cm}
\begin{center}
{\Large \bf
Giant Wilson Loops and AdS$_2$/dCFT$_1$
\par}

\vspace{1.3cm}

\textrm{Simone Giombi$^{\clubsuit}$, Jiaqi Jiang$^{\clubsuit}$, Shota Komatsu$^{\diamondsuit}$}
\\ \vspace{1cm}
\footnotesize{\textit{
$^{\clubsuit}$Department of Physics, Princeton University, Princeton, NJ 08544, USA\\
$^{\diamondsuit}$School of Natural Sciences, Institute for Advanced Study, Princeton, NJ 08540, USA
}  
\vspace{1cm}
}

{\tt sgiombi AT princeton.edu, jiaqij AT princeton.edu, skomatsu AT ias.edu}

\par\vspace{1.5cm}

\textbf{Abstract}\vspace{2mm}
\end{center}
\noindent
The $1/2$-BPS Wilson loop in $\mathcal{N}=4$ supersymmetric Yang-Mills theory is an important and well-studied example of conformal defect. In particular, much work has been done for the correlation functions of operator insertions on the Wilson loop in the fundamental representation. In this paper, we extend such analyses to Wilson loops in the large-rank symmetric and antisymmetric representations, which correspond to probe D3 and D5 branes 
with $AdS_2 \times S^2$ and $AdS_2 \times S^4$ worldvolume geometries, ending at the $AdS_5$ boundary along a one-dimensional contour.
We first compute the correlation functions of protected scalar insertions from supersymmetric localization, and obtain a representation in terms of multiple integrals that are similar to the eigenvalue integrals of the random matrix, but with important differences. Using ideas from the Fermi Gas formalism and the Clustering method, we evaluate their large $N$ limit exactly as a function of the 't Hooft coupling. The results are given by simple integrals of polynomials that resemble the $Q$-functions of the Quantum Spectral Curve, with integration measures depending on the number of insertions. Next, we study the correlation functions of fluctuations on the probe D3 and D5 branes in AdS. We compute a selection of three- and four-point functions from perturbation theory on the D-branes, and show that they agree with the results of localization when restricted to supersymmetric kinematics.  We also explain how the difference of the internal geometries of the D3 and D5 branes manifests itself in the localization computation.

\setcounter{page}{1}
\renewcommand{\thefootnote}{\arabic{footnote}}
\setcounter{footnote}{0}
\setcounter{tocdepth}{3}
\newpage
\tableofcontents

\parskip 5pt plus 1pt   \jot = 1.5ex

\newpage
\section{Introduction}
Wilson loops are among the most fundamental observables in gauge theory. In supersymmetric gauge theories, one can often define a supersymmetric generalization of the Wilson loop that can be computed exactly using supersymmetric localization. Among various supersymmetric Wilson loops, the one that has been studied most intensively is perhaps the $1/2$-BPS Wilson loop \cite{Maldacena:1998im} in $\mathcal{N}=4$ supersymmetric Yang-Mills theory (SYM) in four dimensions. 

The $1/2$-BPS Wilson loop played a pivotal role in the early days of the AdS/CFT correspondence \cite{Maldacena:1997re}. Its expectation value was computed in $\mathcal{N}=4$ SYM, first by resumming a subset of diagrams \cite{Erickson:2000af,Drukker:2000rr} and later rigorously by supersymmetric localization \cite{Pestun:2009nn}. The result, which is a nontrivial function of the coupling constant, reproduces the regularized area of the string worldsheet in AdS in the strong coupling limit \cite{Berenstein:1998ij,Drukker:1999zq}. This agreement was one of the first nontrivial evidence for the holographic duality \cite{Drukker:2000ep}. 
More recently the $1/2$-BPS Wilson loops have gained renewed interest, since they turned out to be ideal testing grounds for various non-perturbative techniques. 

First, the $1/2$-BPS Wilson loop is defined on a circle or a straight line and is known to preserve the $OSp(4^{\ast}|4)$ subgroup of the full superconformal symmetry of $\mathcal{N}=4$ SYM \cite{Gomis:2006sb,Liendo:2018ukf}. In particular, it is invariant under the one-dimensional conformal group $SL(2,R)$  \cite{Drukker:2005af,Drukker:2006xg}, and can be regarded as providing an example of defect conformal field theory (dCFT) \cite{Drukker:2005af, Cooke:2017qgm, Giombi:2017cqn}. From this point of view, important observables to analyze are the correlation functions of insertions on the Wilson loop with or without local operators in the bulk, and much work has been done to compute them at weak and strong coupling \cite{Cooke:2017qgm,Kim:2017sju,Kiryu:2018phb,Giombi:2017cqn}. These correlation functions admit more than one operator product expansions, and one obtains  (defect) crossing equations by equating two different expansions. By applying the idea of the conformal bootstrap and analyzing these crossing equations either numerically or analytically, one can constrain the correlation functions on the Wilson loop without needing to perform direct perturbative computations  \cite{Liendo:2018ukf}.  

Secondly, the spectrum of operators on the $1/2$-BPS Wilson loop in the large $N$ limit can be studied using the integrability methods. This was demonstrated first at weak coupling by mapping the operator to an open spin chain in \cite{Drukker:2006xg}. Subsequently the {\it thermodynamic Bethe ansatz equation}, which determines the spectrum at finite 't Hooft coupling, was written down in \cite{Correa:2012hh,Drukker:2012de}. This was further reformulated into the {\it Quantum Spectral Curve} in \cite{Gromov:2015dfa}, which enabled an efficient numerical computation of the spectrum of non-protected operators. The result for the lightest non-protected operator beautifully interpolates between the answer on the gauge theory at weak coupling and the answer computed from the string worldsheet at strong coupling \cite{Grabner:2020nis}. Furthermore, there are proposals on the integrability description of the correlation functions of insertions on the Wilson loop \cite{Kim:2017phs,Kiryu:2018phb} based on the hexagon formalism \cite{Basso:2015zoa}, which was originally developed to study the correlation functions of single-trace operators.

Thirdly, it was demonstrated in \cite{Giombi:2018qox,Giombi:2018hsx} that the supersymmetric localization, originally applied to the expectation value of the Wilson loop, can be used to compute correlation functions of protected scalar insertions on the Wilson loop. This was achieved by first considering the $1/8$-BPS Wilson loops, which are defined on the $S^2$ subspace and whose expectation values depend on the area of the region inside the loop on $S^2$. By differentiating the expectation value with respect to the area, one can insert operators with the minimal $R$-charge on the Wilson loop \cite{Giombi:2017cqn}. Starting from such minimal-charge operators, we can construct protected scalar operators of arbitrary length by performing the operator product expansion and the Gram-Schmidt orthogonalization. This allows us to compute an infinite set of correlation functions of such operators exactly as a function of the coupling constant. Later this method was generalized to include single-trace operators inserted outside of the Wilson loops. Together, these results provide analytic defect CFT data which can be used in the conformal bootstrap analysis. Furthermore, the planar limit of such correlators is found to be given by simple integrals of polynomials. Rather unexpectedly, these polynomials conicide with (the limit of) the so-called $Q$-functions, which are the basic objects in the Quantum Spectral Curve approach \cite{Gromov:2013pga,Gromov:2014caa,Gromov:2015dfa}. This unexpected connection indicates that the Quantum Spectral Curve might be an useful tool also for the correlation functions\fn{See \cite{Cavaglia:2018lxi,McGovern:2019sdd} for other setups in which the correlation functions, computed by other methods, can be expressed simply in terms of the Q-functions of the Quantum Spectral Curve.}.

Lastly, the AdS/CFT correspondence relates the correlation functions on the $1/2$-BPS Wilson loop to the correlators of the fluctuations on the dual string worldsheet with $AdS_2$ induced geometry. In the large $N$ limit, the fluctuations on the string worldsheet are decoupled from the closed string modes in the bulk of $AdS_5$, and the setup provides a simple example of AdS$_2$/dCFT$_1$ correspondence\fn{This correspondence does not involve gravity on the AdS side and may be viewed as a {\it rigid holography} \cite{Aharony:2015zea}. See \cite{Carmi:2018qzm,Beccaria:2019stp,Beccaria:2019ibr,Beccaria:2019mev,Beccaria:2019dju,Beccaria:2020qtk,Drukker:2020swu} for recent studies of similar rigid holography setups.}. In \cite{Giombi:2017cqn}, a set of four-point functions were computed from perturbation theory on the string worldsheet and various defect CFT data were extracted from the operator product expansions. In special kinematical configurations, the results also reproduced the strong-coupling limit of the correlation functions computed from the localization, thereby providing important consistency checks of both approaches \cite{Giombi:2018qox}. The computation was subsequently generalized to the string worldsheet dual to the ordinary (non-supersymmetric) Wilson loop \cite{Beccaria:2017rbe,Beccaria:2019dws}, and also to the $1/2$-BPS Wilson loop in ABJM theory in \cite{Bianchi:2020hsz}. In addition, the Wilson loops which interpolate between the $1/2$-BPS Wilson loop and the standard Wilson loop were analyzed in \cite{Beccaria:2017rbe}. 

Most of these works discuss the Wilson loop in the fundamental representation. The main aim of this work is to generalize the analysis to the $1/2$-BPS Wilson loops in higher-rank representations. In particular we consider totally symmetric or antisymmetric representations of size of order $N$. These Wilson loops are known to be dual to the D-branes---the D3-branes for the symmetric representations and the D5-branes for the antisymmetric representations---and are analogues of the {\it Giant Gravitons} \cite{Balasubramanian:2001nh,McGreevy:2000cw}, which are D-branes dual to local operators with large $R$-charge of order $N$. For this reason, they are sometimes referred to as {\it Giant Wilson loops}, a terminology we adopt in this paper. Much like the Wilson loop in the fundamental representation, they are examples of (super)conformal defects with the $OSp(4^{\ast}|4)$ symmetry, and can be studied by supersymmetric localization as was demonstrated in \cite{Drukker:2005kx, Okuyama:2006jc, Hartnoll:2006is} for the expectation values (correlation functions of single trace operators in the presence of the Giant Wilson loops were studied in \cite{Giombi:2006de}). 

\begin{figure}[t]
\centering
\begin{minipage}{0.45\hsize}
\centering
\includegraphics[clip,height=4.5cm]{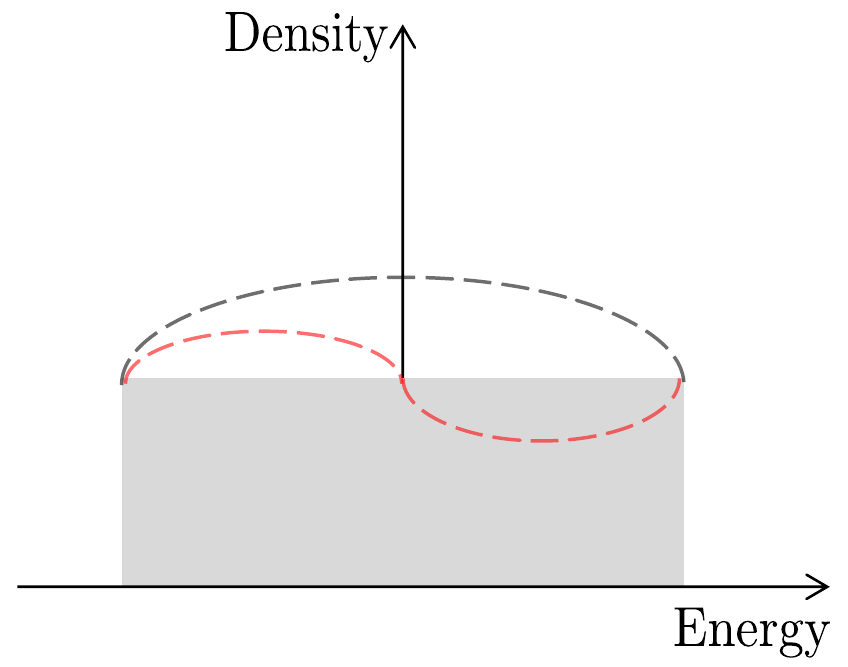}
\subcaption{D5-brane (antisymmetric rep)}
\end{minipage}
\begin{minipage}{0.45\hsize}
\centering
\includegraphics[clip,height=4.5cm]{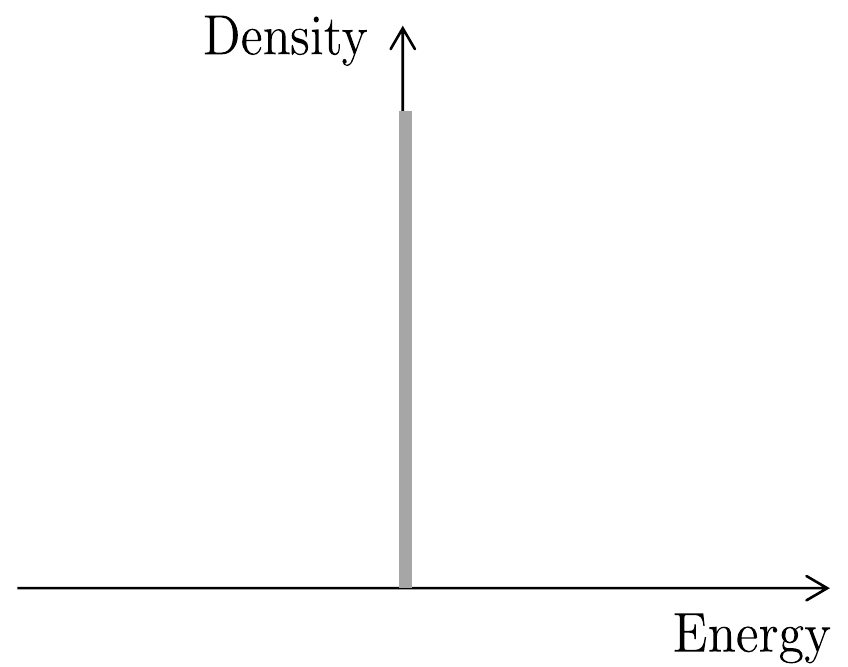}
\subcaption{D3-brane (symmetric rep)}
\end{minipage}
\caption{Kaluza-Klein spectrum and Bose/Fermi distributions. In the localization computation, Kaluza-Klein modes with higher $S^5$ angular momenta correspond to deformations of the density distributions of free Fermi gas (D5-brane) or free Bose gas (D3-brane). The 't Hooft coupling is identified with the inverse temperature $\beta$. (a) For the D5-brane, the distribution has support on a finite interval even at strong coupling, since it corresponds to the zero-temperature limit of the free Fermi gas. We therefore have infinitely many Kaluza-Klein modes corresponding to different deformations of the Fermi distribution, the first two of which are depicted in the figure (dashed curves). (b) For the D3-brane, the distribution has support only at finitely many points at strong coupling owing to the Bose-Einstein condensation. Consequently the number of deformations is finite and the Kaluza-Klein spectrum is truncated.}\label{fig:bosefermi}
\end{figure}

Before discussing the contents of this paper, let us explain a couple of more motivations for studying the Giant Wilson loops. The first motivation is to understand how the structure of the worldvolume geometries of the D-branes is reflected on the gauge theory side. The D3-brane, which is dual to the symmetric Wilson loop, is extended in $AdS_2\times S^3$ subspace inside $AdS_5$ while the D5-brane is extended both in $AdS_5$ and $S^5$ and its worldvolume is given by $AdS_2 \times S^4$. This difference is reflected in the spectrum of the excitations on the D-branes. For the D3-brane we have an infinite tower of Kaluza-Klein modes with higher AdS angular momenta which arise from reducing $S^3$. On the other hand, the D5-brane contains an infinite tower of Kaluza-Klein modes coming from $S^4$, which have higher $S^5$ angular momenta. The existence and the non-existence of such infinite towers of operators are what distinguish the two cases and are clear signatures of the emergent internal geometries of the D-branes. However, at weak coupling on the gauge theory side, it is hard to see such qualitative differences between the antisymmetric and the symmetric representations. In fact, as we discuss in more detail in section \ref{subsec:giant}, both towers seem to exist at weak coupling regardless of the representations. In this  paper we demonstrate, using supersymmetric localization, how one of the two towers on the D3-brane decouples from the rest of the spectrum at strong coupling. This decoupling is realized by a mechanism resembling the Bose-Einstein condensation. See Figure \ref{fig:bosefermi} for a heuristic explanation and section \ref{subsec:symmetricloop} for more details.

Another motivation comes from the relation to the so-called {\it twisted holography} \cite{Mezei:2017kmw,Costello:2017fbo,Ishtiaque:2018str,Costello:2018zrm,Gaiotto:2020vqj}. The twisted holography refers to special examples of the AdS/CFT correspondence in which both the bulk and the boundary theories are topologically (or holomorphically) twisted. Such theories are typically much simpler than the theories relevant for the full-fledged AdS/CFT correspondence, and therefore may provide a good starting point for understanding the duality  in precise details. In \cite{Ishtiaque:2018str}, it was pointed out that there is one such example which involves the topological twist of the D2/D4 brane system: The boundary side is given by two-dimensional BF theory and a product of  Wilson loops in the antisymmetric representations while the bulk side is the holomorphic Chern-Simons theory in four dimensions \cite{Costello:2017dso}. They further showed that the operator algebra living on the Wilson line is isomorphic to the Yangian. This setup is closely related to the Giant Wilson loops in $\mathcal{N}=4$ SYM since the localization relates the $1/2$-BPS Wilson loops to the standard Wilson loops in two-dimensional Yang-Mills theory, whose zero coupling limit is the BF theory. We will not directly address this question in this paper, but we expect that the techniques developed in this paper will be useful for studying such problems.

Let us now describe in more detail the contents of this paper: We first generalize the results in \cite{Giombi:2018qox,Giombi:2018hsx} to the Giant Wilson loops and compute correlation functions of protected scalar insertions by a combination of supersymmetric localization, the operator product expansion and the Gram-Schmidt analysis. The generalization turns out to be nontrivial owing to more complicated structures of the operator spectrum (which we discuss in more detail in section \ref{subsec:eighthBPS}). To overcome this problem, we first consider generalizations of the higher-rank Wilson loops that couple to several different areas. The expectation values of such Wilson loops can be computed by the application of the loop equation in two-dimensional Yang-Mills theory as shown in \cite{Giombi:2020pdd}. The results are given by multiple contour integrals, which are similar but different from the eigenvalue integrals of the matrix models. Owing to this difference, the standard techniques of the matrix models are not directly applicable, but we show how to compute their large $N$ limits by using ideas from the {\it Fermi Gas formalism} \cite{Marino:2011eh} and the {\it Clustering method} \cite{Jiang:2016ulr}. The former was developed originally for the study of the $S^3$ partition function\fn{See \cite{Chester:2020jay,Gaiotto:2020vqj} for recent applications of the Fermi Gas formalism to the computation of the correlation functions of protected operators.} of ABJM theory \cite{Kapustin:2009kz} while the latter was developed for the analysis of the three-point functions in $\mathcal{N}=4$ SYM based on the hexagon formalism \cite{Basso:2015zoa}. Applying these techniques we determine the large $N$ limit of their expectation values and extract the correlation functions of protected scalar insertions. As was the case with the Wilson loop in the fundamental representation, the final results are given by simple integrals of polynomials, which again resemble the $Q$-functions of the Quantum Spectral Curve:
\begin{align}
\langle \!\langle \tilde{\mathcal{O}}_{n_1}\tilde{\mathcal{O}}_{n_2} \rangle \!\rangle&= N\oint d\mu_2 \,Q_{n_1} \left(g(x-x^{-1})\right)Q_{n_2} \left(g(x-x^{-1})\right)\comma\\
\langle \!\langle \tilde{\mathcal{O}}_{n_1}\tilde{\mathcal{O}}_{n_2}\tilde{\mathcal{O}}_{n_3}\rangle \!\rangle&= N\oint d\mu_3 \,Q_{n_1} \left(g(x-x^{-1})\right)Q_{n_2} \left(g(x-x^{-1})\right)Q_{n_3} \left(g(x-x^{-1})\right)\period\nn
\end{align}
One notable difference is that, unlike the results for the Wilson loop in the fundamental representation \cite{Giombi:2018qox}, the measure of the integrals depend on the number of operator insertions. 
This feature seems to be related to the existence of {\it multi-particle operators}, which are the dCFT analogues of the multi-trace operators. 
See section \ref{sec:correlatorfromloc} for more details.

Next, we study the correlation functions of the fluctuations on the D-branes in AdS. In particular we focus on the elementary excitations in the $AdS_5$ and $S^5$ directions. The former corresponds to the so-called displacement operator while the latter corresponds to a single scalar insertion on the Wilson loop. For the D5-brane,  dual to the antisymmetric Wilson loop, we also analyze the correlation functions of higher Kaluza-Klein modes coming from the $S^4$ worldvolume of the D5-brane. These operators carry higher angular momenta on $S^5$ and correspond to protected scalar insertions with higher $R$-charges. In special kinematics where the correlator preserves a fraction of supersymmetry, the results from the D-brane analysis agree, both for D3 and D5 cases, with the strong-coupling limit of the results of supersymmetric localization.

The rest of this paper is organized as follows: In section \ref{sec:setup}, we briefly review the basic facts on the supersymmetric Wilson loops in $\mathcal{N}=4$ SYM including operator insertions and their holographic dual description. We also explain in more detail the puzzles related to the Kaluza-Klein towers, mentioned earlier. Then in section \ref{sec:integral}, we review the mutiple integral represntation of the $1/8$ BPS Wilson loops and derive an expression for the generalized higher-rank loop that couples to different areas. We also explain how to take the large $N$ limit using ideas from the Fermi Gas formalism and the Clustering method. In section \ref{sec:correlatorfromloc}, we use these results to compute the correlation functions of protected operator insertions by applying the Gram-Schmidt analysis. Interestingly, the computation resembles the recent work \cite{Gaiotto:2020vqj} on the protected correlators of supersymmetric gauge theories in three dimensions which are dual to the twisted M-theory. We also make contact with the double-trace deformation of the matrix model studied in \cite{Barbon:1995dx} and discuss the connection to the double-trace deformation in the standard AdS/CFT \cite{Witten:2001ua,Berkooz:2002ug,Hartman:2006dy, Giombi:2011ya}. In section \ref{sec:D5}, we compute the correlation functions of fluctuations on the D5-brane, dual to the Wilson loop in the antisymmetric representation. We compute two-, three- and four-point functions of elementary fluctuations on the D5-brane and also a subset of correlation functions that involve the Kaluza-Klein modes on $S^4$. In section \ref{sec:D3}, we perform a similar analysis for the D3-brane. Finally we conclude and discuss future directions in section \ref{sec:conclusion}. Several appendices are included to explain technical details.

\section{Setup and Generalities\label{sec:setup}}
In this section, we quickly review and summarize the basic facts about the BPS Wilson loops,  their holographic dual descriptions, and their relation to the defect CFT.
\subsection{Giant Wilson loops and holographic dual\label{subsec:giant}}
\paragraph{Higher-rank Wilson loops and D-branes} The $1/2$-BPS Wilson loop in $\mathcal{N}=4$ SYM is the maximally supersymmetric generalization of the ordinary Wilson loop. It can be defined on a straight line or a circle and couples to a single scalar field:
\beq
\mathcal{W}_{R}=\frac{1}{{\rm dim} \,R}{\rm tr}_{R}{\rm P} e^{\oint \left(iA_{\mu}\dot{x}^{\mu}+\Phi^{6}|\dot{x}|\right)d\tau}
\eeq
Here $R$ is the representation of the $U(N)$ gauge group and ${\rm dim}\, R$ is its dimension. In this paper, we consider totally symmetric or antisymmetric representations and take the size of the representation, which is the number of boxes in the Young diagram, to be of order $N$.

In the large $N$ limit, such Wilson loops are known to be dual to D-branes \cite{Drukker:2005kx,Gomis:2006sb,Gomis:2006im,Yamaguchi:2006tq,Hartnoll:2006is}. More precisely the Wilson loop in the large-rank symmetric representation is dual to the D3-brane on the $AdS_2\times S^2$ subspace inside $AdS_5$ \cite{Drukker:2005kx} while the one in the antisymmetric representation is dual to the D5-brane on $AdS_2 \times S^{4}$, where $S^4$ is a subspace inside $S^5$ \cite{Yamaguchi:2006tq}. In both cases, the size of the representation $k$ is related to the fundamental string charge on the D-brane and determines the size of the ``internal space'' of the brane (which is $S^2$ for the symmetric representation and $S^4$ for the antisymmetric representation). The fact that the antisymmetric representation has a cutoff in size translates to the geometric fact that the volume of $S^5$ is finite and the D5-brane has a cutoff in size. 

\paragraph{Defect conformal field theory and classification of operators} Being defined on a circle or a straight line, the $1/2$-BPS Wilson loop preserves a $SL(2,R)$ subgroup of the four-dimensional conformal group \cite{Drukker:2005af,Drukker:2006xg}. Once fermionic symmetries are included, this is extended to the $OSp (4^{\ast}|4)$ 1d (defect) superconformal group \cite{Gomis:2006sb,Giombi:2017cqn,Liendo:2018ukf}. Because of this property, the $1/2$-BPS Wilson loop has been analyzed extensively also from the point of view of the defect CFT \cite{Cooke:2017qgm,Giombi:2017cqn,Kim:2017sju,Liendo:2018ukf,Kiryu:2018phb}. So far, most of the studies have focused on the Wilson loop in the fundamental representation, but the loops in higher representations also provide equally well-defined examples of conformal defects.

From the defect CFT point of view, natural observables are the correlation functions of operators on the defect. As is the case with the fundamental Wilson loop, such operators can be defined by inserting the fields of $\mathcal{N}=4$ SYM inside the Wilson loop trace:
\beq
\langle \!\langle O_1 (\tau_1) \cdots O_m(\tau_m)\rangle\!\rangle\equiv \frac{1}{\langle \mathcal{W}_R\rangle}\left(\frac{1}{{\rm dim}\, R}\left<{\rm tr}_{R}{\rm P}\left[O_1 \cdots O_m e^{\oint \left(iA_{\mu}\dot{x}^{\mu}+\Phi^{6}|\dot{x}|\right)d\tau}\right]\right>\right)\period
\eeq
There is however one important difference between the fundamental Wilson loop and the Wilson loops in higher-rank representations. In the case of the fundamental Wilson loop, there is essentially an unique way to build the insertions $O_j$ from the fundamental fields of $\mathcal{N}=4$ SYM. Namely we take the fields in $\mathcal{N}=4$ SYM and simply multiply them as $N\times N$ matrices,
\beq\label{eq:multifund}
\left(\Phi^2\right)_{ac}\equiv \sum_{b}\left(\Phi\right)_{ab}\left(\Phi\right)_{bc}\period
\eeq 
To express \eqref{eq:multifund} in more group-theoretic terms, it is useful to decompose $\Phi$ into the generators of the fundamental representation $T^{f}_A$ as
\beq
\Phi_{ab}= \sum_{A}\Phi_{A} \left(T_{A}^{f}\right)_{ab}\qquad\quad  (A=1,\ldots, N^2)\comma
\eeq
Then the product \eqref{eq:multifund} can be expressed as
\beq\label{eq:TABC2}
\left(\Phi^2\right)_A\equiv d_{ABC}^{f}\Phi_B\Phi_C\comma
\eeq
where the tensor $d_{ABC}^{f}$ is defined by
\beq\label{eq:TABC1}
T_{A}^{f}T_{B}^{f}=d_{ABC}^{f}T^{f}_C\period
\eeq

On the other hand, for the higher-rank representations, there are two natural approaches to define the insertions. The first approach is to replace \eqref{eq:TABC2} and \eqref{eq:TABC1} with their higher-rank counterparts. Namely we consider
\beq\label{eq:twoparticledef}
\left(\Phi^{[2]}\right)_A\equiv d_{ABC}^{R}\Phi_B\Phi_C\comma
\eeq
where the tensor $d_{ABC}^{R}$ is defined by
\beq
T_{A}^{R}T_{B}^{R}=d_{ABC}^{R}T^{R}_C\comma
\eeq 
and $T_{A}^{R}$'s are the generators in the representation $R$. The operator \eqref{eq:twoparticledef} can be inserted  inside the Wilson loop trace as
\beq
{\rm tr}_{R}{\rm P}\left[\sum_{A}\Phi^{[2]}_AT_{A}^{R}\exp \left(\oint i A_{\mu}\dot{x}^{\mu}+\cdots\right)\right]\period
\eeq 
Since the Wilson loop trace is computed in the representation $R$, such operators arise naturally by bringing together two single insertion of $\Phi$'s on the Wilson loop. 

\begin{figure}[t]
\centering
\begin{minipage}{0.45\hsize}
\centering
\includegraphics[clip,height=1.5cm]{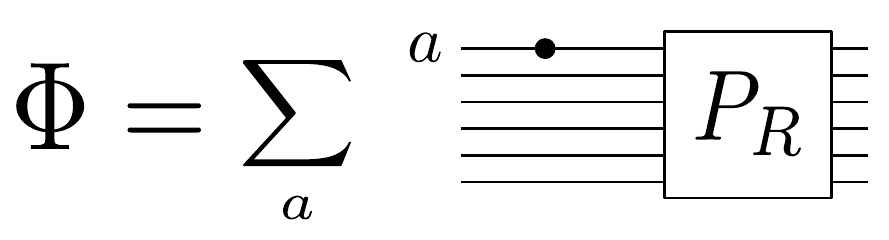}
\end{minipage}
\begin{minipage}{0.45\hsize}
\centering
\includegraphics[clip,height=1.5cm]{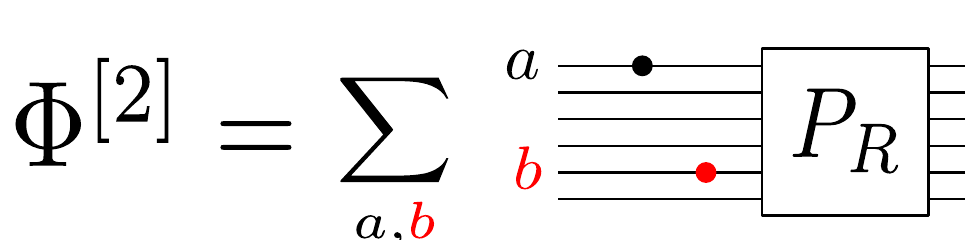}
\end{minipage}
\begin{minipage}{0.33\hsize}
\centering
\includegraphics[clip,height=1.5cm]{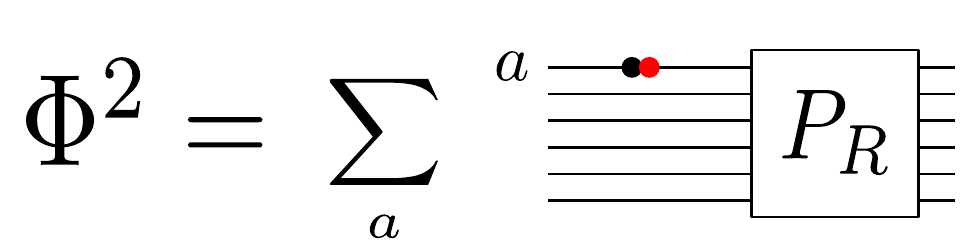}
\end{minipage}
\caption{Single-particle insertions and two-particle insertions on the higher-rank Wilson loop. The Wilson loop in the higher-rank representation can be thought of as a collection of fundamental Wilson loops (black straight lines in the figure) joined together by a projector (denoted by $P_R$ in the figure). In this representation, the insertion of a single scalar $\Phi$ is a sum over insertions of $\Phi$ (denoted by a dot) on each fundamental loop. To insert two $\Phi$'s, there are two possibilities: The first possibility is to simply consider a product of two $\Phi$'s and is given by a double sum $\Phi^{[2]}$. The other possibility is to insert $\Phi^2$ to each fundamental loop. The former operator ($\Phi^{[2]}$) is a two-particle operator while the latter ($\Phi^2$) is a single-particle operator.}\label{fig:singlemulti}
\end{figure}

The second approach is to use the multiplication rule for the fundamental Wilson loop and then insert the product inside the Wilson loop trace. Namely we take \eqref{eq:TABC2} and insert it as
\beq
{\rm tr}_{R}{\rm P}\left[\sum_{A}\Phi^{2}_AT_{A}^{R}\exp \left(\oint i A_{\mu}\dot{x}^{\mu}+\cdots\right)\right]\period
\eeq
Obviously, the two insertions $\Phi^{[2]}$ and $\Phi^{2}$ are different (except in the case of the fundamental representation). To understand their physical meaning, it is useful to represent the higher-rank Wilson loop as a collection of fundamental Wilson loops joined together by a projector to the representation $R$ (see Figure \ref{fig:singlemulti}). In this representation, the insertion of a single field $\Phi$ corresponds to a sum over insertions of $\Phi$ onto each constituent fundamental loop. Now, if we bring together two of such insertions, we obtain $\Phi^{[2]}$, which is given by a double sum as depicted in Figure \ref{fig:singlemulti}. In this case, the two insertions of $\Phi$ generally live on different fundamental loops as depicted in the figure. On the other hand, the insertion of $\Phi^2$ corresponds to directly inserting two $\Phi$'s onto each constituent fundamental loop.

This representation also provides a holographic interpretation of these operators. As mentioned above, the Giant Wilson loop is dual to a D-brane and the excitations on the brane are described by open strings attached to it. Combined with the fact that each fundamental Wilson loop represents a single string worldsheet, this suggests that the operator $\Phi^{[2]}$ corresponds to excitations of two separate strings, while the operator $\Phi^2$ corresponds to a single string excitation with higher mass. This interpretation will be justified in our paper through the comparison of the localization computation and the D-brane analysis. We will find that $\Phi^2$ and its higher charge analogs are related to ``single-particle" excitations on the D-branes, while insertions like $\Phi^{[2]}$ to multi-particle ones. In the rest of this paper, we call the operator ($\Phi^{[2]}$) a two-particle operator/insertion while we call ($\Phi^2$) a single-particle operator/insertion.  
\paragraph{Protected scalar operators and displacement operator} The main subject of this paper is the calculation of correlation functions of certain protected operator insertions on the Wilson loop. In particular, we focus on two important class of operators.

The first set of operators are made out of five scalar fields $\Phi_a$ ($a=1,\ldots, 5$)
\beq\label{eq:defOLu}
\mathcal{O}_L(\tau,{\bf u})\equiv ({\bf u}\cdot \Phi)^{L}(\tau)\comma
\eeq 
where ${\bf u}$ is a five-dimensional null vector satisfying $({\bf u}\cdot {\bf u})=0$. These operators belong to a short multiplet of the defect superconformal group and have protected scaling dimension $\Delta=L$ \cite{Giombi:2017cqn,Liendo:2018ukf}. The correaltion functions of such operators are constrained by the conformal symmetry and the R-symmetry. In particular, the two- and the three-point functions are fixed up to overall constants $n_{L_1}$ and $c_{L_1,L_2,L_3}$,
\begin{align}
\langle \!\langle \mathcal{O}_{L_1}(\tau_1,{\bf u}_1)\mathcal{O}_{L_2}(\tau_2,{\bf u}_2)\rangle \!\rangle&=n_{L_1}\frac{\delta_{L_1,L_2}({\bf u}_1\cdot {\bf u}_2)^{L_1}}{(2\sin \frac{\tau_{12}}{2})^{2L_1}}\comma\\
\langle \!\langle \mathcal{O}_{L_1}(\tau_1,{\bf u}_1)\mathcal{O}_{L_2}(\tau_2,{\bf u}_2)\mathcal{O}_{L_3}(\tau_3,{\bf u}_3)\rangle \!\rangle&=c_{L_1,L_2,L_3}\frac{({\bf u}_1\cdot {\bf u}_2)^{L_{12|3}}({\bf u}_2\cdot {\bf u}_3)^{L_{23|1}}({\bf u}_3\cdot {\bf u}_1)^{L_{31|2}}}{(2\sin \frac{\tau_{12}}{2})^{2L_{12|3}}(2\sin \frac{\tau_{23}}{2})^{2L_{23|1}}(2\sin \frac{\tau_{31}}{2})^{2L_{31|2}}}\comma\nn
\end{align}
with $\tau_{ij}\equiv \tau_i-\tau_j$ and $L_{ij|k}\equiv (L_i+L_j-L_k)/2$. Here we wrote the results for the correlators on the circular loop. The results for the straight line Wilson loop can be obtained by a simple replacement
\beq
2\sin \frac{\tau_{ij}}{2} \quad \mapsto \quad |\tau_i-\tau_j|\period
\eeq
On the other hand, the four-point functions are expressed in terms of the conformal and the R-symmetry cross ratios as
\beq
\begin{aligned}
&\langle \!\langle \mathcal{O}_{L_1}(\tau_1,{\bf u}_1)\mathcal{O}_{L_2}(\tau_2,{\bf u}_2)\mathcal{O}_{L_3}(\tau_3,{\bf u}_3)\mathcal{O}_{L_4}(\tau_4,{\bf u}_4)\rangle \!\rangle=\comma\\
&=\frac{1}{(2\sin \frac{\tau_{12}}{2})^{L_1+L_2}(2\sin \frac{\tau_{34}}{2})^{L_3+L_4}}
\biggl(\frac{\sin \frac{\tau_{24}}{2}}{\sin \frac{\tau_{14}}{2}}\biggr)^{L_1-L_2}
\biggl(\frac{\sin \frac{\tau_{14}}{2}}{\sin \frac{\tau_{13}}{2}}\biggr)^{L_3-L_4}
G(\chi,\vu)
\end{aligned}
\eeq
The function $G(\chi,\vu)$ can be further expressed as
\be
G(\chi,\vu)=(\vu_1\cdot\vu_4)^{L_4-E}(\vu_1\cdot\vu_3)^{L_3-E}
(\vu_1\cdot\vu_2)^{L_2}(\vu_3\cdot\vu_4)^{E}\mathcal{G}(\chi,\xi,\zeta)
\ee
with $2E\equiv L_2+L_3+L_4-L_1$. 
The $\chi$, $\xi$ and $\zeta$ are the cross ratios defined as
\be
\chi \equiv \frac{\sin \frac{\tau_{12}}{2}\sin \frac{\tau_{34}}{2}}{\sin \frac{\tau_{13}}{2}\sin \frac{\tau_{24}}{2}},\quad
\xi\equiv\frac{(\mathbf{u}_1\cdot\mathbf{u}_3) (\mathbf{u}_2\cdot\mathbf{u}_4)}{(\mathbf{u}_1\cdot\mathbf{u}_2)( \mathbf{u}_3\cdot\mathbf{u}_4)},\quad
\zeta\equiv\frac{(\mathbf{u}_1\cdot\mathbf{u}_4)( \mathbf{u}_2\cdot\mathbf{u}_3)}{(\mathbf{u}_1\cdot\mathbf{u}_2) (\mathbf{u}_3\cdot\mathbf{u}_4)}.
\ee
Note that on the straightline, the cross ratio is given by
\beq
\chi\equiv \frac{\tau_{12}\tau_{34}}{\tau_{13}\tau_{24}}\period
\eeq
Although the functional form of $\mathcal{G}$ cannot be fixed purely from the symmetry, the superconformal symmetry imposes the Ward identity \cite{Liendo:2018ukf}
\beq
\begin{aligned}\label{eq:superconformal_id}
&\left.\left(\del_{z_1}+\frac{1}{2}\del_{\chi}\right)\mathcal{G}\left(\chi,\frac{1}{z_1z_2},\frac{(1-z_1)(1-z_2)}{z_1 z_2}\right)\right|_{z_1=\chi}=0\comma\\
 &\left.\left(\del_{z_2}+\frac{1}{2}\del_{\chi}\right)\mathcal{G}\left(\chi,\frac{1}{z_1z_2},\frac{(1-z_1)(1-z_2)}{z_1 z_2}\right)\right|_{z_2=\chi}=0\period
\end{aligned}
\eeq
We will later check that the correlators computed on the D-brane side indeed satisfy these identities.

The other set of operators that we discuss in this paper are the displacement operators $\mathbb{F}_{t j}\equiv i F_{t j}+D_j\Phi^{6}$ along the directions $j=1,2,3$ transverse to the Wilson loop \cite{Giombi:2017cqn,Liendo:2018ukf}. They have the protected dimension $\Delta=2$ and the transverse spin $S=1 $. These operators correspond to infinitesimal deformations of the Wilson loop orthogonal to the contour. They are in the same ultrashort multiplet as $\mathcal{O}_1$ and together give eight bosonic operators (which on the D-brane side correspond to certain combinations of the fluctuations in the eight directions transverse to $AdS_2$ and of the worldvolume gauge field excitations). 
\paragraph{Comparison of the protected spectrum at weak and strong coupling} In addition to $\mathcal{O}_1$ and $\mathbb{F}_{t j}$, there is an infinite set of protected single-particle operators with higher $R$-charge $\mathcal{O}_L$ $(L\geq 2)$. For the D5-brane, which is dual to the antisymmetric loop, there are natural candidates of their holographic dual: Since the D5-brane is extended in $S^4$ inside $S^5$, it has infinitely many Kaluza-Klein modes upon reducing to $AdS_2$ \cite{Harrison:2011fs,Faraggi:2011ge}. They have integer angular momenta (dual to $R$-charges) and are natural candidates for $\mathcal{O}_L$.
 
The situation is quite different for the D3-brane. Since the D3-brane is point-like on $S^5$, it does not have the Kaluza-Klein modes with higher angular momenta on $S^5$ \cite{Faraggi:2011bb}. The only excitations that have higher angular momenta are then multi-particle states. However, from the discussions above, we expect that $\mathcal{O}_L$ is dual to a single-particle state. This poses a sharp puzzle: On the gauge theory side, we have an infinite set of protected operator $\mathcal{O}_L$'s but they seem to be absent on the D-brane side. One of the aim of this paper is to resolve this apparent puzzle: We perform the explicit computation based on the supersymmetric localization and show that the operators $\mathcal{O}_L$ with $L\geq 2$ do exist in the spectrum of the Wilson loop defect CFT dual to the D3-brane, but their couplings to $\mathcal{O}_1$ are exponentially suppressed at strong coupling. This explains why all these higher charge operators could not be seen on the D-brane side. At the mathematical level, this decoupling is realized by a mechanism analogous to the Bose-Einstein condensation as we see in section \ref{subsec:symmetricloop}.

Note that a similar puzzle exists also for the higher transverse spin operators that arise from products of the displacement operator $\mathbb{F}_{t j}$. The D3-brane dual to the symmetric representation is extended in the $S^2$ subspace inside $AdS_5$. Therefore, it has infinitely many single-particle excitations on $AdS_2$ that have higher $AdS$ angular momenta \cite{Faraggi:2011bb}. Natural candidates for such operators on the gauge theory side are products of the displacement operators $\left(\mathbb{F}_{t j}\right)^{S}$ inserted on the Wilson loop, which indeed exist at weak coupling. On the other hand, such excitations are absent in the D5-brane since it is not extended in the directions transverse to $AdS_2$ inside $AdS_5$. Therefore we again have an apparent paradox, now with the roles of the D3-brane and the D5-brane exchanged. However, this puzzle is not as sharp as the one mentioned earlier since the operators $\left(\mathbb{F}_{t j}\right)^{S}$ are not protected and they can disappear from the spectrum at strong coupling  simply by acquiring infinite anomalous dimensions. In addition, since they are not protected, they cannot be studied by the localization analysis which we perform in this paper. It would be an interesting future problem to understand the fate of these operators at strong coupling using other nonperturbative techniques such as integrability or conformal bootstrap.
\subsection{$1/8$ BPS Wilson loops and topological sector\label{subsec:eighthBPS}}
The defect CFT defined by the $1/2$-BPS Wilson loop contains a supersymmetric subsector whose correlation functions are position-independent \cite{Giombi:2009ds,Giombi:2012ep,Giombi:2018qox,Giombi:2018hsx,Giombi:2020pdd,Wang:2020seq}. For the Wilson loops in the fundamental representation, such correlators were computed exactly using the supersymmetric localization\fn{Recently the localization computation \cite{Nekrasov:2002qd,Pestun:2007rz} was extended to a large class of observables that include various kinds of defects and the correlation functions on $\mathbb{R}\mathbb{P}^4$ in \cite{Wang:2020seq,Wang:2020jgh}. The formalism was then applied to the D5-brane defect one-point functions in \cite{Komatsu:2020sup}.} in \cite{Giombi:2018qox,Giombi:2018hsx}. The results provide non-perturbative defect CFT data, which are important inputs for the conformal bootstrap analysis \cite{Billo:2016cpy,Liendo:2018ukf}. 

\paragraph{1/8 BPS Wilson loops and 2d Yang-Mills}One of the goals of this paper is to extend the aforementioned analysis to the Wilson loops in higher-rank representations. For this purpose, it is useful to first consider a broader class of supersymmetric Wilson loops which are $1/8$ BPS. They can be defined on a arbitrary contour $C$ on a $S^2$ subspace inside $R^4$ (or $S^4$) in the following way:
\beq
\mathcal{W}_{1/8}=\frac{1}{N}{\rm tr}_{R}\,{\rm P} \exp \left[\oint_{C}\left(iA_j+\epsilon_{kjl}x^{k}\Phi^{l}\right)dx^{j}\right]\qquad (i,j,k=1,2,3)\,.
\eeq
Here $x_i$'s are the embedding coordinates of $S^2$ of unit radius, $x_1^{2}+x_2^2+x_3^{2}=1$. Thanks to the specific choice of the coupling to the scalars $\Phi_i$'s, they preserve four supercharges in general. If the contour is placed along the great circle of $S^2$, it preserves sixteen supercharges and becomes half-BPS.

An advantage of studying this specific class of supersymmetric Wilson loops is their equivalence to the two-dimensional Yang-Mills theory (2d YM) in the zero-instanton sector: It was first conjectured based on perturbation theory and AdS/CFT \cite{Drukker:2007yx,Drukker:2007qr} and later derived from the supersymmetric localization \cite{Pestun:2009nn} that the expectation value of the $1/8$ BPS Wilson loops coincides with that of the standard Wilson loops in 2d YM\fn{The equivalence to the 2d YM was later tested extensively against various perturbative computations \cite{Giombi:2009ms,Giombi:2009ek,Giombi:2009ds,Giombi:2012ep,Bassetto:2009rt,Bassetto:2009ms,Bonini:2014vta,Bonini:2015fng}.} defined on the same contour,
\beq
\mathcal{W}_{1/8}\qquad \longleftrightarrow \qquad \mathcal{W}_{\rm 2dYM}\equiv \frac{1}{N}{\rm tr}_{R}{\rm P}\exp \left(\oint_{C}i A_j dx^{j}\right)\comma
\eeq
under the identification of the coupling constants,
\beq
g_{\rm 2d}^2=-\frac{g_{\rm YM}^2}{2\pi}\period
\eeq

\begin{figure}[t]
\centering
\includegraphics[clip,height=3.5cm]{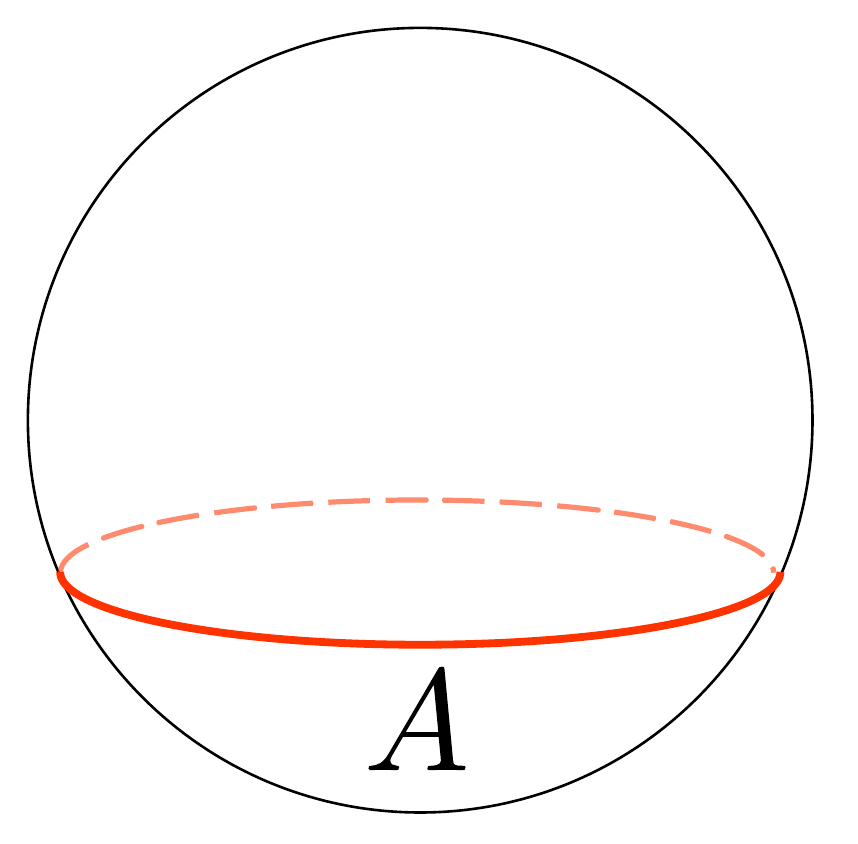}
\caption{$1/8$ BPS Wilson loop on $S^2$. The Wilson loop (denoted by a thick red curve) divides the $S^2$ into two regions, one with area $A$ and the other with area $4\pi-A$.}\label{fig:loop1}
\end{figure}

Based on this equivalence, the expectation values of the $1/8$ BPS Wilson loops can be computed exactly and expressed in terms of simple matrix integrals which we review in section \ref{sec:integral}. Solving the matrix models in the large $N$ limit, one obtains the following expressions for the Wilson loops in the antisymmetric ($\mathcal{W}_{{\sf A}_k}$) or the symmetric ($\mathcal{W}_{{\sf S}_k}$) representations in the planar limit,
\beq
\begin{aligned}
\left.\langle \mathcal{W}_{{\sf A}_k}\rangle\right|_{N\to \infty} 
=&\oint \frac{dz}{2\pi i z^{k+1}}\exp\left[\frac{2N}{\pi}\int_{-1}^{1} ds\sqrt{1-s^2}\log \left(1+ze^{-\sqrt{\lambda(1-a^2)} s}\right)\right]\comma\\
\left.\langle \mathcal{W}_{{\sf S}_k}\rangle\right|_{N\to \infty} 
=&\oint \frac{dz}{2\pi i z^{k+1}}\exp\left[-\frac{2N}{\pi}\int_{-1}^{1} ds\sqrt{1-s^2}\log \left(1-ze^{-\sqrt{\lambda (1-a^2)} s}\right)\right]\comma
\label{W-LargeN}
\end{aligned}
\eeq
where $\lambda =g_{\rm YM}^2 N$ is the 't Hooft coupling, $A=2\pi(1+a)$ is the area of the region inside the Wilson loop on $S^2$ (see Figure \ref{fig:loop1}), and $k$ is the size of the representation.
They are related to the results for the $1/2$-BPS Wilson loops computed in \cite{Hartnoll:2006is} by a simple rescaling of the coupling constant, $\lambda \to \lambda (1-a^2)$.

\paragraph{Topological correlators on the Wilson loop} In addition to the expectation values of the Wilson loops, there are other observables that preserve a fraction of supersymmetry and therefore can be computed by 2d YM. The ones relevant in this paper are the following correlation functions of scalar fields inside a Wilson loop trace,
\beq
\mathcal{W}[\NO{\tilde{\Phi}^{L_1}}\NO{\tilde{\Phi}^{L_2}}\cdots \NO{\tilde{\Phi}^{L_n}}]\equiv \frac{1}{N}{\rm tr}_{R}{\rm P}\left[\NO{\tilde{\Phi}^{L_1}(\tau_1)} \cdots \NO{\tilde{\Phi}^{L_n} (\tau_n)}e^{\oint_{C}\left(iA_j+\epsilon_{kjl}x^{k}\Phi^{l}\right)dx^{j}}\right]\period
\eeq
Here $\tilde{\Phi}$ is a position-dependent linear combination of the scalars
\beq
\tilde{\Phi}(x)=x_1\Phi^{1}+x_2\Phi^2+x_3\Phi^3 +i \Phi_4\comma
\eeq
and $\tilde{\Phi}^{L}$ is a single-particle insertion made out of $L$ such fields. We used a normal ordering symbol $\NO{\bullet}$ to emphasize the absence of the self-contractions within each operator. One important feature of these correlation functions is their position-independence, which follows from the fact that the spatial translation of $\tilde{\Phi}(x)$ is $Q$-exact \cite{Giombi:2009ds,Wang:2020seq}. In the rest of this paper, we often denote these operators by
\beq
\tilde{\mathcal{O}}_L\equiv\,\, :\!\!\tilde{\Phi}^{L}\!\!:\period
\eeq
When the Wilson loop is circular and preserves the $1/2$-BPS supersymmetry, they can be obtained from the scalar insertions $\mathcal{O}_L$ in \eqref{eq:defOLu} by setting the polarization ${\bf u}=(\cos \tau,\sin\tau,0,0)$, where $\tau\in [0,2\pi]$ is the position of the operator on the circle. This connection allows us to extract the defect CFT data from the topological correlators, see e.g.~section 2.3 of \cite{Giombi:2018hsx} for more details.

The simplest class of such correlators are the correlation functions of the insertions of a single scalar. They are known to correspond to the insertions of a dual field strength of the two-dimensional Yang-Mills theory \cite{Pestun:2009nn}, 
\beq
\tilde{\Phi}\quad \leftrightarrow\quad i\ast F_{\rm 2d}\comma
\eeq
which in turn is related to an infinitesimal deformation of the contour of the Wilson loop. Thanks to this correspondence, we can compute the correlators of multiple $\tilde{\Phi}$'s by taking the area derivatives of the Wilson loop expectation value,
\beq
\langle \mathcal{W}[\,\underbracket{\tilde{\Phi}\cdots \tilde{\Phi}}_{n}\,]\rangle=\frac{\del^{n}\langle \mathcal{W}\rangle}{(\del A)^{n}}\period
\eeq
For the fundamental Wilson loops, it was demonstrated in \cite{Giombi:2018qox} that the insertion of higher-charge operators $\mathcal{W}_{f}[\NO{\tilde{\Phi}^{L}}]$ can also be related to the area derivatives. The basic idea of the computation is as follows: By taking the $n$-th area derivatives, one can insert $n$ scalars $\tilde{\Phi}$ on the Wilson loop. Since the correlation functions do not depend on the positions of the insertion, we can bring all the scalars close to a single point and build the insertion of $\tilde{\Phi}^{L}$,
\beq
(\del_A)^{n} \sim \tilde{\Phi}^{n}\period
\eeq
 However, the insertion constructed in this way would contain self-contractions and is not normal-ordered. In order to define the normal-ordered operators $\NO{\tilde{\Phi^{n}}}$, we then perform the Gram-Schmidt orthogonalization.  

\begin{figure}[t]
\centering
\includegraphics[clip, height=2.5cm]{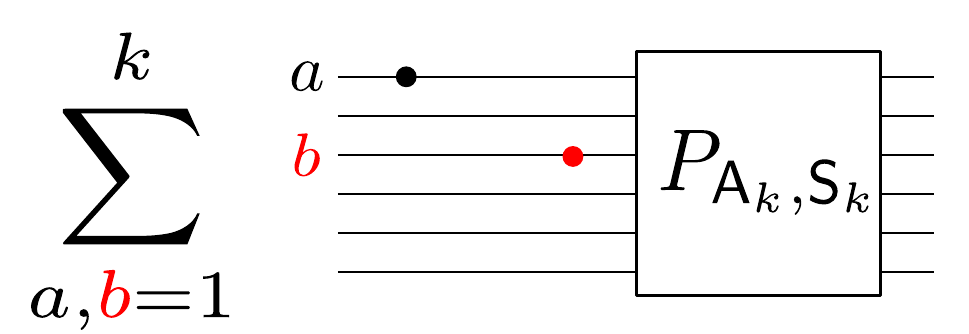}
\caption{The operator that can be obtained by brining two $\tilde{\Phi}'s$ together. It consists of $k^2$ terms and only $k$ of them contains two $\tilde{\Phi}$'s on the same fundamental loop. Therefore, the leading term in the OPE is given by a two-particle operator, not a single-particle operator of $\tilde{\Phi}^2$.}\label{fig:singlepGS}
\end{figure}

Unfortunately, these procedures do not work straightforwardly for the Giant Wilson loops. Although it is still true that a single area derivative $\del_{A}$ corresponds to a single-particle insertion $\tilde{\Phi}$, we cannot get a single-particle insertion of $\tilde{\Phi}^{L}$ just by bringing together $L$ $\tilde{\Phi}$'s. To understand this, it is again useful to represent the  Giant Wilson loop as a collection of $k$ fundamental Wilson loops joined together by a projector to a particular representation (see Figure \ref{fig:singlepGS}). In this representation, the insertion of $\tilde{\Phi}$ on the Giant Wilson loop is given by a sum of $k$ terms, each of which corresponds to an insertion of $\tilde{\Phi}$ to one of the $k$ fundamental loops. Now, if we bring two $\tilde{\Phi}$'s together, we then get $k^2$ terms. Among these $k^2$ terms, $k$ of them contain two insertions of $\tilde{\Phi}$'s on the same fundamental Wilson loop and correspond to single-particle insertions of $\tilde{\Phi}^2$. However, their contributions are always suppressed as compared to the other $k(k-1)$ terms when $k$ is of order $N$. This is completely analogous to the operator product expansion (OPE) of single-trace operators in the large $N$ CFTs, where the leading terms in the OPE in the large $N$ limit are given by double-trace operators and the contributions from single-trace operators are suppressed by powers of $1/N$.

In the following two sections, we develop techniques to overcome this problem. The idea is to consider a generalization of the Giant Wilson loop, to be called the ``generalized'' higher-rank Wilson loop, in which each constituent fundamental loop is coupled to a different area ($A_j$, $j=1,\ldots, k$). We can then define the following area derivative,
\beq\label{eq:secondordercorrect}
\sum_{j=1}^{k}(\del_{A_j})^n\comma
\eeq
which consists only of $k$ terms and directly inserts $\tilde{\Phi}^{n}$ to each fundamental loop. Although the insertion $\tilde{\Phi}^{n}$ is not normal-ordered in general, this can be remedied by the application of the Gram-Schmidt process.

Note that \eqref{eq:secondordercorrect} is genuinely different from taking multiple area derivatives of the standard higher-rank Wilson loop which couples to a single area, since that would amount to considering
\beq
(\del_{A})^n\sim \left(\sum_{j=1}^{k}\del_{A_j}\right)^n\comma
\eeq
and corresponds to a multi-particle insertion if $k$ is of order $N$.
\section{Multiple Integral Representation of the 1/8 BPS Wilson Loops\label{sec:integral}}
In this section, we discuss a representation \cite{Giombi:2012ep, Giombi:2020pdd} for the expectation value of the BPS Wilson loop in which the area dependence appears only in the exponent\footnote{In a more standard representation \cite{Drukker:2007yx,Drukker:2007qr}, the expectation value is given by a ratio of two different partition functions, with and without an insertion of the Wilson loop, and each partition function is a nontrivial function of the area.}. Using such a representation and the loop equation in 2d Yang-Mills, new results for intersecting Wilson loops were derived in \cite{Giombi:2020pdd}. Below we review its derivation for the fundamental Wilson loops and generalize it to the higher-rank Wilson loops. We also explain how to analyze the large $N$ limit systematically using ideas from  the Fermi Gas approach \cite{Marino:2011eh} and the Clustering method \cite{Jiang:2016ulr}. After that, we extend those results to the case of generalized higher-rank Wilson loops that couple to different areas. We will then use this construction in section \ref{sec:correlatorfromloc} to derive exact results for defect CFT correlators on the higher-rank Wilson loops. 

\subsection{Partition function and fundamental loops\label{subsec:partition}}
The correlation functions of non-intersecting $1/8$ BPS Wilson loops defined on $S^2$ subspace of $R^4$ (or $S^4$) can be computed by a multi-matrix model given in (3.30) of \cite{Giombi:2012ep}. After appropriate rescaling of the matrices, the action of the matrix model reads
\beq\label{eq:matrixaction}
S =\sum_{\Sigma_m} \left[\frac{2\pi A_{\sigma_m}}{g_{\rm YM}^2}{\rm tr}\left(B^2_{\Sigma_m}\right)-i\sum_{j\in \del \Sigma_m}s_j^{(m)}{\rm tr}\left(X_j B_{\Sigma_m}\right)\right]\period
\eeq
Here $\Sigma_m$'s denote different regions on $S^2$ bordered by the Wilson loops, $g_{\rm YM}$ is the gauge coupling of $\mathcal{N}=4$ SYM and $s_j^{(m)}$ are the orientation factors which take $\pm 1$ depending on the relative orientation of the loop and the boundary $\del \Sigma_m$. To compute the expectation values of the Wilson loops, we simply evaluate the expectation values of ${\rm tr}_{R}(e^{\epsilon X})$ where $\epsilon$ is given by
\beq
\epsilon\equiv \frac{g_{\rm YM}^2}{4\pi}=\frac{\lambda}{4\pi N}=\frac{4\pi g^2}{N}\period
\eeq
Here $\lambda =g_{\rm YM}^2 N$ is the standard 't Hooft coupling constant while
\beq
g^2\equiv \frac{\lambda}{16\pi^2}\comma
\eeq
 is the convention for the coupling constant commonly used in the integrability literature. 

The action \eqref{eq:matrixaction} can be viewed as a matrix-model analogue of the BF-theory representation of the 2d Yang-Mills theory: Namely, we can derive \eqref{eq:matrixaction} from the action of the 2d Yang-Mills by identifying $B_{\Sigma_m}$ and $X_{j}$ with constant modes of $B$ and $\ast F$ respectively. See \cite{Giombi:2012ep,Giombi:2020pdd} for more details. 

When there is only one fundamental Wilson loop, the action simplifies to
\beq\label{eq:actionfund}
S=\frac{2\pi A_0}{g_{\rm YM}^2} {\rm tr}(B_0^2)+\frac{2\pi A_1}{g_{\rm YM}^2} {\rm tr}(B_1^2)-i{\rm tr}\left(X(B_0-B_1)\right)\period
\eeq
Here $A_{0}$ and $A_1$ are the areas of the two regions separated by the Wilson loop. In the convention of Figure \ref{fig:loop1}, they read $A_0=4\pi -A$ and $A_1=A$.
In \cite{Giombi:2012ep}, this matrix model was solved by first integrating out $B$ fields and reducing it to a Gaussian matrix model. To derive the representation in \cite{Giombi:2020pdd}, we instead integrate out $X$ and reduce it to a matrix model of the $B$ fields. The integration of the $X$ field can be performed by the use of the Harish-Chandra-Itzykson-Zuber integral,
\beq
\int d\Omega\,\, e^{i {\rm tr}\left[\Omega^{\dagger}A\Omega B\right]}=\frac{\det e^{ia_ib_j}}{\Delta (a)\Delta(b)}\comma
\eeq 
where $A$ and $B$ are diagonal matrices with eigenvalues $a_j$'s and $b_j$'s, $\Omega$ is an element of the unitary group, and $\Delta$ is a Vandermonde factor $\Delta (a)\equiv \prod_{i<j}(a_i-a_j)$. Applying the formula, we obtain the following eigenvalue integrals for the expectation value of the fundamental Wilson loop:
\beq\label{eq:fundamentalinsertion}
\begin{aligned}
\langle \mathcal{W}_f\rangle=\frac{\int d^{N}a \,d^{N}b \,d^{N}x \,\,\mu(a,b,x)\,\frac{1}{N}\sum_{k}e^{\epsilon x_k}}{\int d^{N}a \,d^{N}b \,d^{N}x\,\, \mu(a,b,x) } \comma
\end{aligned}
\eeq
where the measure $\mu(a,b,x)$ is given by 
\beq
\begin{aligned}
\label{calI}
\mu(a,b,x) &= \Delta (a)\Delta (b)\det e^{i x_i a_j}\det e^{-i x_k b_l} \,\,e^{-\frac{2\pi}{g_{\rm YM}^2}\sum_{j} (A_0a_j^2+A_1b_j^2)}\period
\end{aligned}
\eeq

\paragraph{Partition function} Let us first consider the partition function without operator insertion
\beq
Z=\int d^{N}a \,d^{N}b \,d^{N}x\,\, \mu(a,b,x) \period
\eeq
By expanding the determinant
\beq\label{eq:expanddet}
\begin{aligned}
\det e^{ia_ix_j}=\sum_{\sigma \in S_N}(-1)^{\sigma}\prod_{j}e^{ia_{\sigma_j}x _j}\comma\qquad \det e^{-ib_kx_l}=\sum_{\sigma^{\prime} \in S_N}(-1)^{\sigma^{\prime}}\prod_{j}e^{-ib_{\sigma^{\prime}_j}x _j}\comma
\end{aligned}
\eeq
and performing the integral, we get the Gaussian matrix model
\beq
\begin{aligned}
Z&=\!\!\sum_{\sigma,\sigma^{\prime} \in S_{N}}(-1)^{\sigma+\sigma^{\prime}}\!\!\int d^{N} a \,d^{N}b\,\,\left(\prod_{j}2\pi \delta (a_{\sigma_j}-b_{\sigma^{\prime}_j})\right)\Delta (a)\Delta(b)e^{-\frac{2\pi }{g_{\rm YM}^2}\sum_{j}(A_0a_j^2+A_1b_j^2)}\comma\\
&=(2\pi)^{N}\left(N!\right)^2\int d^{N}a\,\Delta^2 (a)e^{-\frac{8\pi^2 }{g_{\rm YM}^2}\sum_{j}a_j^2}\period
\end{aligned}
\eeq
In the second line, we used $A_0+A_1=4\pi$, and the permutation symmetry of the Vandermonde factor $\Delta (a_{\sigma})=(-1)^{\sigma}\Delta (a)$.

\paragraph{Fundamental Wilson loop} We now consider the insertion of a fundamental Wilson loop \eqref{eq:fundamentalinsertion}. This can be evaluated in a similar manner by expanding the determinants as in \eqref{eq:expanddet} and integrating out $x_i$'s. The only difference is that one of the delta function now gets shifted by $-i\epsilon$ because of the insertion $e^{\epsilon x_k}$. As a result we get
\beq\label{eq:singlefundsum}
\begin{aligned}
\langle \mathcal{W}_f\rangle=&\frac{(2\pi)^{N}(N!)^2}{Z }\int d^{N} a \,d^{N}b\,\Delta (a)\Delta (b)e^{-\frac{2\pi}{g_{\rm YM}^2}\sum_{j} (A_0a_j^2+A_1b_j^2)}\\
&\times \frac{1}{N}\sum_k\delta (a_{k}-b_{k}-i\epsilon)\left(\prod_{j\neq k}\delta (a_{j}-b_{j})\right)\\
=&\frac{(2\pi)^{N}(N!)^2}{Z N}\int d^{N}a\,\, \Delta^2 (b)e^{-\frac{8\pi^2}{g_{\rm YM}^2}\sum_j (a_j)^2}\sum_k e^{i A_1 (a_k-\frac{i\epsilon}{2})}\prod_{j\neq k}\frac{a_k-a_j-i\epsilon}{a_k-a_j}\period
\end{aligned}
\eeq
Next we rewrite the sum $\sum_k$ in terms of a contour integral
\beq\nn
\begin{aligned}
\langle \mathcal{W}\rangle&=\frac{(2\pi)^{N}(N!)^2}{Z}\int d^{N}a\,\, \Delta^2 (a)e^{-\frac{8\pi^2}{g_{\rm YM}^2}\sum_j (a_j)^2}\left[\oint_{\mathcal{C}}\frac{du}{8\pi^2 g^2}e^{i A_1(u-\frac{i\epsilon}{2})}\prod_{j}\frac{u-a_j-i\epsilon}{u-a_j}\right]\period
\end{aligned}
\eeq
Here the integration contour $\mathcal{C}$ encircles all the eigenvalues $b_k$'s.
This can be further re-expressed as an expectation value of an operator
\beq\label{eq:deffA}
f_{A}(u)\equiv e^{i A(u-\frac{i\epsilon}{2})}\det \left[\frac{u-M-i\epsilon}{u-M}\right]\period
\eeq
in the Gaussian matrix model with the action $S_M:=8\pi^2{\rm tr}\left(M^2\right)/g_{\rm YM}^2$:
\beq\label{eq:singleresult}
\langle \mathcal{W}_f\rangle=\left<\oint_{\mathcal{C}} \frac{du}{8\pi^2g^2} f_{A_1}(u)\right>_M\period
\eeq
Here and below $\left<\bullet\right>_M$ denotes the following expectation value
\beq
\left<\bullet \right>_M:=\frac{\int [dM]\,\bullet\, e^{-S_M}}{\int [dM]\,e^{-S_M}}\period
\eeq
The representation \eqref{eq:singleresult} is exact at finite $N$. 

\paragraph{Large $N$ limit} In the large $N$ limit, $\epsilon \equiv \frac{4\pi g^2}{N}$ becomes small. In this limit we can approximate the expectation value of the determinant \eqref{eq:deffA} as
\beq
\left< \det \left[1-\frac{i \epsilon}{u-M}\right]\right> \sim\exp\left[-i\frac{4\pi g^2}{N}\left<{\rm tr}\frac{1}{u-M} \right>\right]=e^{-4\pi i g^2 G(u)}\comma
\eeq
with $G(u)$ being the planar resolvent
\beq
G(u)=\frac{1}{2g^2}\left(u-\sqrt{u^2-4g^2}\right)=\frac{i}{ g x(u)}\period
\eeq
Here $x (u)$ is the Zhukovsky variable defined by
\beq\label{eq:defzhukowsky}
u=-ig(x-1/x)\quad \iff \quad x(u)=i\frac{u+\sqrt{u^2-4g^2}}{2g}\period
\eeq

\begin{figure}[t]
\centering
\includegraphics[clip, height=3.5cm]{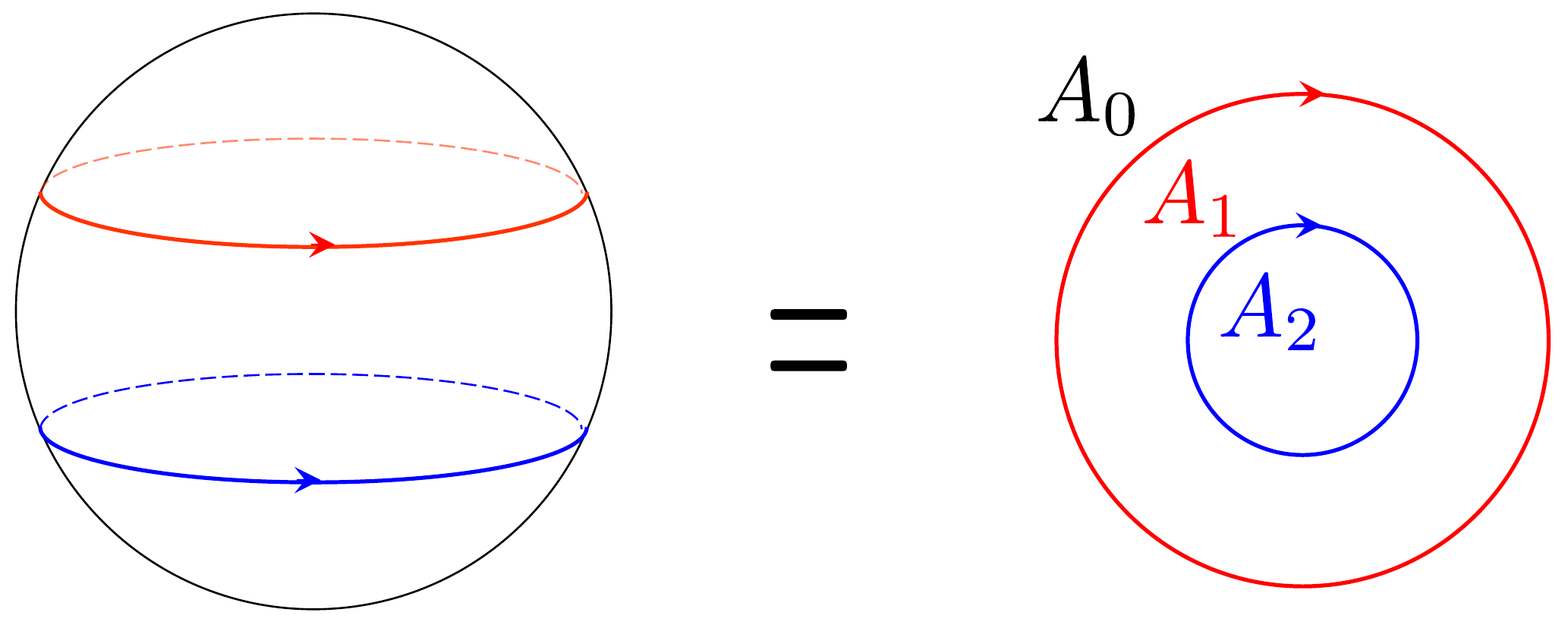}
\caption{The correlation function of multiple fundamental Wilson loops. When viewed from the south pole, it can be represented as multiple concentric loops as shown on the right hand side of the figure. We count the loops from the north pole to the south pole and denote the area inside the $j$-th loop by $A_j$.}\label{fig:multiloop}
\end{figure}

As a result, we get\fn{Note that $e^{\frac{g_{\rm YM}^2 A_2}{8\pi^2}}\sim 1$ in the large $N$ limit.}
\beq
\langle \mathcal{W}_f\rangle_{N\to \infty}=\oint\frac{du}{8\pi^2 g^2}\mathfrak{f}_{A_1}(u)=\frac{1}{4\pi g}\oint \frac{dx (x+x^{-1})}{2\pi i x}\mathfrak{f}_{A_1}(u)\comma
\eeq
with
\beq\label{eq:deffrakf}
\mathfrak{f}_{A}(u)\equiv e^{iA u-4\pi i g^2 G(u)}=e^{2g\pi (x+1/x)}e^{2 g a (x-1/x)}\comma\qquad \qquad a\equiv \frac{A-2\pi}{2}\period
\eeq
This reproduces the integral representation given in \cite{Giombi:2018qox}.
\paragraph{Multiple fundamental Wilson loops} As shown in \cite{Giombi:2020pdd}, the integral representation \eqref{eq:singleresult} can be extended to the correlation function of multiple fundamental Wilson loops with the same orientation. Each Wilson loop $\mathcal{W}_j$ divides the sphere into two regions and we denote the area of the lower region by $A_j$ (see Figure \ref{fig:multiloop}).  Since the derivation is given in \cite{Giombi:2020pdd}, here we simply quote the result:
\beq\label{eq:genmultiply}
\left< \prod_{j=1}^{n}\mathcal{W}_j\right>=\left<\oint_{\mathcal{C}_1\prec\cdots \prec\mathcal{C}_n}\prod_{j=1}^{n}\frac{du_j f_{A_j}(u_j)}{8\pi^2 g^2}  \prod_{j<k}\bDelta (u_j,u_k)\right>_{M}\period
\eeq
Here $\bar{\Delta}(u,v)$ is given by
\beq
\bar{\Delta}(u,v)\equiv \frac{(u-v)^2}{(u-v)^2+\epsilon^2}\comma
\eeq
and the notation $\mathcal{C}_1 \prec\mathcal{C}_2$ means that the contour $\mathcal{C}_1$ is inside the contour $\mathcal{C}_2$ and they are far apart from each other. We will use this representation when we discuss the generalized higher-rank Wilson loops in section \ref{subsec:generalized}.

\subsection{Antisymmetric representation\label{subsec:antisymintegral}}
We now generalize the integral representation to the Wilson loop in the $k$-th antisymmetric representation. At the level of the eigenvalue integrals, we simply need to replace $\sum_ie^{x_i}/N$ in \eqref{eq:fundamentalinsertion} with
\beq
 \frac{\sum_je^{\epsilon x_j}}{N}\quad \mapsto \quad \frac{1}{d_{{\sf A}_k}}\sum_{1\leq j_1<\cdots <j_k\leq N} e^{\epsilon\sum_{a=1}^{k}x_{j_a}}\comma
 \eeq
 where $d_{{\sf A}_k}$ is the dimension of the $k$-th antisymmetric representation $d_{{\sf A}_k}\equiv \frac{N!}{k!(N-k)!}$. The derivation in the previous subsection can be applied almost straightforwardly to this case, the only difference being that $k$ (instead of one) eigenvalues of $B_1$ get shifted by $-i\epsilon$. As a result we obtain
 \beq\label{eq:summationAk}
 \begin{aligned}
 \langle \mathcal{W}_{{\sf A}_k}\rangle=&\frac{(2\pi)^{N}(N!)^2}{Z d_{{\sf A}_k}}\int d^{N}a\,\Delta^2 (b)e^{-\frac{8\pi^2}{g_{\rm YM}^2}\sum_j (a_j)^2}\\
 &\times 
\!\!\sum_{\substack{\alpha_0\cup \alpha_1 = \{1,\cdots, N\}\\|\alpha_1|=k}}e^{ i A_1 \sum_{j\in \alpha_1}(a_j-i\frac{\epsilon}{2})}\prod_{\substack{n\in \alpha_1\\m\in \alpha_0}}\frac{a_n-a_m-i\epsilon}{a_n-a_m}\comma
 \end{aligned}
 \eeq
 where the sum is over all possible ways of partitioning $\{1,\ldots, N\}$ into two subsets $\alpha_0$ and $\alpha_1$ with $N-k$ and $k$ elements respectively ($|\alpha_0|=N-k$, $|\alpha_1|=k$). Physically $\alpha_1$ corresponds to shifted eigenvalues while $\alpha_0$ corresponds to those that are not shifted.  The summation in \eqref{eq:summationAk} resembles the sum over partitions that arises in the hexagon approach to the three-point functions \cite{Basso:2015zoa}; see for instance (3.9) and (3.10) in \cite{Jiang:2016ulr}. Just as in that case, we can express it as multiple contour integrals,
 \beq\nonumber
\sum_{\substack{\alpha_0 \cup \alpha_1= \{1,\cdots, N\}\\|\alpha_1|=k}}e^{i A_1 \sum_{l\in \alpha_1}(a_l-i\frac{\epsilon}{2})}\prod_{\substack{n\in \alpha_1\\m\in \alpha_0}}\frac{a_n-a_m-i\epsilon}{a_n-a_m}=\frac{N^{k}}{k!}\oint_{\mathcal{C}}\prod_{j=1}^{k}\frac{du_j F (u_j)}{8\pi^2 g^2}\prod_{n<m}\bDelta (u_n,u_m)\comma
 \eeq
 with
 \beq
 F(u)\equiv e^{iA_1 (u-\frac{i\epsilon}{2})}\prod_{j}\frac{u-a_j-i\epsilon}{u-a_j}\period
 \eeq
 We can then rewrite \eqref{eq:summationAk} as an expectation value in the Gaussian matrix model:
 \beq\label{eq:antisymfinal}
 \langle \mathcal{W}_{{\sf A}_k}\rangle=\left<\frac{N^{k}}{d_{A_k}k!}\oint_{\mathcal{C}} \prod_{j=1}^{k}\frac{du_j f_{A_1} (u_j)}{8\pi^2 g^2}\prod_{n<m}^{k}\bDelta (u_n,u_m)\right>_M
 \eeq
 Note that, although the integrand coincides with the one for the correlator of multiple fundamental loops \eqref{eq:genmultiply}, the integration contours are different: Unlike in \eqref{eq:genmultiply}, the integration contours $\mathcal{C}$ in \eqref{eq:antisymfinal} are all on top of each other. If one tries to deform these contours into the ones in \eqref{eq:genmultiply}, there will be additional contributions from the poles in the interaction term $\bar{\Delta}(u_n,u_m)$, which make the result different from \eqref{eq:genmultiply}.
 \paragraph{Generating function and Fredholm determinant} To analyze the large $N$ limit of the Wilson loop in a large-rank representation, it is often convenient to  consider the generating function of all the antisymmetric representations,
 \beq\label{eq:generatingantisym}
 \mathcal{Z}_{\rm anti}(z)\equiv \sum_{k=0}^{N}z^{k}d_{A_k}\mathcal{W}_{{\sf A}_k}\comma
 \eeq
 from which one can recover the result for a fixed representation by
 \beq
 \langle \mathcal{W}_{{\sf A}_k}\rangle=\frac{1}{d_{{\sf A}_k}}\oint \frac{dz}{2\pi i z^{k+1}}\langle \mathcal{Z}_{\rm anti}(z)\rangle\period
 \eeq
 From \eqref{eq:antisymfinal}, we can derive an integral representation for the generating function
 \beq\label{eq:Zinteractinganti}
 \langle \mathcal{Z}_{\rm anti}(z)\rangle=\left<\sum_{k=0}^{\infty}\frac{z^{k}N^{k}}{k!}\oint_{\mathcal{C}} \prod_{j=1}^{k}\frac{du_j f_{A_1} (u_j)}{8\pi^2 g^2}\prod_{n<m}\bDelta (u_n,u_m)\right>_{M}\period
 \eeq
 Here we extended the upper bound of the summation from $N$ to $\infty$ without changing the result: Owing to the factor $(u_n-u_m)^2$ in $\bar{\Delta}(u_n,u_m)$, all the integration variables need to take different values. However since the integrals of $u_m$'s  contain only $N$ distinct poles, the terms with $k>N$ all vanish. 
  
 We can further simplify this expression by rewriting the interaction term using the Cauchy determinant identity\fn{The Cauchy determinant identity is given by
 \beq
 \frac{\prod_{i<j}(x_i-x_j)(y_i-y_j)}{\prod_{i,j}(x_i-y_j)}=\det \frac{1}{x_i-y_j}\period
 \eeq
 }
 \beq
 \prod_{n<m}\frac{(u_n-u_m)^2}{(u_n-u_m)^2+\epsilon^2}=\det \left(\frac{i \epsilon}{u_n-u_m+i\epsilon}\right)\period 
 \eeq
 We then get
 \beq\label{eq:antiexpanded}
 \langle \mathcal{Z}_{{\rm anti}}(z)\rangle=\left<\sum_{k=0}^{\infty}\frac{z^{k}}{k!}\oint_{\mathcal{C}}\prod_{j=1}^{k}\frac{du_j f_{A_1} (u_j)}{2\pi}\det \left(\frac{i}{u_n-u_m+i\epsilon}\right)\right>_M\period
 \eeq
This can be identified with the expansion of the following Fredholm determinant\fn{One can verify this by expanding \eqref{eq:antifredholm} and comparing it with \eqref{eq:antiexpanded}. See also \cite{Jiang:2016ulr} for details of the identification.}
 \beq\label{eq:antifredholm}
 \langle \mathcal{Z}_{\rm anti}(z)\rangle =\left<{\rm Det} \left(1+z\mathcal{K}\right)\right>_M\comma
 \eeq
 where ${\rm Det}$ denotes the Fredholm determinant and $\mathcal{K}$ is an integral operator defined by
 \beq\label{eq:finiteNkernel}
 \mathcal{K}\cdot h (u)\equiv f_{A_1}(u) \oint_{\mathcal{C}} \frac{dv}{2\pi} \frac{i h(v)}{u-v+i\epsilon}\period
 \eeq
 
The Fredholm determinant---or equivalently a grand canonical partition function of a free fermion---shows up in various other contexts; to name a few, the sphere partition function of ABJM theory \cite{Marino:2011eh}, the topological string on a toric Calabi-Yau manifold \cite{Grassi:2014zfa,Codesido:2015dia,Codesido:2016ixn}, the $g$-functions in integrable theories \cite{Dorey:2004xk,Pozsgay:2010tv,Kostov:2018dmi,Kostov:2019fvw,Kostov:2019sgu,Caetano:2020dyp} and the correlation functions in $\mathcal{N}=4$ SYM \cite{Jiang:2016ulr,Bettelheim:2014gma,Fleury:2016ykk,Basso:2017khq,Coronado:2018ypq,Bargheer:2019exp,Kostov:2019stn,Kostov:2019auq,Belitsky:2019fan,Belitsky:2020qrm,Jiang:2019xdz,Jiang:2019zig,Komatsu:2020sup}. In particular, the relation to the Fredholm determinant proved to be useful for analyzing nonperturbative corrections to the sphere partition function of ABJM theory \cite{Marino:2011eh}. It would be interesting to perform a similar analysis to \eqref{eq:antifredholm} and compute nonperturbative corrections to the expectation values of the Wilson loop (see \cite{Fiol:2013hna,Gordon:2017dvy,Okuyama:2017feo,CanazasGaray:2018cpk} for related works).

\paragraph{Large $N$ limit from Clustering} We now consider the large $N$ limit of \eqref{eq:Zinteractinganti}. The first step is to evaluate the expectation value in the Gaussian matrix model $\langle \bullet \rangle_{M}$ in the large $N$ limit. Since the matrix $M$ is contained only in the factor $f_{A}(u)$, this simply amounts to perform the replacement \cite{Giombi:2020pdd},
\beq
\langle \prod_{j} f_{A_j} (u_j)\rangle_M \quad \overset{N\to \infty}{\mapsto} \quad \prod_j\mathfrak{f}_{A_j}(u_j)\comma
\eeq
with $\mathfrak{f}_{A}$ given by \eqref{eq:deffrakf}. We then get the following multiple integral representation for the large $N$ generating function
\beq\label{eq:largeNanti1}
\langle \mathcal{Z}_{{\rm anti}}(z)\rangle_{N\to \infty}=\sum_{k=0}^{\infty}\frac{z^{k}}{k!}\oint_{\mathcal{C}}\prod_{j=1}^{k}\frac{du_j \mathfrak{f}_{A_1} (u_j)}{2\pi}\det \left(\frac{i}{u_n-u_m+i\epsilon}\right)\period
\eeq
The next task is to take the large $N$ limit of the integrals \eqref{eq:largeNanti1}. This is more complicated than taking the large $N$ limit of standard matrix models: The main difficulty comes from the poles $\frac{i}{u_i-u_j+i\epsilon}$ inside the Cauchy determinant, which pinch the integration contours of $u_j$'s in the limit $N\to \infty$ and make the integrals singular. 

There are two known methods to deal with this problem. The first method is the Fermi Gas approach used extensively in the analysis of the ABJM matrix model \cite{Marino:2011eh}. It is based on the observation that the multiple integrals \eqref{eq:largeNanti1} can be regarded as a partition function of a free fermion system. Under this identification, $\epsilon$ plays the role of the Planck constant and the limit $\epsilon\to 0$ corresponds to the semi-classical limit. Then the large $N$ limit of $\langle \mathcal{Z}_{\rm anti}(t)\rangle$ is given by the semi-classical free energy of the free fermion. The second method is the Clustering method developed in \cite{Jiang:2016ulr} and used in the analysis of the strong-coupling limit of the correlation functions in $\mathcal{N}=4$ SYM \cite{Jiang:2016ulr,Bargheer:2019exp}. The basic strategy of the method is to first deform the contours so that every contour is far apart from each other. This produces extra terms which come from poles that cross the contours. After that, we can straightforwardly take the large $N$ limit without worrying about the contour pinching. Below we present a simple derivation of the large $N$ limit combining the ideas of both approaches.

\begin{figure}[t]
\centering
\includegraphics[clip,height=4.5cm]{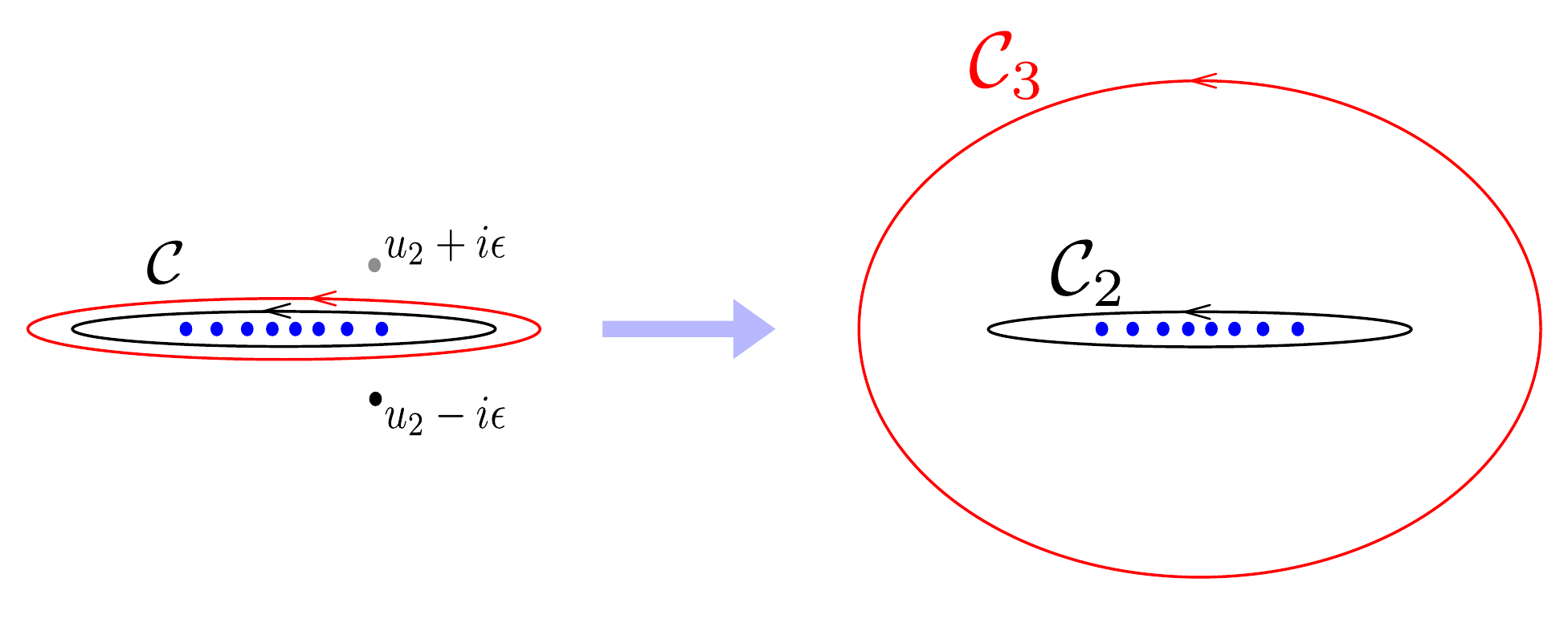}
\caption{The deformation of the contours and the Clustering mechanism. To evaluate $\mathcal{Z}_{\rm anti}(z)$ in the large $N$ limit, we first deform the original contours $\mathcal{C}$ so that each contour is far apart from each other ($\mathcal{C}_1\prec\mathcal{C}_2\prec\mathcal{C}_3$). Upon doing so, the contour picks up contributions from poles at $u_j=u_k- i\epsilon$. Such contributions are important to obtain the correct large $N$ answer.}\label{fig:contourdefcluster}
\end{figure}

The first step is to use the Fredholm determinant representation \eqref{eq:antifredholm} and express $\log \langle \mathcal{Z}_{{\rm anti}}(z)\rangle_{N\to \infty}$ as
\beq
\log \langle \mathcal{Z}_{{\rm anti}}(z)\rangle =-\sum_{k=1}^{\infty}\frac{(-z)^{k}}{k}I_k\comma \qquad I_k\equiv \langle {\rm Tr}\left(\mathcal{K}^{k}\right)\rangle_{M}\comma
\eeq
where ${\rm Tr}$ is the operator trace
\beq
{\rm Tr}\left(\mathcal{K}^{k}\right)=\oint_{\mathcal{C}} \left(\prod_{j=1}^{k}\frac{du_j}{2\pi}\right) \frac{if_{A_1}(u_1)}{u_1-u_2+i\epsilon}\frac{if_{A_1}(u_2)}{u_2-u_2+i\epsilon}\cdots \frac{if_{A_1}(u_k)}{u_k-u_1+i\epsilon}\period 
\eeq
The next step is to deform the contours so that they are far separated from each other. To illustrate the idea in a concrete example, let us consider $I_3$,
\beq
I_3=\left<\oint_{\mathcal{C}} \frac{du_1du_2du_3}{(2\pi)^3} \frac{if_{A_1}(u_1)}{u_1-u_2+i\epsilon}\frac{if_{A_1}(u_2)}{u_2-u_3+i\epsilon}\frac{if_{A_1}(u_3)}{u_3-u_1+i\epsilon}\right>_{M}\period
\eeq
We first deform the contour of $u_3$ from $\mathcal{C}$ to a lager contour $\mathcal{C}_3$ which is far separated from $\mathcal{C}$. Upon doing so, the contour crosses the poles at $u_3 =u_1-i\epsilon$ and $u_3=u_2+i\epsilon$ (see Figure \ref{fig:contourdefcluster}). The residues from these poles are proportional to
\beq\label{eq:residuecompare}
\begin{aligned}
u_3&=u_1-i\epsilon:\,\, f_{A_1}(u_1)f_{A_1} (u_1-i\epsilon)\propto \det \left[\frac{u_1-M-2i\epsilon}{u_1-M}\right]\comma\\
u_3&=u_2+i\epsilon:\,\, f_{A_1}(u_2)f_{A_1} (u_2+i\epsilon)\propto \det \left[\frac{u_2-M-i\epsilon}{u_2-M+i\epsilon}\right]\period
\end{aligned}
\eeq
This shows that the residue for $u_3=u_2+i\epsilon$ is nonsingular inside the contour $\mathcal{C}$ of $u_2$. We thus conclude that the contribution from the pole at $u_3=u_2+i\epsilon$ vanishes and can be neglected. Continuing in this fashion, we can rewrite $I_3$ as
\beq
\begin{aligned}
I_3=&\left<\oint_{\mathcal{C}_1\prec\mathcal{C}_2\prec\mathcal{C}_3} \frac{du_1du_2du_3}{(2\pi)^3} \frac{if_{A_1}(u_1)}{u_1-u_2+i\epsilon}\frac{if_{A_1}(u_2)}{u_2-u_3+i\epsilon}\frac{if_{A_1}(u_3)}{u_3-u_1+i\epsilon}\right>_{M}\\
&+2\left<\oint_{\mathcal{C}_1\prec\mathcal{C}_2} \frac{du_1du_2}{(2\pi)^2} \frac{if_{A_1}(u_1)}{u_1-u_2+i\epsilon}\frac{if_{A_1}(u_2)f_{A_1}(u_2-i\epsilon)}{u_2-u_1+2i\epsilon}\right>_{M}\\
&+\left<\oint_{\mathcal{C}_1} \frac{du_1}{2\pi} \frac{f_{A_1}(u_1)f_{A_1}(u_1-i\epsilon)f_{A_1}(u_1-2i\epsilon)}{3\epsilon}\right>_{M}
\end{aligned}
\eeq
Among these terms, the last term is dominant in the large $N$ limit since it is proportional to $1/\epsilon$. Collecting all such terms from $I_k$'s and replacing $f_{A}$'s with its large $N$ counterpart $\mathfrak{f}_{A}$'s, we arrive at the following expression,
\beq
\begin{aligned}
\log \langle \mathcal{Z}_{{\rm anti}}(z)\rangle_{N\to \infty} =-\sum_{k=1}^{\infty}\frac{(-z)^{k}}{ \epsilon k^2} \oint_{\mathcal{C}}\frac{du}{2\pi}\left(\mathfrak{f}_{A_1}(u)\right)^{k}\period
\end{aligned}
\eeq
The sum can be performed explicitly and the result reads
\beq\label{eq:antisymmetricdilog}
\begin{aligned}
\left.\langle \mathcal{Z}_{\rm anti}(z)\rangle\right|_{N\to \infty} =&\exp\left[-\frac{N}{4\pi g^2}\oint \frac{du}{2\pi} {\rm Li}_2\left(-z\, \mathfrak{f}_{A_1}(u)\right)\right]\period
\end{aligned}
\eeq

To make contact with the results in the literature \cite{Hartnoll:2006is}, we perform the integration by parts and replace the dilogarithm with its derivative. After a further change of the integration variable
\beq\label{eq:changeofintegration}
u=\frac{2g}{\sqrt{1-a^2}}\left(\sqrt{1-s^2}-i a s\right)\qquad\quad  (a=\tfrac{A_1-2\pi}{2\pi})\comma
\eeq 
we get
\beq
\begin{aligned}
\left.\langle \mathcal{Z}_{\rm anti}(z)\rangle\right|_{N\to \infty} =&\exp\left[\frac{N}{\pi}\oint ds\left(\sqrt{1-s^2}-i a s\right)\log \left(1+ze^{-4\pi g\sqrt{1-a^2} s}\right)\right]\\
=&\exp\left[\frac{2N}{\pi}\int_{-1}^{1} ds\sqrt{1-s^2}\log \left(1+ze^{-4\pi g\sqrt{1-a^2} s}\right)\right]\period
\end{aligned}
\eeq
Upon setting $a=0$ ($A_1=2\pi$), this reproduces the result obtained in \cite{Hartnoll:2006is} for the half-BPS circular Wilson loop. For general $a$, it provides the $1/8$-BPS generalization of their result.

\subsection{Symmetric representation}
We now analyze the Wilson loop in the $k$-th symmetric representation. The result again has a structure similar to multiparticle integrals in the hexagon approach \cite{Basso:2015zoa}, but with an important modification that the integral now contains terms that resemble the contributions from {\it bound states} in \cite{Basso:2015zoa}.

\paragraph{Integral representation} For the $k$-th symmetric representation, we replace $\sum_j e^{\epsilon x_j}/N$ in \eqref{eq:fundamentalinsertion}  with
\beq\label{eq:replrulesym}
\frac{\sum_j e^{\epsilon x_j}}{N}\quad \mapsto \quad \frac{1}{d_{{\sf S}_k}}\sum_{1\leq j_1\leq \cdots \leq j_k\leq N} e^{\epsilon\sum_{a=1}^{k}x_{j_a}}\comma
\eeq
with $d_{{\sf S}_k}$ being the dimension of the $k$-th symmetric representation, $d_{{\sf S}_k}\equiv \frac{(N+k)!}{N! k!}$. Unlike the antisymmetric representation, the same eigenvalues $x_k$ can appear several times in the exponent in \eqref{eq:replrulesym}. If $x_k$ appears $s$ times, the corresponding eigenvalue of $B_1$ ($b_k$) will be shifted by $-i s\epsilon$. Taking this into account and following the derivation in section \ref{subsec:partition}, we obtain
\beq\label{eq:symmetricsum}
\begin{aligned}
&\langle \mathcal{W}_{{\sf S}_k}\rangle=\frac{(2\pi)^{N}(N!)^2}{Z d_{{\sf S}_k}}\int d^{N}a\Delta^2 (a) e^{-\frac{8\pi^2}{g_{\rm YM}^2}\sum_j (a_j)^2}\\
&\times \sum_{\substack{\cup_{s=0}^{\infty} \alpha_s =\{1,\ldots, N\}\\\sum_{s=0}^{\infty}s |\alpha_s|=k}}\left(\prod_{s=0}^{\infty}e^{i s A_1\left(\sum_{j\in \alpha_s} a_j-is\frac{\epsilon }{2}\right) }\right)\left(\prod_{ s^{\prime}<s}\prod_{\substack{n\in \alpha_s\\m\in \alpha_{s^{\prime}}}}\frac{a_n-a_m-i \epsilon(s-s^{\prime})}{a_n-a_m}\right)
\end{aligned}
\eeq
Here $\alpha_s$ is a set of eigenvalues which are shifted by $-i\epsilon s$. The summation is over all possible ways of partitioning integers $\{1,\ldots, N\}$ into subsets $\{\alpha_0,\alpha_1,\ldots\}$ under the condition 
\beq
 \sum_{s=0}^{\infty} s|\alpha_s|=k\comma
\eeq
with $|\alpha_s|$ being the number of elements in $\alpha_s$.

We can recast the summation \eqref{eq:symmetricsum} into multiple integrals by introducing integration variables for each of the elements in $\alpha_s$ with $s\geq 1$:
\beq
\begin{aligned}
& \sum_{\substack{\cup_{s=0}^{\infty} \alpha_s =\{1,\ldots, N\}\\\sum_{s=0}^{\infty}s |\alpha_s|=k}}\left(\prod_{s=0}^{\infty}e^{i s A_1\left(\sum_{l\in \alpha_s} a_l-i\frac{\epsilon s}{2}\right) }\right)\left(\prod_{0\leq s^{\prime}<s\leq \infty}\prod_{\substack{n\in \alpha_s\\m\in \alpha_{s^{\prime}}}}\frac{a_n-a_m-i \epsilon(s-s^{\prime})}{a_n-a_m}\right)=\\
 &\sum_{\substack{\{n_1,n_2,\ldots\}\\\sum_{s=1}^{\infty} sn_s=k}}\left(\prod_{s=1}^{\infty}\frac{N^{n_s}}{n_s!}\oint_{\mathcal{C}} \prod_{m=1}^{n_s}\frac{du_{s,m}F_s (u_{s,m})}{8\pi^2 g }\frac{\prod_{m<l}^{n_s}\bar{\Delta}_{s,s}(u_{s,m},u_{s,l})}{s}\right)\\
 &\times \left(\prod_{1\leq s^{\prime}<s}\,\,\prod_{m=1}^{n_s}\prod_{m^{\prime}=1}^{n_{s^{\prime}}}\bar{\Delta}_{ s,s^{\prime}}(u_{s,m},u_{s^{\prime},m^{\prime}})\right)\period
\end{aligned}
\eeq
Here the integration variables $u_{s,m}$ $(m=1,\ldots, n_s)$ correspond to eigenvalues shifted by $-i s \epsilon$, and the summation is over all possible sets of integers $\{n_1,n_2,\ldots\}$ satisfying $\sum_{s}s n_s=k$. $F_s$ and $\bar{\Delta}_{s,s^{\prime}}$ are defined by
\begin{align}
F_s (u)&\equiv e^{i A_1(u-is\frac{ \epsilon}{2})}\prod_{j}\frac{u-a_j-is \epsilon}{u-a_j}\comma\\
\bar{\Delta}_{s,s^{\prime}}(u,v)&\equiv\frac{(u-v)(u-v+i\epsilon(s-s^{\prime}))}{(u-v+i\epsilon s )(u-v-i\epsilon s^{\prime})}\period
\end{align}
This can be further rewritten as an expectation value in the Gaussian matrix model,
\beq\label{eq:symmetricintegralfin}
\begin{aligned}
\langle \mathcal{W}_{{\sf S}_k}\rangle=&\left< \frac{1}{d_{{\sf S}_k}}\sum_{\substack{\{n_1,n_2,\ldots\}\\\sum_{s=1}^{\infty} sn_s=k}}\left(\prod_{s=1}^{\infty}\frac{N^{n_s}}{n_s!}\oint_{\mathcal{C}} \prod_{m=1}^{n_s}\frac{du_{s,m}f_{A_1, s} (u_{s,m})}{8\pi^2 g^2 }\frac{\prod_{m<l}^{n_s}\bar{\Delta}_{s,s}(u_{s,m},u_{s,l})}{s}\right)\right.\\
&\left.\times \left(\prod_{1\leq s^{\prime}<s}\,\,\prod_{m=1}^{n_s}\prod_{m^{\prime}=1}^{n_{s^{\prime}}}\bar{\Delta}_{ s,s^{\prime}}(u_{s,m},u_{s^{\prime},m^{\prime}})\right)\right>_M\comma
\end{aligned}
\eeq
with
\beq\label{eq:boundf}
f_{A_1,s}(u)\equiv e^{i s A_1(u-is\frac{\epsilon}{2})}\det \left[\frac{u-M-is\epsilon}{u-M}\right]\period
\eeq

It is worth noting that there is again a striking resemblance with the multiparticle integrals in the hexagon approach. For instance, the integration variable $u_{s,\bullet}$'s correspond to the bound states made out of $s$ elementary particles, and the relation between $f_{A_1,s}(u)$ and $f_{A_1}(u)$
\beq
f_{A_1,s}(u)=\prod_{k=0}^{s-1}f_{A_1}(u -i k \epsilon)\comma
\eeq
parallels the relation between the form factors for elementary particles and bound states \cite{Basso:2015zoa}.

\paragraph{Generating function and Fredholm determinant} As is the case with the antisymmetric loop, it is useful to consider the generating function
\beq\label{eq:generatingsym}
\mathcal{Z}_{\rm sym}(z)\equiv \sum_{k=0}^{\infty}z^{k}d_{{\sf S}_k}\mathcal{W}_{S_k}\period
\eeq
The integral representation for $\langle \mathcal{Z}_{{\sf S}_k}(z)\rangle$ can be derived from \eqref{eq:symmetricintegralfin}:
\beq
\begin{aligned}
\langle \mathcal{Z}_{\rm sym}(z)\rangle=&\left<\sum_{\{n_1,n_2,\ldots\}}\left(\prod_{s=1}^{\infty}\frac{(z^{s}N)^{n_s}}{n_s!}\oint_{\mathcal{C}} \prod_{m=1}^{n_s}\frac{du_{s,m}f_{A_1, s} (u_{s,m})}{8\pi^2 g }\frac{\prod_{m<l}^{n_s}\bar{\Delta}_{s,s}(u_{s,m},u_{s,l})}{s}\right)\right.\\
&\left.\times \left(\prod_{1\leq s^{\prime}<s}\,\,\prod_{m=1}^{n_s}\prod_{m^{\prime}=1}^{n_{s^{\prime}}}\bar{\Delta}_{ s,s^{\prime}}(u_{s,m},u_{s^{\prime},m^{\prime}})\right)\right>_M
\end{aligned}
\eeq
To proceed, we rewrite the interaction terms using the Cauchy identity
\beq
\begin{aligned}
\left(\prod_{s=1}^{\infty}\frac{\prod_{m<l}^{n_s}\bar{\Delta}_{s,s}(u_{s,m},u_{s,l})}{s}\right)\left(\prod_{1\leq s^{\prime}<s}\,\,\prod_{m=1}^{n_s}\prod_{m^{\prime}=1}^{n_{s^{\prime}}}\bar{\Delta}_{ s,s^{\prime}}(u_{s,m},u_{s^{\prime},m^{\prime}})\right)=\det \left(\frac{i\epsilon}{\tilde{z}_I-z_J}\right)\comma
\end{aligned}
\eeq
with $z$ and $\tilde{z}$ given by
\beq
\begin{aligned}
z_I &:=\{u_{1,1}, \ldots, u_{1,n_1}, u_{2,1},\ldots, u_{2,n_2}, u_{3,1},\ldots\}\comma\\
\tilde{z}_I&:=\{u_{1,1}+i\epsilon,\ldots, u_{1,n_1}+i\epsilon, u_{2,1}+2i\epsilon, \ldots, u_{2,n_2}+2 i\epsilon, u_{3,1}+3i\epsilon,\ldots \}\period
\end{aligned}
\eeq
Using this expression, one can rewrite \eqref{eq:symmetricintegralfin} as the Fredholm determinant
\beq
\begin{aligned}\label{eq:generatingsymrep}
\langle \mathcal{Z}_{\rm sym}(z)\rangle= \left<{\rm Det}\left(1+\sum_{s=1}^{\infty}z^s \mathcal{K}_s\right) \right>_M\comma
\end{aligned}
\eeq
with
\beq
\mathcal{K}_s \cdot h(u)\equiv f_{A_1, s}(u)\oint_{\mathcal{C}} \frac{dv}{2\pi}\frac{i h(v)}{u-v+i s\epsilon }\period
\eeq
A notable difference from the antisymmetric loops is that it involves an (infinite) sum of operators. Similar structures appeared in \cite{Jiang:2016ulr} as the contributions from the mirror particles to the hexagon form factors, and also in \cite{Codesido:2015dia,Codesido:2016ixn} in the context of topological strings on toric Calabi-Yau threefolds whose mirror curves have higher genus.

\paragraph{Large $N$ limit from Clustering} The large $N$ limit of \eqref{eq:generatingsymrep} can be analyzed again using ideas from the Fermi Gas approach and the Clustering method\footnote{See \cite{Jiang:2016ulr} for details of the derivation.}: We first use the Fredholm determinant representation \eqref{eq:generatingsymrep} to write down the expansion of $\log \langle \mathcal{Z}_{\rm sym}(z)\rangle$. We then deform the contours and collect the terms that dominate in the large $N$ limit. We then replace $f_{A_1, s}$ with their large $N$ expressions, $f_{A_1,s} \sim (\mathfrak{f}_{A_1})^s$. As a result we obtain
\beq
\log \langle \mathcal{Z}_{{\rm sym}}(z)\rangle_{N\to \infty} =\sum_{k=1}^{\infty}\frac{z^{k}}{ \epsilon k^2} \oint_{\mathcal{C}}\frac{du}{2\pi}\left(\mathfrak{f}_{A_1}(u)\right)^{k}\period
\eeq
Performing the sum explicitly, we get
\beq\label{eq:symmetricdilog}
\begin{aligned}
\left.\langle \mathcal{Z}_{\rm sym}(z)\rangle\right|_{N\to \infty} =&\exp\left[\frac{N}{4\pi g^2}\oint \frac{du}{2\pi} {\rm Li}_2\left(z\, \mathfrak{f}_{A_1}(u)\right)\right]\period
\end{aligned}
\eeq

To compare with the result in the literature \cite{Hartnoll:2006is}, we again perform the integration by parts and change the variable \eqref{eq:changeofintegration}.  This leads to 
\beq
\begin{aligned}
\left.\langle \mathcal{Z}_{\rm sym}(t)\rangle\right|_{N\to \infty} =&\exp\left[-\frac{N}{\pi}\oint ds\sqrt{1-s^2}\log \left(1-ze^{-4\pi g \sqrt{1-a^2}s}\right)\right]\\
=&\exp\left[-\frac{2N}{\pi}\int_{-1}^{1} ds\sqrt{1-s^2}\log \left(1-ze^{-4\pi g \sqrt{1-a^2}s}\right)\right]\comma
\label{Ws-vev}
\end{aligned}
\eeq 
which is in agreement with \cite{Hartnoll:2006is} after setting $a=0$.
\subsection{Generalized higher-rank loops from the loop equation\label{subsec:generalized}}
\begin{figure}[t]
\centering
\includegraphics[clip,height=4cm]{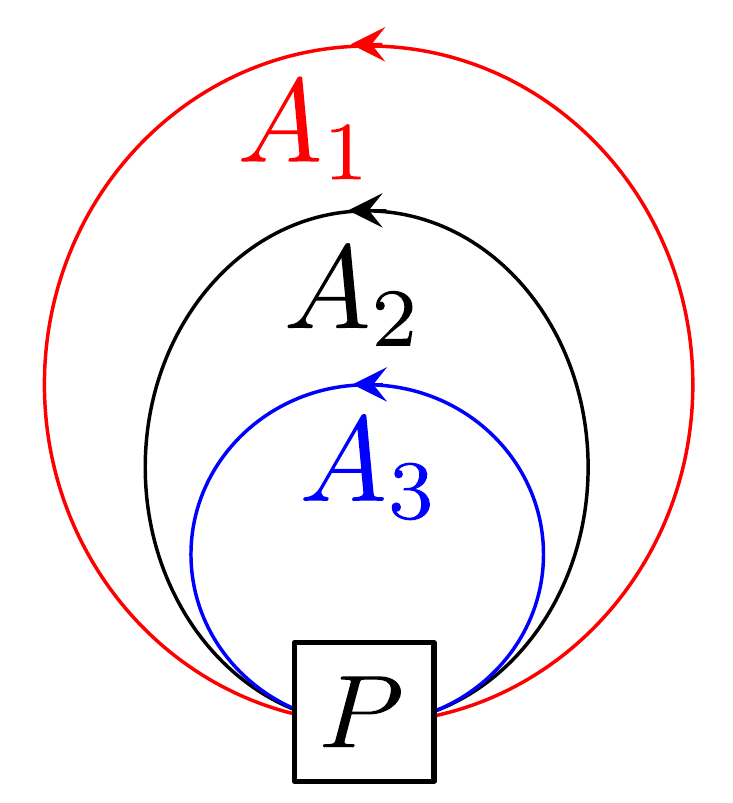}
\caption{The generalized higher-rank Wilson loop: It is defined by joining together multiple fundamental Wilson loops with different areas by a projector $P$ to a particular representation.}\label{fig:defgeneralized}
\end{figure}

We now consider a generalization of the higher-rank Wilson loops that couples to different areas. It is defined by joining together multiple fundamental Wilson loops with different areas by a projector to a higher-rank representation. See Figure \ref{fig:defgeneralized}. Using the loop equation, the expectation value for such intersecting loops can be obtained, and 
the result for the ordinary higher-rank Wilson loops can be recovered in the 
limit where the areas coincide. 

Before proceeding, let us first give some motivation. Recall that the expectation value for the anti-symmetric loop takes the following form,
\beq
\langle \mathcal{W}_{{\sf A}_k}\rangle=\left<\frac{N^{k}}{d_{A_k}k!}\oint_{\mathcal{C}} \prod_{j=1}^{k}\frac{du_j f_{A} (u_j)}{8\pi^2 g^2}\prod_{n<m}^{k}\bDelta (u_n,u_m)\right>_M
\eeq 
where $A$ is the area of the region inside the Wilson loop.
It is then natural to consider a small generalization of this formula in which different integration variables are coupled to different area:
\beq\label{eq:whatwewant}
\langle \mathcal{W}_{{\sf A}_k}^{\{A_j\}}\rangle\overset{\displaystyle ?}{=}\left<\frac{N^{k}}{d_{A_k}k!}\oint_{\mathcal{C}} \prod_{j=1}^{k}\frac{du_j f_{\red{A_j}} (u_j)}{8\pi^2 g^2}\prod_{n<m}^{k}\bDelta (u_n,u_m)\right>_M\period
\eeq
Of course, at this point this is just a mathematical generalization of the formula. In fact we will later see that the formula \eqref{eq:whatwewant} does not give the expectation value of the Wilson loop depicted in Figure \ref{fig:defgeneralized}. The goal of this subsection is to re-analyze the higher-rank Wilson loop from the loop equation and provides a physical derivation of the correct formula.
\paragraph{Rank-2 antisymmetric loop} Let us first consider the simplest example; the rank-2 antisymmetric loop. As is well-known, the standard rank-2 antisymmetric Wilson loop can be viewed as a linear combination of the doubly-wound Wilson loop and a product of two coincident fundamental loops:
\beq\label{eq:antidecomp}
\mathcal{W}_{{\sf A}_2}=\frac{1}{2d_{{\sf A}_2}}\left(N^2 \mathcal{W}_{f} \mathcal{W}_{f}-N \mathcal{W}_{\rm double}\right)\comma
\eeq
where $\mathcal{W}_f$ is the fundamental Wilson loop and $\mathcal{W}_{\rm double}$ is the doubly-wound Wilson loop, which corresponds to the insertion of ${\rm tr}\left(e^{2\epsilon X}\right)/N$ in the matrix model \eqref{eq:actionfund}. Physically, this relation follows from the fact that the rank-2 antisymmetric loop can be obtained by inserting a projector to a product of two fundamental loops. The projector consists of two terms; one is proportional to the identity operator and the other reconnects the two fundamental loops. These two terms correspond to the two terms on the right hand side of \eqref{eq:antidecomp}.

\begin{figure}[t]
\centering
\includegraphics[clip,height=3.5cm]{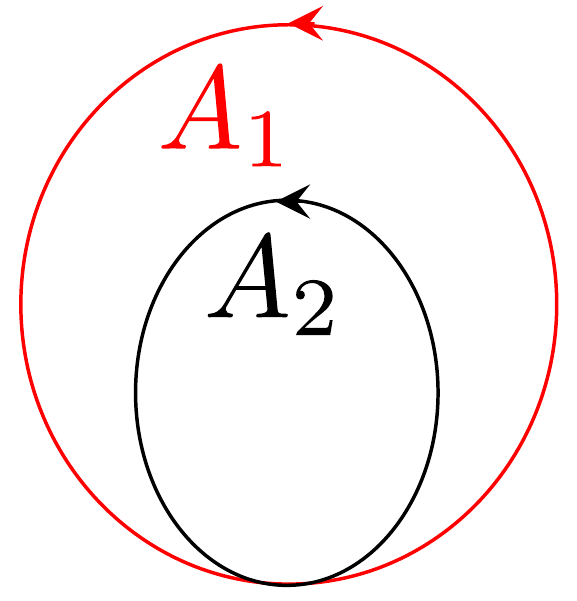}
\caption{The definition of the intersecting Wilson loop with areas $A_1$ and $A_2$. $A_1$ is the area inside the outer loop while $A_2$ is the area inside the inner loop.}\label{fig:defintersecting}
\end{figure}

The relation can be readily generalized to the generalized rank-2 antisymmetric loop. In that case, we start from two fundamental loops with different areas and insert the projector. We then get the relation
\beq
\mathcal{W}_{{\sf A}_2}^{\{A_1,A_2\}}=\frac{1}{2d_{{\sf A}_2}}\left(N^2\mathcal{W}_{A_1}\mathcal{W}_{A_2}-N\mathcal{W}_{A_1,A_2}\right)\comma
\eeq
where $\mathcal{W}_{A_j}$ is the fundamental Wilson loop with area $A_j$ and $\mathcal{W}_{A_1,A_2}$ is the self-intersecting Wilson loop depicted in Figure \ref{fig:defintersecting}. As shown in \cite{Giombi:2020pdd}, the expectation value of the intersecting Wilson loop can be computed by the application of the loop equation, which in this case reduces to
\beq
(\del_{A_1}-\del_{A_2})\mathcal{W}_{A_1,A_2} = -4\pi g \mathcal{W}_{A_1}\mathcal{W}_{A_2}\period
\eeq
Using the integral representation for multiple fundamental Wilson loops \eqref{eq:genmultiply}, we can solve this equation as follows:
\beq\label{eq:contourseparatesingle}
\langle \mathcal{W}_{A_1,A_2}\rangle=\left<4\pi g^2 i\oint_{\mathcal{C}_1\prec\mathcal{C}_2} \frac{du_1}{8\pi^2 g^2}\frac{du_2}{8\pi^2 g^2} \bar{\Delta} (u_1,u_2) \frac{f_{A_1}(u_1)f_{A_2}(u_2)}{u_1-u_2}\right>_M\period
\eeq
Combining this with $\langle \mathcal{W}_{A_1}\mathcal{W}_{A_2}\rangle$ given by \eqref{eq:genmultiply}, we obtain 
\beq\label{eq:genrank2new}
\langle \mathcal{W}_{{\sf A}_2}^{\{A_1,A_2\}}\rangle=\frac{N^2}{2d_{{\sf A}_2}}\oint_{\mathcal{C}_1\prec\mathcal{C}_2}\frac{du_1}{8\pi^2g^2}\frac{du_2}{8\pi^2g^2} f_{A_1}(u_1)f_{A_2}(u_2)\frac{u_1-u_2}{u_1-u_2+i\epsilon}\period
\eeq

\begin{figure}[t]
\centering
\includegraphics[clip,height=6cm]{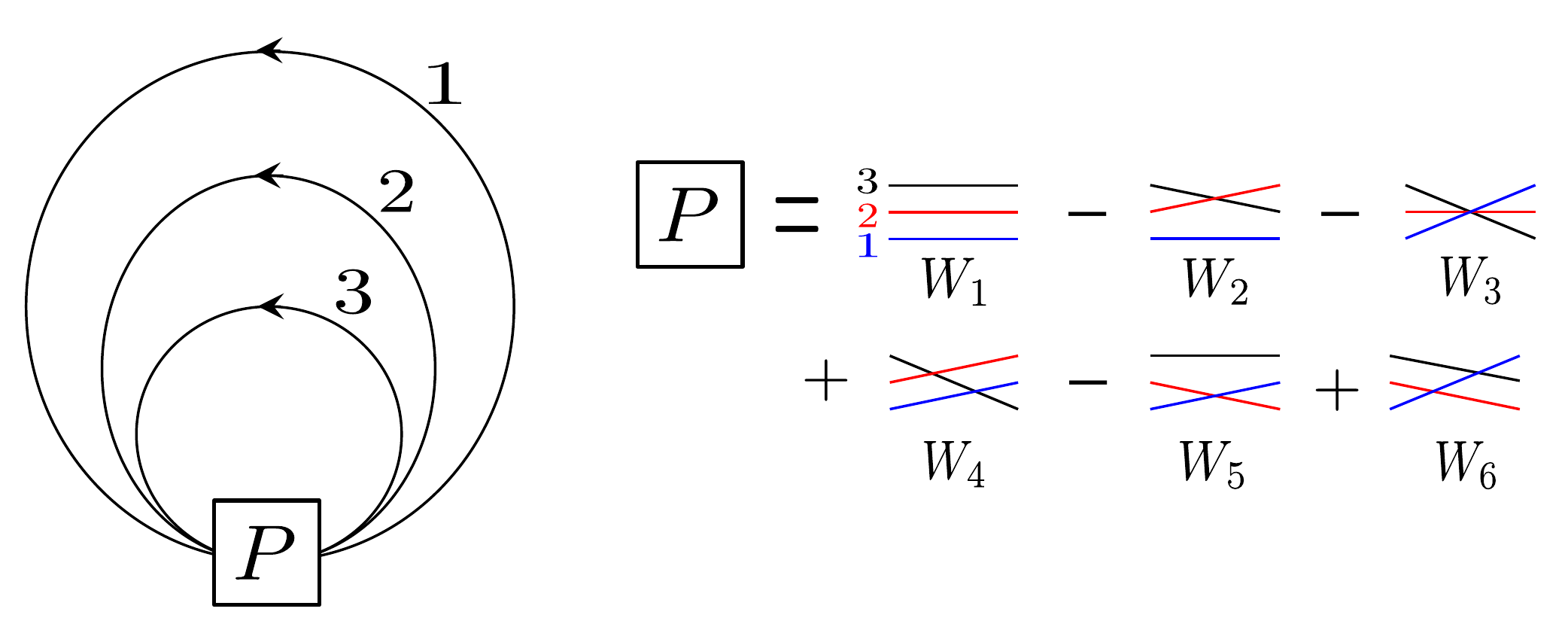}
\caption{The rank-3 antisymmetric Wilson loop and its generalization can be obtained by taking a linear combination of the intersecting loops $W_1$-$W_6$. Each $W_j$ corresponds to an element of the permutation group $S^3$.}
\label{fig:generalized3}
\end{figure}

To make contact with the integral representation obtained in section \ref{subsec:antisymintegral}, we deform the contours $\mathcal{C}_j$'s and bring them on top of each other. As mentioned already several times, such a deformation normally produces extra terms coming from the poles in the interaction term. However, because the interaction term in \eqref{eq:genrank2new} is given by $(u_1-u_2)/(u_1-u_2+i\epsilon)$ instead of $\bar{\Delta}(u_1,u_2)$, it turns out that there are no such extra contributions\fn{This is basically  because the residue at the pole $u_2=u_1+i\epsilon$ is proportional to $\det (u-M-i\epsilon)/(u-M+i\epsilon)$, which is nonsingular inside the integration contour. See the discussion around \eqref{eq:residuecompare}.}. We can therefore simply replace $\mathcal{C}_j$'s with $\mathcal{C}$:
\beq
\langle \mathcal{W}_{{\sf A}_2}^{\{A_1,A_2\}}\rangle=\frac{N^2}{2d_{{\sf A}_2}}\oint_{\mathcal{C}}\frac{du_1}{8\pi^2g^2}\frac{du_2}{8\pi^2g^2} f_{A_1}(u_1)f_{A_2}(u_2)\frac{u_1-u_2}{u_1-u_2+i\epsilon}\period
\eeq
If we further set $A_1=A_2$ and symmetrize the integrand with respect to $u_1\leftrightarrow u_2$, we recover the expression given in \eqref{eq:antisymfinal}:
\beq
\langle \mathcal{W}_{{\sf A}_2}\rangle=\frac{N^2}{2d_{{\sf A}_2}}\oint_{\mathcal{C}}\frac{du_1}{8\pi^2g^2}\frac{du_2}{8\pi^2g^2} f_{A_1}(u_1)f_{A_1}(u_2)\bar{\Delta}(u_1,u_2)\period
\eeq
\paragraph{Rank-3 antisymmetric loop} Let us next consider a slightly more complicated case, the rank-3 antisymmetric loop. It can be represented as a sum of 6 different Wilson loops, each of which corresponds to an element of the permutation group $S_3$. The relevant loop equations for computing such loops are presented in \cite{Kazakov:1980zj}. In the notations of figure 6 in \cite{Kazakov:1980zj}, the relation between the elements of the permutation and the Wilson loop is given by
\beq\label{eq:permrank3}
\begin{aligned}
&W_{1}:\{1,2,3\}\comma\quad W_{2}:\{1,3,2\}\comma\quad W_{3}:\{3,2,1\}\comma\\
&W_{4}:\{2,3,1\}\comma\quad W_5: \{2,1,3\}\comma\quad W_6:\{3,1,2\}\period
\end{aligned}
\eeq
See also Figure \ref{fig:generalized3}.

Note that \cite{Kazakov:1980zj} does not discuss $W_6$ since it is related to $W_4$ by the spacetime parity and its expectation value is identical to that of $W_4$. Solving the loop equations presented in \cite{Kazakov:1980zj}, we obtain the following results for their expectation values (here we used the same overall normalization $1/N^{3}$ for all the six loops):
\beq\nn
\begin{aligned}
W_1&=\left<\oint_{\mathcal{C}_1\prec \mathcal{C}_2 \prec\mathcal{C}_3}\prod_{j=1}^{3}\frac{du_j}{8\pi^2 g^2}f_{A_j}(u_j)\prod_{1\leq s<t\leq 3}\bar{\Delta}(u_s,u_t)\right>_M\comma\\
W_2&=\left<\oint_{\mathcal{C}_1\prec \mathcal{C}_2 \prec\mathcal{C}_3}\prod_{j=1}^{3}\frac{du_j}{8\pi^2 g^2}f_{A_j}(u_j)\left(\frac{i\epsilon}{u_2-u_3}\right)\prod_{1\leq s<t\leq 3}\bar{\Delta}(u_s,u_t)\right>_M\comma\\
W_3&=\left<\oint_{\mathcal{C}_1\prec \mathcal{C}_2 \prec\mathcal{C}_3}\prod_{j=1}^{3}\frac{du_j}{8\pi^2 g^2}f_{A_j}(u_j)\left(\frac{i\epsilon}{u_1-u_3}\left(1+\frac{i\epsilon}{u_1-u_2}\frac{i\epsilon}{u_2-u_3}\right)\right)\prod_{1\leq s<t\leq 3}\bar{\Delta}(u_s,u_t)\right>_M\comma\\
W_4&=\left<\oint_{\mathcal{C}_1\prec \mathcal{C}_2 \prec\mathcal{C}_3}\prod_{j=1}^{3}\frac{du_j}{8\pi^2 g^2}f_{A_j}(u_j)\left(\frac{i\epsilon}{u_1-u_2}\frac{i\epsilon}{u_2-u_3}\right)\prod_{1\leq s<t\leq 3}\bar{\Delta}(u_s,u_t)\right>_M\comma\\
W_5&=\left<\oint_{\mathcal{C}_1\prec \mathcal{C}_2 \prec\mathcal{C}_3}\prod_{j=1}^{3}\frac{du_j}{8\pi^2 g^2}f_{A_j}(u_j)\left(\frac{i\epsilon}{u_1-u_2}\right)\prod_{1\leq s<t\leq 3}\bar{\Delta}(u_s,u_t)\right>_M\comma\\
W_6&=\left<\oint_{\mathcal{C}_1\prec \mathcal{C}_2 \prec\mathcal{C}_3}\prod_{j=1}^{3}\frac{du_j}{8\pi^2 g^2}f_{A_j}(u_j)\left(\frac{i\epsilon}{u_1-u_2}\frac{i\epsilon}{u_2-u_3}\right)\prod_{1\leq s<t\leq 3}\bar{\Delta}(u_s,u_t)\right>_M\period
\end{aligned}
\eeq
The generalized antisymmetric loop is given by a linear combination of these Wilson loops with appropriate signs,
\beq
\begin{aligned}
\langle \mathcal{W}_{{\sf A}_3}^{\{A_1,A_2,A_3\}}\rangle&=\frac{N^3}{d_{{\sf A}_3}3!}\left(W_1-W_2-W_3+W_4-W_5+W_6\right)\period
\end{aligned}
\eeq
It turns out that the integrands combine nicely and give
\beq
\langle \mathcal{W}_{{\sf A}_3}^{\{A_1,A_2,A_3\}}\rangle=\frac{N^3}{d_{{\sf A}_3}3!}\left<\oint_{\mathcal{C}_1\prec\mathcal{C}_2\prec\mathcal{C}_3} \prod_{j=1}^{3}\frac{du_j}{2\pi}f_{A_j}(u_j)\prod_{1\leq s<t\leq 3}\frac{u_s-u_t}{u_s-u_t+i\epsilon}\right>_M\period
\eeq
As is the case with the rank-2 antisymmetric loop, we can deform all the contours to $\mathcal{C}$ without producing extra contributions. If the areas are identical $(A_1=A_2=A_3)$, we can further symmetrize the integrand with respect to the permutation of $u_j$'s and reproduce the expression \eqref{eq:antisymfinal}.
\paragraph{General cases} Repeating the same procedures for general $k$-th antisymmetric Wilson loop, we find that the result is similar but different from what we expected \eqref{eq:whatwewant}. Namely we have
\beq
\langle \mathcal{W}_{{\sf A}_k}^{\{A_j\}}\rangle=\left<\frac{N^{k}}{d_{A_k}k!}\oint_{\mathcal{C}_1\prec\cdots \prec\mathcal{C}_k} \prod_{j=1}^{k}\frac{du_j f_{\red{A_j}} (u_j)}{8\pi^2 g^2}\prod_{n<m}^{k}\frac{u_n-u_m}{u_n-u_m+i\epsilon}\right>_M\period
\eeq
Here we separated contours from each other but we can deform them to $\mathcal{C}$ without producing extra terms.

We can perform the same analysis also for the $k$-th symmetric Wilson loop. Since the computation is similar, here we just present the final result:
\beq
\langle \mathcal{W}_{{\sf S}_k}^{\{A_j\}}\rangle=\left<\frac{N^{k}}{d_{A_k}k!}\oint_{\mathcal{C}_1\prec\cdots \prec\mathcal{C}_k} \prod_{j=1}^{k}\frac{du_j f_{\red{A_j}} (u_j)}{8\pi^2 g^2}\prod_{n<m}^{k}\frac{u_n-u_m}{u_n-u_m-i\epsilon}\right>_M\period
\eeq
Note that the result is very similar to the one for the antisymmetric Wilson loop; the only modification is the sign in front of $i\epsilon$ in the interaction term. However, owing to this change of signs, it will produce extra contributions when we deform the contours and bring them on top of each other. This is the reason why the formula for the (standard) symmetric loop  \eqref{eq:symmetricintegralfin} is much more complicated than the one for the antisymmetric loop \eqref{eq:antisymfinal}.
\section{Topological Correlators on the Giant Wilson Loops\label{sec:correlatorfromloc}}
\subsection{Deformed partition function}
Having computed the expectation values of the generalized higher-rank Wilson loops, we can now consider the area derivative \eqref{eq:secondordercorrect}
\beq\label{eq:derwewant}
\sum_{j=1}^{k}(\del_{A_j})^{n}\comma
\eeq
which directly inserts $n$ $\tilde{\Phi}$ fields to each constituent fundamental Wilson loop.

At the level of the integral representation derived in the previous section, the action of \eqref{eq:derwewant} translates to the insertion of
\beq
\sum_{j=1}^{k}\left(\del_{A_j}\right)^{n} \quad \mapsto \quad \sum_{j=1}^{k}(iu_j)^{n}\period
\eeq
To analyze the integral with such insertions, it is convenient to deform the integrals by exponentiating the insertions. This corresponds to changing the factor $f_{A}(u)$ to
\beq\label{eq:newsource}
f_{A}(u)e^{\sum_{j=2}^{\infty}t_j(i u)^{j}} \period
\eeq
As mentioned in the previous section, it is often convenient to consider the generating function in order to analyze the Wilson loops in the (anti)symmetric representations. In such cases, it is convenient to absorb the chemical potential $z$ in the generating functions \eqref{eq:generatingantisym} and \eqref{eq:generatingsym} into $f_{A}(u)$, and write
\beq\label{eq:newsource}
f_{A} (u)\quad \mapsto \quad \tilde{f}_{\bf t}(u)\equiv e^{\sum_{j=0}^{\infty}t_j(i u)^{j}}e^{2\pi i u+ \frac{\epsilon A}{2}}\det \left[\frac{u-M-i\epsilon}{u-M}\right]\comma
\eeq
where $t_0$ and $t_1$ are given by $e^{t_0}\equiv z$ and $t_1\equiv A-2\pi$.

An advantage of this reformulation is that we can insert $\tilde{\Phi}^{n}$ (without normal ordering) simply by the first-order derivative of $t_n$:
\beq
\sum_{j=1}^{k}\left(\del_{A_j}\right)^{n} \quad \mapsto \quad \frac{d}{dt_n}\period
\eeq
In the rest of this paper, we use a simplified notation
 \beq
 d_{n}\equiv \frac{d}{dt_n}\period 
 \eeq
 
 After the deformation \eqref{eq:newsource}, the expectation value of the generating function for the antisymmetric representations can be expressed as
 \beq\label{eq:generatingdeformedanti}
 \langle \tilde{\mathcal{Z}}_{\rm anti}\rangle =\left< {\rm Det}\left(1+\tilde{\mathcal{K}}\right)\right>_M\comma
 \eeq
 where the Fredholm kernel $\tilde{\mathcal{K}}$ reads
 \beq
 \tilde{\mathcal{K}}\cdot h(u)\equiv \tilde{f}_{{\bf t}}(u)\oint_{\mathcal{C}}\frac{dv}{2\pi i}\frac{i h(v)}{u-v+i\epsilon}\period
 \eeq
 Similarly the generating function for the symmetric representations is given by
 \beq
\begin{aligned}\label{eq:generatingdeformedsym}
\langle \tilde{\mathcal{Z}}_{\rm sym}\rangle= \left<{\rm Det}\left(1+\sum_{s=1}^{\infty} \tilde{\mathcal{K}}_s\right) \right>_M\comma
\end{aligned}
\eeq
with
\beq
\tilde{\mathcal{K}}_s \cdot h(u)\equiv \tilde{f}_{{\bf t}, s}(u)\oint_{\mathcal{C}} \frac{dv}{2\pi}\frac{i h(v)}{u-v+i s\epsilon }\period
\eeq
Here $\tilde{f}_{{\bf t}}$ is given by
\beq
\tilde{f}_{{\bf t},s}(u)=\prod_{k=0}^{s-1}\tilde{f}_{{\bf t}}(u -i k \epsilon)\period
\eeq
\paragraph{Large N limit} The large $N$ limits of the generating functions \eqref{eq:generatingdeformedanti} and \eqref{eq:generatingdeformedsym} can be computed in a similar manner to the undeformed case. As a result, the large $N$ free energies
\beq
\left.\langle \tilde{\mathcal{Z}}_{{\rm anti}}(z)\rangle\right|_{N\to \infty} =e^{NF_{\rm anti}(t_0,t_1,\ldots)}\comma\qquad\left.\langle \tilde{\mathcal{Z}}_{{\rm sym}}(z)\rangle\right|_{N\to \infty}=e^{NF_{\rm sym}(t_0,t_1,\ldots)}\comma
\eeq
are given by
\begin{align}\label{eq:integralFanti}
&F_{{\rm anti}}(t_0,t_1,\ldots)=\frac{-1}{4\pi g^2}\oint \frac{du}{2\pi}{\rm Li}_2 \left(-\tilde{\mathfrak{f}}(u)\right)=\frac{-1}{4\pi g}\oint\frac{dx (1+x^{-2})}{2\pi i }{\rm Li}_2\left(-\tilde{\mathfrak{f}}(u)\right)\comma\\
&F_{{\rm sym}}(t_0,t_1,\ldots)=\frac{1}{4\pi g^2}\oint \frac{du}{2\pi}{\rm Li}_2 \left(\tilde{\mathfrak{f}}(u)\right)=\frac{1}{4\pi g}\oint\frac{dx (1+x^{-2})}{2\pi i }{\rm Li}_2\left(\tilde{\mathfrak{f}}(u)\right)\comma\label{eq:integralFsym}
\end{align}
where $\tilde{\mathfrak{f}} (u)$ is 
\beq
\tilde{\mathfrak{f}}(u)\equiv e^{2\pi i u-4\pi i g^2 G(u)}e^{\sum_{j=0}^{\infty}t_j (iu)^{j}}=e^{2\pi g (x+1/x)}e^{\sum_j t_j (g(x-1/x))^{j}}\period
\eeq

To compute the correlation functions on the Wilson loop with a fixed representation of size $k$, we further need to perform the integral of $t_0$,
\beq\label{eq:t0integral}
\langle \tilde{\mathcal{W}}\rangle=\int dt_0 \,\,e^{NJ (\kappa;t_0,\ldots)}\comma
\eeq
with 
\beq\label{eq:Jdef}
J(\kappa;t_0,\ldots)\equiv F(t_0,\ldots)-\kappa t_0\qquad (\kappa\equiv \tfrac{k}{N})\period
\eeq
Here we dropped the subscripts (${\rm anti}$ or ${\rm sym}$) to simplify the notation. In the large $N$ limit, the integral \eqref{eq:t0integral} can be approximated by the saddle point, which is determined by
\beq\label{eq:Jsaddlepoint}
\del_{t_0} J =0\quad \iff \quad \del_{t_0} F(t_0,t_1,\ldots)=\kappa\period
\eeq
The equation \eqref{eq:Jsaddlepoint} determines $t_0$ as a function of other $t_n$ ($n\geq 1$) and $\kappa$. Plugging in the saddle-point value of $t_0$ to \eqref{eq:Jdef}, we get a large $N$ approximation for the deformed Wilson loop with a fixed representation,
\beq\label{eq:WtilNinf}
\left.\langle \tilde{\mathcal{W}}\rangle\right|_{N\to \infty}=e^{N \tilde{F} (\kappa;t_1,t_2,\cdots)}\comma
\eeq
where $\tilde{F}$ is the saddle-point value of $J$, which is now a function of $\kappa$ and $t_n$ with $n\geq 1$ (but not of $t_0$).

\paragraph{Correlators} From  the deformed Wilson loop \eqref{eq:WtilNinf}, we compute the correlators of un-normal-ordered single-particle insertions $\tilde{\Phi}^{n}$'s by differentiating with respect to the coupling constants $t_n$'s. For instance, the two-point functions $\tilde{\Phi}^{n}$'s are given by
 \beq
 \langle\!\langle \tilde{\Phi}^{n}\tilde{\Phi}^{m}\rangle\!\rangle=\left.N^2 d_n \tilde{F}d_m\tilde{F}+N d_n d_m\tilde{F}\right|_{t_{n\geq 2}=0}\period
 \eeq
 Among these two terms on the right hand side, the first term is a product of one-point functions and must be eliminated in order to define normal-ordered operators. This can be achieved by subtracting the identity operators\fn{A similar analysis was performed in \cite{Rodriguez-Gomez:2016ijh} for correlation functions of single-trace operators in large $N$ SCFTs.} as
 \beq
 \tilde{\Phi}^{n}\quad \mapsto \quad \tilde{\Phi}^{n}-\left(Nd_{n}\tilde{F}\right){\bf 1}\period
 \eeq
 After doing so, we get a simpler formula
 \beq\label{eq:unnormalorderedtwopt}
 \langle \!\langle \tilde{\Phi}^{n} \tilde{\Phi}^{m} \rangle\!\rangle=\left.Nd_{n} d_{m} \tilde{F}\right|_{t_{n\geq 2}=0}\period
 \eeq
 As is clear from the formula, this is basically equivalent to considering the connected two-point functions.
 
 In what follows, we use this representation \eqref{eq:unnormalorderedtwopt} of the two-point functions. However we should keep in mind that the operators $\tilde{\Phi}^{n}$ are still not normal-ordered since we only resolved the mixing with the identity operators so far. To define the normal-ordered operators, we need to perform the Gram-Schmidt orthogonalization as in \cite{Giombi:2018qox}.
\subsection{Diagrammatic rules and ``wormholes"}
Before discussing the Gram-Schmidt orthogonalization, let us derive useful expressions for derivatives of the free energy $\tilde{F}$. The free energy $\tilde{F}$ has two sources of $t_n$ dependence: First it contains explicitly $t_{n\geq 1}$ as a deformation parameter as can be seen from \eqref{eq:Jdef}. Second the saddle-point value of $t_0$ depends implicitly on $t_{n\geq 1}$ through the saddle-point equation \eqref{eq:Jsaddlepoint}. Thus we can decompose $d_n=\frac{d}{dt_n}$ into two parts as
\beq
d_{n}=\frac{d}{dt_n} =\del_{n}+\del_{n}t_0\del_{0}\period
\eeq 
Here $\del_n \equiv \del_{t_n}$ means taking a partial derivative with respect to $t_n$ by treating all the $t_n$'s---including $t_0$---as independent variables, while $d_{n}$ means computing a derivative by taking into account the implicit dependence of $t_0$ on $t_n$.

The factor appearing in the second term $\del_{n} t_0$ can be expressed in terms of the deformed free energy $\tilde{F}$ by differentiating the saddle point equation \eqref{eq:Jsaddlepoint} $\del_{t_0} F=\kappa$ by $t_n$:
\beq
\del_n\del_{0}F+\del_n t_0 \del_0^2 F=0\quad \iff \quad \del_n t_0=-\frac{\del_n\del_{0}F}{\del_0^2F}\period
\eeq
Therefore, we can rewrite $d_{n}$ as
\beq\label{eq:defofdtn}
d_n=\del_{n}-\frac{\del_n\del_{0}F}{\del_0^2F} \del_0\period
\eeq
The relation allows us to rewrite derivatives of the Legendre-transformed free energy $\tilde{F}$ in terms of derivatives of the original free energy $F$.

\begin{figure}[t]
\centering
\begin{minipage}{0.6\hsize}
\centering
\includegraphics[clip,height=3cm]{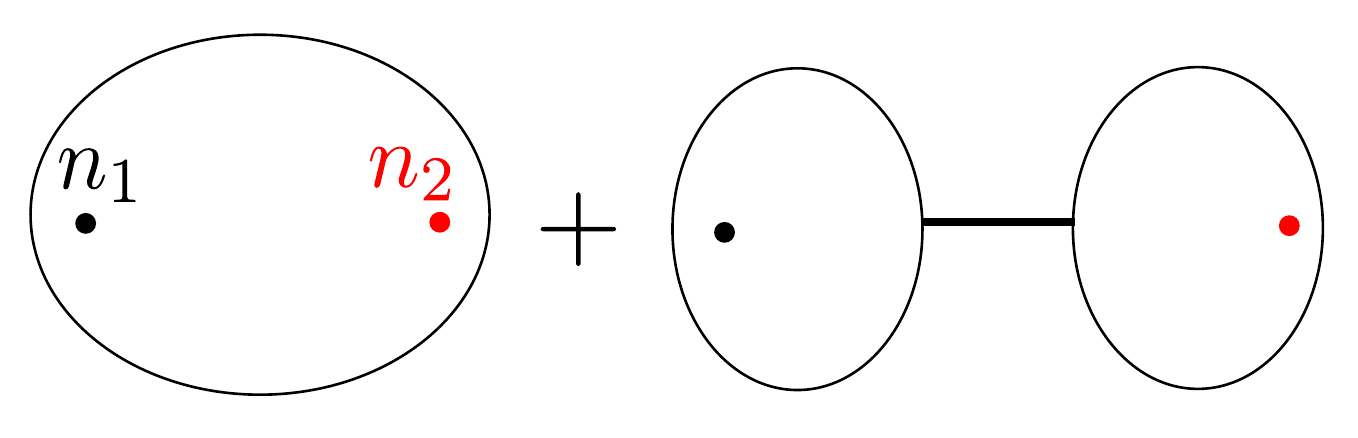}
\subcaption{Two-point function}
\vspace{30pt}
\end{minipage}
\begin{minipage}{0.6\hsize}
\centering
\includegraphics[clip,height=4cm]{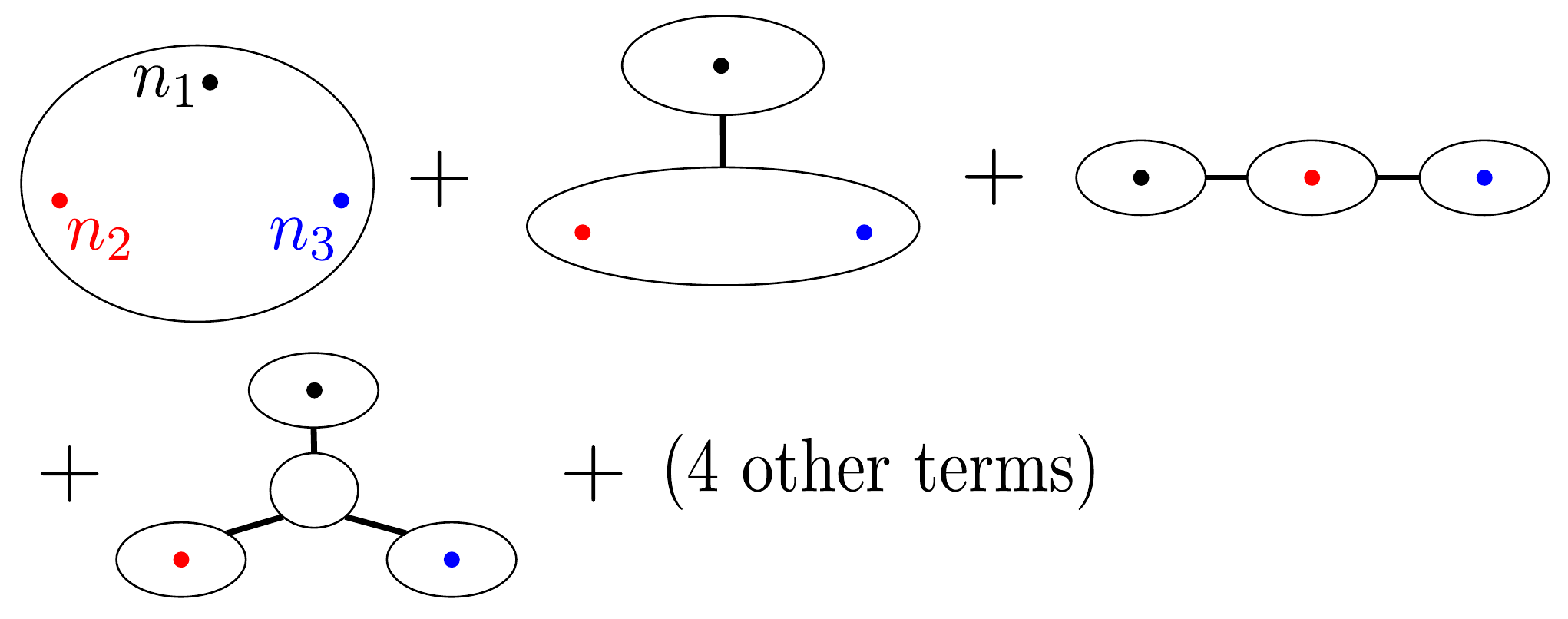}
\subcaption{Three-point function}
\end{minipage}
\caption{The diagrammatic rule to compute $\left(\prod_j d_{n_j}\right)\tilde{F}$. The result is given by a sum of products of disconnected orrelators (denoted by spheres with punctures) joined together by wormholes (denoted by thick black lines). In (b), there are four other diagrams that can be obtained from the second and the third diagrams by the permutation of punctures.}\label{fig:wormhole1}
\end{figure}

\paragraph{Diagrammatic rules, double-trace deformation and wormholes}It turns out that the relation between $d_n$ and $\del_n$ \eqref{eq:defofdtn} is precisely the same as the relation between derivatives of the coupling constants in a standard matrix model and a double-trace deformed matrix model, discussed in \cite{Barbon:1995dx}. As was discussed there, there is a simple diagrammatic rule to relate $\left(\prod_j d_{n_j}\right) \tilde{F}$ and $\left(\prod_{j}\del_{n_j}\right)F$. Roughly speaking, it expresses $\left(\prod_{j} d_{n_j}\right)\tilde{F}$ as a sum of products of disconnected correlators connected by ``wormholes'' (see \cite{Barbon:1995dx} for details). Applying the rule we get the following results for two- and three-point functions (see also Figure \ref{fig:wormhole1}):

\begin{align}\label{eq:twoderivatives}
&d_{n_1}d_{n_2}\tilde{F}=\langle n_1,n_2\rangle -\frac{\langle n_1,0\rangle\langle 0,n_2\rangle}{\langle 0,0\rangle}\comma\\
&\begin{aligned}\label{eq:threederivatives}
&d_{n_1}d_{n_2}d_{n_3}\tilde{F}=\langle n_1,n_2,n_3 \rangle-\frac{\langle n_1,0\rangle \langle 0,n_2,n_3\rangle}{\langle 0,0\rangle}-\frac{\langle n_2,0\rangle \langle 0,n_3,n_1\rangle}{\langle 0,0\rangle}-\frac{\langle n_3,0\rangle \langle 0,n_1,n_2\rangle}{\langle 0,0\rangle}\\
&+\frac{\langle n_1,0 \rangle\langle 0,n_2,0 \rangle \langle 0,n_3\rangle}{(\langle 0,0\rangle)^2}+\frac{\langle n_2,0 \rangle\langle 0,n_3,0 \rangle \langle 0,n_1\rangle}{(\langle 0,0\rangle)^2}+\frac{\langle n_3,0 \rangle\langle 0,n_1,0 \rangle \langle 0,n_2\rangle}{(\langle 0,0\rangle)^2}\\
&-\frac{\langle n_1,0\rangle\langle n_2,0\rangle\langle n_3,0\rangle\langle 0,0,0 \rangle}{(\langle 0,0 \rangle)^3}\comma
\end{aligned}
\end{align} 
with
\beq\label{eq:deflanglebracket}
\langle n_1,n_2,\cdots,n_m\rangle \equiv \del_{n_1}\cdots \del_{n_m}F\period
\eeq
Here a wormhole corresponds to the insertion of a factor 
\beq
-\frac{\langle\bullet, 0 \rangle\langle 0,\bullet \rangle}{\langle 0,0\rangle}\comma
\eeq 
in the correlator.
For instance, the first line for $d_{n_1}d_{n_2}d_{n_3}\tilde{F}$ correspond to the diagrams with $0$ and $1$ wormholes while the second and the third lines correspond to the diagrams with $2$ and $3$ wormholes respectively. 

This diagrammatic rule is {\it similar but different} from the rule of computing the correlators in the double-trace-deformed AdS/CFT \cite{Hartman:2006dy, Giombi:2011ya}: In AdS/CFT, the double-trace deformation changes the boundary condition for one of the fields (to be denoted by $\varphi$) in AdS \cite{Witten:2001ua,Berkooz:2002ug}, and  modifies its bulk-to-bulk propagator. Therefore, whenever $\varphi$ shows up as an intermediate state in the Witten diagrams, we need to add additional contributions which convert the bulk-to-bulk propagators of $\varphi$ from the original one to the new one. Although such additional contributions seem similar to the extra terms on the right hand sides of \eqref{eq:twoderivatives} and \eqref{eq:threederivatives}, there is one important difference: In the AdS/CFT setup, such additional contributions show up only for the four- and higher-point functions since there will be no intermediate particle exchanges for the two- and three-point functions. In contrast, here we have extra terms already for the two and the three-point functions. We will later show that this apparent difference is because of the mixing of operators and once we resolve the mixing using the Gram-Schmidt process, the results take exactly the same form as the correlation functions in the double-trace-deformed AdS/CFT.

Similarly we can compute the four derivatives but the expression becomes more complicated:
\beq\label{eq:fourderivative}
\begin{aligned}
&d_{n_1}d_{n_2}d_{n_3}d_{n_4}F=\langle n_1,n_2,n_3,n_4 \rangle-\frac{\langle n_1,0\rangle \langle 0,n_2,n_3,n_4\rangle}{\langle 0,0\rangle}-\frac{\langle n_2,0\rangle \langle 0,n_3,n_4,n_1\rangle}{\langle 0,0\rangle}\\
&-\frac{\langle n_3,0\rangle \langle 0,n_4,n_1,n_2\rangle}{\langle 0,0\rangle}-\frac{\langle n_4,0\rangle \langle 0,n_1,n_2,n_3\rangle}{\langle 0,0\rangle}\\
&-\frac{\langle n_1,n_2,0\rangle\langle 0,n_3,n_4\rangle}{\langle 0,0\rangle}-\frac{\langle n_1,n_3,0\rangle\langle 0,n_2,n_4\rangle}{\langle 0,0\rangle}-\frac{\langle n_1,n_4,0\rangle\langle 0,n_2,n_3\rangle}{\langle 0,0\rangle}\\
&+\text{(terms with more than one wormholes)}
\end{aligned}
\eeq
Note that the relations \eqref{eq:twoderivatives}, \eqref{eq:threederivatives} and \eqref{eq:fourderivative} are derived originally to $n_i>0$, but they can be applied also for $n_i=0$: One can check explicitly that all these formulae vanish when we set one of $n_i$'s to zero. This is consistent with the fact that $d_0$ identically vanishes owing to its definition \eqref{eq:defofdtn}. This property plays an important role when deriving integral representations for the normal-ordered correlators in section \ref{subsec:qfunction}.

\paragraph{Integral representation} These diagrammatic rules allow us to express the correlators in terms of the partial derivatives $\langle n_1,\ldots, n_m\rangle=\del_{n_1}\cdots \del_{n_m}F$, which in turn can be computed from the integral representations for the free energy $F$ \eqref{eq:integralFanti} and \eqref{eq:integralFsym}. For both antisymmetric and symmetric representations, the results can be expressed compactly as
\begin{align}
&\langle n_1,n_2\rangle =\oint d\mu_2 \left(g(x-x^{-1})\right)^{n_1+n_2}\comma\qquad \langle n_1,n_2,n_3\rangle =\oint d\mu_3 \left(g(x-x^{-1})\right)^{n_1+n_2+n_3}\comma\nn\\
&\langle n_1,n_2,n_3,n_4\rangle =\oint d\mu_4 \left(g(x-x^{-1})\right)^{n_1+n_2+n_3+n_4}\comma \label{eq:integralelementaryderivatives}
 \end{align}
where the measures $d\mu_{2,3,4}$ are given by
\beq\label{eq:measureantisym}
\begin{aligned}
\text{\bf antisymmetric:}\quad &d\mu_{2}=\frac{1}{4\pi g}\frac{dx (1+x^{-2})}{2\pi i}\frac{1}{1+e^{-2\pi g (x+1/x)-t_0}}\comma\\
&d\mu_3 =\frac{1}{16\pi g}\frac{dx (1+x^{-2})}{2\pi i}\frac{1}{\left(\cosh \left(g\pi (x+1/x)+\frac{t_0}{2}\right)\right)^2}\comma\\
&d\mu_4=-\frac{1}{16\pi g}\frac{dx (1+x^{-2})}{2\pi i}\frac{\sinh \left(g\pi (x+1/x)+\frac{t_0}{2}\right)}{\left(\cosh \left(g\pi (x+1/x)+\frac{t_0}{2}\right)\right)^3}\comma
\end{aligned}
\eeq
\beq\label{eq:measuresymm}
\begin{aligned}
\text{\bf symmetric:}\quad &d\mu_{2}=-\frac{1}{4\pi g}\frac{dx (1+x^{-2})}{2\pi i}\frac{1}{1-e^{-2\pi g (x+1/x)-t_0}}\comma\\
&d\mu_3 =\frac{1}{16\pi g}\frac{dx (1+x^{-2})}{2\pi i}\frac{1}{\left(\sinh \left(g\pi (x+1/x)+\frac{t_0}{2}\right)\right)^2}\comma\\
&d\mu_4=-\frac{1}{16\pi g}\frac{dx (1+x^{-2})}{2\pi i}\frac{\cosh \left(g\pi (x+1/x)+\frac{t_0}{2}\right)}{\left(\sinh \left(g\pi (x+1/x)+\frac{t_0}{2}\right)\right)^3}\period
\end{aligned}
\eeq
\subsection{Gram-Schmidt analysis and Q-functions\label{subsec:qfunction}} We now define the normal-ordered operators $\tilde{\mathcal{O}}_L\equiv\, :\!\!\!\tilde{\Phi}^{L}\!\!\!:$, whose two-point functions are diagonal. As is the case with the fundamental Wilson loop \cite{Giombi:2018qox,Giombi:2018hsx}, this can be achieved by the application of the Gram-Schmidt orthogonalization.
As a result of a direct application of the Gram-Schmidt process, we obtain\footnote{See \cite{Gerchkovitz:2016gxx,Baggio:2016skg,Rodriguez-Gomez:2016ijh,Rodriguez-Gomez:2016cem,Billo:2017glv,Ishtiaque:2017trm,Chen:2017fvl,Bourget:2018obm,Bourget:2018fhe,Grassi:2019txd} for applications of the Gram-Schmidt orthogonalization to SCFTs.}
\beq\label{eq:normalordereddef}
\tilde{\mathcal{O}}_{L}\,\,=\frac{1}{D_{L}}\left|\begin{array}{cccc}d_{1}d_{1}\tilde{F}&d_{1}d_{2}\tilde{F}&\cdots& d_{1}d_{L}\tilde{F}\\d_{2}d_{1}\tilde{F}&d_{2}d_{2}\tilde{F}&\cdots &d_{2}d_{L}\tilde{F}\\\vdots&\vdots&\ddots&\vdots\\d_{L-1}d_{1}\tilde{F}&d_{L-1}d_{2}\tilde{F}&\cdots &d_{L-1}d_{L}\tilde{F}\\
\tilde{\Phi}&\tilde{\Phi}^{2}&\cdots &\tilde{\Phi}^{L}\end{array}\right|\comma
\eeq
with
\beq
D_{L}=\left|\begin{array}{ccc}d_{1}d_{1}\tilde{F}&\cdots& d_{1}d_{L-1}\tilde{F}\\d_{2}d_{1}\tilde{F}&\cdots &d_{2}d_{L-1}\tilde{F}\\\vdots&\ddots&\vdots\\d_{L-1}d_{1}\tilde{F}&\cdots &d_{L-1}d_{L-1}\tilde{F}\end{array}\right|\period
\eeq

It turns out that the expression \eqref{eq:normalordereddef} can be rewritten purely in terms of the partial derivatives $\langle n_1, n_2\rangle$ given in  \eqref{eq:deflanglebracket}: 
\beq\label{eq:normalordered2}
\tilde{\mathcal{O}}_{L}=\frac{1}{\tilde{D}_{L}}\left|\begin{array}{cccc}\langle 0,0\rangle&\langle 0,1\rangle&\cdots& \langle 0, L\rangle\\\langle 1,0\rangle&\langle 1,1\rangle&\cdots &\langle 1,L\rangle\\\vdots&\vdots&\ddots&\vdots\\\langle L-1,0\rangle&\langle L-1,1\rangle&\cdots &\langle L-1,L\rangle\\  (\!( \tilde{\Phi}^{0} )\!)\left(=0\right)&\tilde{\Phi}&\cdots &\tilde{\Phi}^{L}\end{array}\right|\comma
\eeq
with
\beq\label{eq:DtilL}
\tilde{D}_{L}=\left|\begin{array}{ccc}\langle 0,0\rangle &\cdots& \langle 0,L-1\rangle\\\langle 1,0\rangle&\cdots &\langle 1,L-1\rangle\\\vdots&\ddots&\vdots\\\langle L-1,0\rangle&\cdots &\langle L-1,L-1\rangle\end{array}\right|\period
\eeq
Here the lower-left corner of \eqref{eq:normalordered2} is $0$ but we denoted it by $(\!( \tilde{\Phi}^{0} )\!)$ for a reason that becomes clear below.
The equivalence between the two expressions, \eqref{eq:normalordereddef} and \eqref{eq:normalordered2}, can be proven in the following way: We start from \eqref{eq:normalordered2} and subtract $(\del_0 \del_n F)/(\del_0^2F)$ times the first columns in  from the $n$-th columns. After that, we subtract $(\del_0 \del_n F)/(\del_0^2F)$ times the first rows from the $n$-th rows and rewrite them using the relation between $\del_n$ and $d_n$ given by \eqref{eq:defofdtn}. Performing the same manipulation to \eqref{eq:DtilL}, we can show the equivalence between \eqref{eq:normalordereddef} and \eqref{eq:normalordered2}.

Now using the expression \eqref{eq:normalordered2}, we can compute the correlation functions of normal ordered operators $\langle \!\langle \prod_{k}\tilde{\mathcal{O}}_{L_k}\rangle\!\rangle$ in the following steps:
\begin{enumerate}
\item We first express each nornal-ordered operator $\tilde{\mathcal{O}}_{L_k}$ as a sum of un-normal-ordered operators $\tilde{\Phi}^{L}$'s using \eqref{eq:normalordered2}. 
\item We next replace a product of un-normal-ordered operators $\prod_{k}\tilde{\Phi}^{L_k}$ with $\left(\prod_{k}d_{L_k}\right)\tilde{F}$. In particular, we also replace $(\!(\tilde{\Phi}^{0})\!)(=0)$ with $d_0$.  This is a consistent manipulation since $d_0$ identically vanishes because of its definition \eqref{eq:defofdtn}, and it allows us to treat all $d_L$'s in a uniform way.
\item We then decompose each $\left(\prod_{k}d_{L_k}\right)\tilde{F}$ into $\del_{L_k}F$'s using \eqref{eq:twoderivatives}, \eqref{eq:threederivatives} and \eqref{eq:fourderivative}. After that, we evaluate them using the integral representations \eqref{eq:integralelementaryderivatives}: Namely we replace each $\del_{L_k}$ with the insertion of the monomial $(g(x-x^{-1}))^{L_k}$ in the integral representations.
\end{enumerate}
To express the results obtained by these procedures, it is convenient to define a polynomial
\beq\label{eq:defQfunctions}
Q_{L}(X)\equiv \frac{1}{\tilde{D}_{L}}\left|\begin{array}{cccc}\langle 0,0\rangle&\langle 0,1\rangle&\cdots& \langle 0, L\rangle\\\langle 1,0\rangle&\langle 1,1\rangle&\cdots &\langle 1,L\rangle\\\vdots&\vdots&\ddots&\vdots\\\langle L-1,0\rangle&\langle L-1,1\rangle&\cdots &\langle L-1,L\rangle\\ 1&X&\cdots &X^{L}\end{array}\right|\comma
\eeq
and introduce the notation,
\beq
\begin{aligned}
\,\,[n_1,\ldots, n_m]&\equiv \oint d\mu_m \prod_{k=1}^{m}Q_{n_k} \left(g(x-x^{-1})\right)\period
\end{aligned}
\eeq
We can then express the two- and the three-point functions as
\begin{align}
&N^{-1}\langle \!\langle \tilde{\mathcal{O}}_{n_1}\tilde{\mathcal{O}}_{n_2} \rangle \!\rangle=[ n_1,n_2] -\frac{[ n_1,0][ 0,n_2]}{[ 0,0]}\comma\nn\\
&N^{-1}\langle \!\langle \tilde{\mathcal{O}}_{n_1}\tilde{\mathcal{O}}_{n_2}\tilde{\mathcal{O}}_{n_3}\rangle \!\rangle=[ n_1,n_2,n_3 ]-\frac{[ n_1,0] [ 0,n_2,n_3]}{[ 0,0]}-\frac{[ n_2,0] [ 0,n_3,n_1]}{[ 0,0]}-\frac{[ n_3,0] [ 0,n_1,n_2]}{[ 0,0]}\nn\\
&+\frac{[ n_1,0 ][ 0,n_2,0 ] [ 0,n_3]}{([ 0,0])^2}+\frac{[ n_2,0 ][ 0,n_3,0 ] [ 0,n_1]}{([ 0,0])^2}+\frac{[ n_3,0 ][ 0,n_1,0 ] [ 0,n_2]}{([ 0,0])^2}\nn\\
&-\frac{[ n_1,0][ n_2,0][ n_3,0][ 0,0,0 ]}{([ 0,0 ])^3}\comma\label{eq:correlatorcomplicated}
\end{align}

\begin{figure}[t]
\centering
\includegraphics[clip, height=3cm]{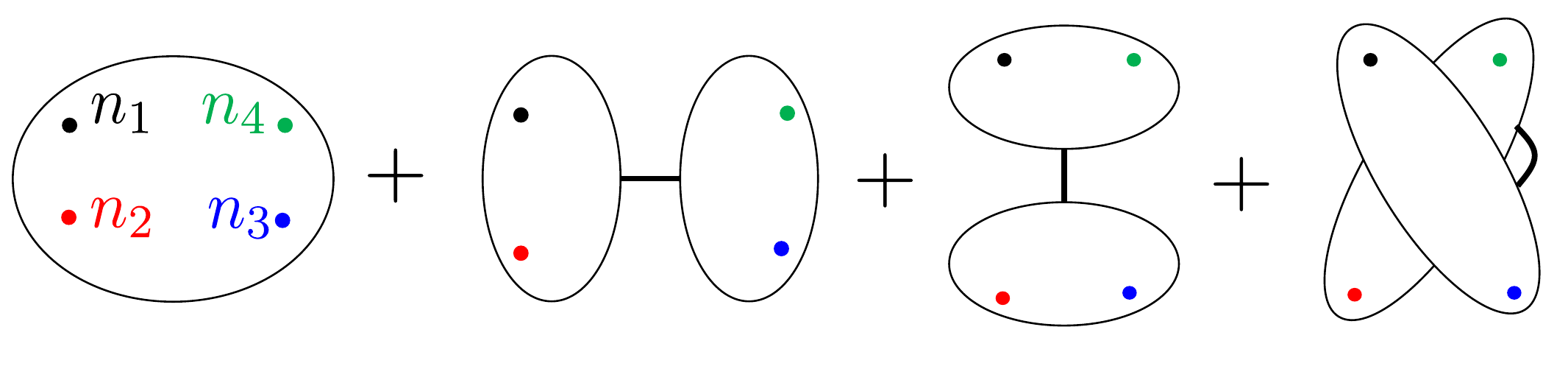}
\caption{The diagrammatic rule to compute the four-point function after the Gram-Schmidt orthogonalization. As in the standard rule for computing the correlators in the double-trace-deformed AdS/CFT, we correct correlators only when the deformed operator is exchanged in the diagram. For the four-point function, there are three possibilities (s-, t-, and u-channels) and we need to subtract them to compute the correct four-point function. Here again the thick black lines denote the wormholes.}\label{fig:wormhole2}
\end{figure}

These expressions can be further simplified by using the following fact: By construction \eqref{eq:defQfunctions}, the Gram-Schmidt process gives an orthogonal basis of functions $Q_{n}$'s under the measure $d\mu_2$. This means $[n_j,0]=[0,n_j]=0$ for all $n_j>0$. Because of this, all the extra terms in \eqref{eq:correlatorcomplicated} vanish and we simply have
\beq
N^{-1}\langle \!\langle \tilde{\mathcal{O}}_{n_1}\tilde{\mathcal{O}}_{n_2} \rangle \!\rangle=[ n_1,n_2] \comma\qquad
N^{-1}\langle \!\langle\tilde{\mathcal{O}}_{n_1}\tilde{\mathcal{O}}_{n_2} \tilde{\mathcal{O}}_{n_3}\rangle \!\rangle=[ n_1,n_2,n_3 ]\period 
\eeq
They can be expressed more explicitly as integrals of the polynomials $Q_n$:
\begin{align}
\langle \!\langle \tilde{\mathcal{O}}_{n_1}\tilde{\mathcal{O}}_{n_2} \rangle \!\rangle&= N\oint d\mu_2 \,Q_{n_1} \left(g(x-x^{-1})\right)Q_{n_2} \left(g(x-x^{-1})\right)\comma\label{eq:correlatorfinal}\\
\langle \!\langle\tilde{\mathcal{O}}_{n_1}\tilde{\mathcal{O}}_{n_2}\tilde{\mathcal{O}}_{n_3}\rangle \!\rangle&= N\oint d\mu_3 \,Q_{n_1} \left(g(x-x^{-1})\right)Q_{n_2} \left(g(x-x^{-1})\right)Q_{n_3} \left(g(x-x^{-1})\right)\period\nn
\end{align}
The computation can be readily generalized to  the four-point functions, but the result takes a more complicated form. For instance the analogue of \eqref{eq:correlatorcomplicated} reads
\begin{align}
\begin{aligned}
&N^{-1}\langle \!\langle\tilde{\mathcal{O}}_{n_1}\tilde{\mathcal{O}}_{n_2}\tilde{\mathcal{O}}_{n_3}\tilde{\mathcal{O}}_{n_4}\rangle \!\rangle=[ n_1,n_2,n_3,n_4 ]-\frac{[ n_1,0] [ 0,n_2,n_3,n_4]}{[ 0,0]}-\frac{[ n_2,0] [ 0,n_3,n_4,n_1]}{[ 0,0]}\\
&-\frac{[ n_3,0] [ 0,n_4,n_1,n_2]}{[ 0,0]}-\frac{[ n_4,0] [ 0,n_1,n_2,n_3]}{[ 0,0]}\\
&\red{-\frac{[ n_1,n_2,0][ 0,n_3,n_4]}{[ 0,0]}-\frac{[ n_1,n_3,0][ 0,n_2,n_4]}{[ 0,0]}-\frac{[ n_1,n_4,0][ 0,n_2,n_3]}{[ 0,0]}}+\cdots\comma
\end{aligned}
\end{align}
where $+\cdots$ denotes terms with more than one wormholes. Importantly, the three terms written in the last line do not include a factor $[n_j,0](=0)$. We therefore need to keep those terms when writing down an integral representation and the result reads (see also Figure \ref{fig:wormhole2})
\begin{align}
&\langle \!\langle \tilde{\mathcal{O}}_{n_1}\tilde{\mathcal{O}}_{n_2}\tilde{\mathcal{O}}_{n_3}\tilde{\mathcal{O}}_{n_4}\rangle \!\rangle=\nn\\
&N\oint d\mu_{4} \,Q_{n_1}Q_{n_2} Q_{n_3}Q_{n_4}-N\frac{\left(\oint d\mu_3\,Q_{n_1}Q_{n_2}Q_0\right)\left(\oint d\mu_3\,Q_{0}Q_{n_3}Q_{n_4}\right)}{\oint d\mu_2\, Q_0 Q_0}\label{eq:correlatorfinal2}\\
&-N\frac{\left(\oint d\mu_3\,Q_{n_1}Q_{n_3}Q_0\right)\left(\oint d\mu_3\,Q_{0}Q_{n_2}Q_{n_4}\right)}{\oint d\mu_2\, Q_0 Q_0}-N\frac{\left(\oint d\mu_3\,Q_{n_1}Q_{n_4}Q_0\right)\left(\oint d\mu_3\,Q_{0}Q_{n_2}Q_{n_3}\right)}{\oint d\mu_2\, Q_0 Q_0}\period\nn
\end{align}
The expressions \eqref{eq:correlatorfinal} and \eqref{eq:correlatorfinal2} are the precise analogues of the correlation functions in the double-trace-deformed AdS/CFT. Namely there are no corrections for the two- and the three-point functions while the four- and higher-point functions receive corrections whenever the deformed operators are exchanged.

The integral representations similar to \eqref{eq:correlatorfinal} were obtained for the correlation functions on the fundamental Wilson loop \cite{Giombi:2018qox,Giombi:2018hsx}. There the polynomials $Q_n$'s were unexpectedly related to the $Q$-functions in the Quantum Spectral Curve approach \cite{Gromov:2013pga,Gromov:2014caa,Gromov:2015dfa}, which is the most efficient method to compute the operator spectrum in planar $\mathcal{N}=4$ SYM. The appearance of the $Q$-functions in the integral representations  was taken as a strong hint that the Quantum Spectral Curve can be applied not only to the spectrum but also to the correlation functions. Here again we are seeing the same structure. However, there are also notable differences. 

First unlike the case of the fundamental Wilson loop where the measure $d\mu$ was the same for all the topological correlators, here the measures $d\mu_{2,3,4}$ depend on the number of operators. This seems to be related to the difference of the structures of the operator product expansions in the large $N$ limit. In the case of the fundamental Wilson loop, the operators corresponding to the $Q$-functions form a closed subsector of OPE in the large $N$ limit. In particular, there is one-to-one correspondence between the OPE of the operators and the multiplication of the $Q$-functions. To realize such a structure in the integral representation, the measure {\it need to be} the same for all the correlation functions. On the other hand, the situation is quite different for the Giant Wilson loops: The single-particle operators, which correspond to the $Q$-functions, do not form a subsector of OPE since their OPEs necessarily contain the multi-particle operators even in the large $N$ limit. Therefore we do not expect the measures to be the same\fn{Put differently, the measure $d\mu_3$ can be thought of as an ``effective measure'' which one obtains after subtracting the effects of the two-particle operators, although we do not know how to make this statement more precise.} and that is indeed realized in the formulae \eqref{eq:correlatorfinal}. This structure of the OPE is common also to the single-trace operators in the large $N$ limit. Also there, the OPE of two single-trace operators is not closed, and contains a double-trace operator. This suggests that the measures for the correlation functions of single-trace operators may also depend on the number of operators.

Second the Quantum Spectral Curve for the Giant Wilson loop has not been formulated yet. At least for the Giant Wilson loop in the antisymmetric representation, which is dual to D5-brane, there is already evidence that the problem is integrable \cite{Correa:2013em}, and our observation suggests that the formulation in terms of the Quantum Spectral Curve should be possible. The situation is less clear for the Giant Wilson loop in the symmetric representation since the dual D3-brane is not in the classification of integrable boundaries at strong coupling \cite{Dekel:2011ja}. Nevertheless, our formula is still applicable and the result takes a form reminiscent of integrals of $Q$-functions. It would be interesting to study the integrability properties of these Giant Wilson loops at weak coupling, and  if they turn out to be integrable, write down the Quantum Spectral Curve.

\subsection{Antisymmetric loop at strong coupling}
We now explicitly evaluate the topological correlators on the antisymmetric Wilson loop at strong coupling. In order to compare with the D-brane analysis in section \ref{sec:D5}, we focus on the special case of the $1/2$-BPS Wilson loop by setting $A=2\pi$ (or equivalently $t_1=0$). 
\paragraph{Saddle point and measure at strong coupling}
We first consider the saddle point equation \eqref{eq:Jsaddlepoint}, which can be expressed using the integral representation as
\beq\label{eq:logfreefermi}
\kappa =\frac{1}{4\pi g} \oint \frac{dx(1+x^{-2})}{2\pi i}\log \left(1+e^{2\pi g(x+1/x)+t_0}\right)\period
\eeq
In the limit $g\to \infty$, this equation can be solved explicitly once we rewrite $t_0$ as
\beq\label{eq:saddlet0}
t_0 \equiv -4\pi g \cos \theta_k\period
\eeq
We then get the following saddle point equation at strong coupling which determines $\theta$ as a function of $\kappa=k/N$:
\beq
\kappa\overset{g\to\infty}{=} \frac{\theta_k-\cos\theta_k\sin\theta_k}{\pi }\,\,\left(=\frac{1}{4\pi g}\int^{\theta_k}_{-\theta_k} \frac{d\alpha \cos \alpha}{\pi} 4\pi g (\cos \alpha-\cos \theta_k)\right) \period
\eeq
As we see later in section \ref{sec:D5}, the parameter $\theta_k$ determines the size of the D5-brane on $S^5$ while here it governs the size of the Fermi-distribution in \eqref{eq:logfreefermi}. A similar qualitative relation seems to exist also for the symmetric loop and the D3-brane as we see in the next subsection.

Plugging in the saddle point value of $t_0$ \eqref{eq:saddlet0} to \eqref{eq:measureantisym} and taking the $g\to \infty$ limit, we get the following expressions for the measures:
\beq
\begin{aligned}
\oint d\mu_{2}\overset{g\to\infty}{=}&\frac{1}{4\pi g}\int_{-\theta_k}^{\theta_k}\frac{d\alpha \cos \alpha}{\pi }\comma\qquad
\oint d\mu_3 \overset{g\to\infty}{=}\frac{1}{16\pi^2 g^2}\int\frac{d\alpha \cos\alpha}{\pi }\delta (\cos\alpha-\cos\theta_k)\comma\\
\oint d\mu_4\overset{g\to\infty}{=}&\frac{1}{64\pi^4 g^3}\int \frac{d\alpha \cos \alpha}{\pi}\delta^{\prime}(\cos\alpha-\cos\theta_k)\period
\end{aligned}
\eeq
Here we used the identity
\beq
\lim_{\Lambda\to \infty}\frac{\Lambda}{\cosh (2\Lambda x)^2}=\delta (x)\period
\eeq
Changing the variable from $\alpha$ to $y =\frac{\sin\alpha}{\sin\theta}$, we obtain
\beq\label{eq:everythingintermsofy}
\begin{aligned}
\oint d\mu_{2}\overset{g\to\infty}{=}&\frac{\sin\theta_k}{4\pi^2 g}\int_{-1}^{1}dy\comma\qquad \oint d\mu_3 \overset{g\to\infty}{=}\frac{1}{16\pi^3 g^2}\frac{\cos\theta_k}{\sin\theta_k}\int dy \left(\delta (y-1)+\delta (y+1)\right)\comma\\
\oint d\mu_4\overset{g\to\infty}{=}&-\frac{\cos\theta_k}{64\pi^5 g^3\sin^3\theta_k}\int dy  \frac{\sqrt{1-y^2\sin^2\theta_k }}{y}\left(\delta^{\prime} (y-1)+\delta^{\prime} (y+1)\right) \period
\end{aligned}
\eeq
Here we rewrote the derivative of the delta function using the following identity (where $f(y)\equiv \sqrt{1-y^2\sin^2\theta_k}$):
\beq
\begin{aligned}
&\delta^{\prime}(\cos \alpha-\cos \theta_k)=\delta^{\prime} (f(y)-\cos \theta_k)=\frac{d}{df}\delta (f (y)-\cos \theta_k)\\
&=\frac{1}{f^{\prime}(y)}\frac{d}{dy}\left[ \frac{1}{f^{\prime}(1)}\left(\delta (y-1)+\delta (y+1)\right)\right]
\end{aligned}
\eeq
\paragraph{Q-functions}
The next step is to compute the $Q$-functions at strong coupling. Although the $Q$-functions were originally defined by the Gram-Schmidt determinants \eqref{eq:defQfunctions}, one can compute them more directly by requiring the orthogonality under the two-point measure
\beq
\oint d\mu_2 \,\,Q_{n} Q_{m} \propto \delta_{nm}\comma
\eeq
and imposing that it is a polynomial in $X$ of degree $n$:
\beq\label{eq:Qasympt}
Q_n (X) = X^{n}+\cdots\period
\eeq
In terms of the variable $y$ introduced in \eqref{eq:everythingintermsofy}, the condition \eqref{eq:Qasympt} reads
\beq
Q_n =\left(2i g \sin\theta_k y\right)^{n}+\cdots\period
\eeq

It is known that the orthogonal polynomials with the measure $\int^{1}_{-1}dy$ are the Legendre polynomial $P_n (y)$:
\beq\label{eq:legendre}
\begin{aligned}
&\int_{-1}^{1}P_{n}(y)P_{m}(y)=\frac{\delta_{nm}}{n+\frac{1}{2}}\comma\qquad P_n (+1)=1\comma \quad P_n (-1)=(-1)^{n}\comma\\
&P_n (y)=\frac{1}{2^{n}}\sum_{k=0}^{n}\left(\frac{\Gamma (n+1)}{\Gamma (k+1)\Gamma (n-k+1)}\right)^{2}(y-1)^{n-k}(y+1)^{k}\period
\end{aligned}
\eeq
From a comparison of the leading coefficients, we conclude that the $Q$-function at strong coupling is given by
\beq
Q_n =\frac{\left(i g\sin \theta_k\right)^{n}\sqrt{\pi}\,\,\Gamma (n+1)}{\Gamma (n+\frac{1}{2})}P_n (y)\period
\eeq

Note that \eqref{eq:logfreefermi} has the same structure as the free energy of free Fermi gas. From this point of view, each $Q_n$ corresponds to a different way of deforming the Fermi distribution. At strong coupling, the Fermi distribution has finite support and therefore there exist infinitely many different deformations labelled by the integer $n$. These deformations correspond to the Kaluza-Klein modes with higher $S^5$ angular momenta. As we see in section \ref{subsec:symmetricloop}, the situation is quite different for the D3-brane which is described by free Bose gas. See also Figure \ref{fig:bosefermi}.
\paragraph{Two-, three- and four-point functions}
Having identified the $Q$-function with the Legendre polynomial, it is by now a trivial exercise to compute the two- and the three-point functions. The two-point function can be computed by using the first equation in \eqref{eq:legendre}, and the result reads
\beq
\langle \!\langle \tilde{\mathcal{O}}_{n}\tilde{\mathcal{O}}_{m} \rangle \!\rangle=N \delta_{n m}\frac{(i g \sin\theta_k)^{n+m}\sin\theta_k}{4\pi g}\frac{\left(\Gamma(n+1)\right)^2}{\Gamma (n+\frac{1}{2})\Gamma (n+\frac{3}{2})}\period
\eeq
On the other hand, the three-point functions are given by integrals with the measure $d\mu_3$. Since $d\mu_3$ is a delta function, we simply need to evaluate the product of the Legendre polynomials at $y=\pm 1$. We then get
\beq
\langle \!\langle\tilde{\mathcal{O}}_{n}\tilde{\mathcal{O}}_{m}\tilde{\mathcal{O}}_{l}\rangle \!\rangle=N\frac{[1+(-1)^{n+m+l}](ig \sin\theta_k)^{n+m+l}\cot\theta_k}{16\pi^{\frac{3}{2}}g^2}\frac{\Gamma (n+1)\Gamma (m+1)\Gamma (l+1)}{\Gamma (n+\frac{1}{2})\Gamma (m+\frac{1}{2}) \Gamma (l+\frac{1}{2})}\period
\eeq
Combining the two results, we obtain the following expression for the normalized three-point functions:
\beq
\begin{aligned}
&\frac{\langle \!\langle \tilde{\mathcal{O}}_{n}\tilde{\mathcal{O}}_{m}\tilde{\mathcal{O}}_{l}\rangle \!\rangle}{\sqrt{\langle \!\langle \tilde{\mathcal{O}}_{n}\tilde{\mathcal{O}}_{n} \rangle \!\rangle\langle \!\langle \tilde{\mathcal{O}}_{m}\tilde{\mathcal{O}}_{m} \rangle \!\rangle\langle \!\langle \tilde{\mathcal{O}}_{l}\tilde{\mathcal{O}}_{l} \rangle \!\rangle}}=\\
&\frac{(1+(-1)^{n+m+l})}{2}\frac{\cos\theta_k}{(\sin\theta_k)^{\frac{5}{2}}}\sqrt{\frac{(n+\frac{1}{2}) (m+\frac{1}{2})(l+\frac{1}{2})}{gN}}\period
\end{aligned}
\eeq
We can also compute the four-point functions using the formula \eqref{eq:correlatorfinal2} and the measure \eqref{eq:measureantisym}. Here we show a sample of results which we later compare with the D-brane computation:
\beq
\begin{aligned}
&\frac{\langle \!\langle  \tilde{\mathcal{O}}_1\tilde{\mathcal{O}}_1\tilde{\mathcal{O}}_1\tilde{\mathcal{O}}_1\rangle\!\rangle}{\langle \!\langle \tilde{\mathcal{O}}_1\tilde{\mathcal{O}}_1\rangle\!\rangle^2}
=-\frac{9}{8gN}\frac{1}{\sin^3\theta_k},\\
&\frac{\langle \!\langle \tilde{\mathcal{O}}_1\tilde{\mathcal{O}}_1\tilde{\mathcal{O}}_2\tilde{\mathcal{O}}_2\rangle\!\rangle}{\langle \!\langle \tilde{\mathcal{O}}_1\tilde{\mathcal{O}}_1\rangle\!\rangle 
\langle \!\langle \tilde{\mathcal{O}}_2\tilde{\mathcal{O}}_2\rangle\!\rangle}
=\frac{15}{4 g N \sin^3 \theta_k}\left(3\cot^2\tk-\frac{1}{2}\right)\period
\end{aligned}
\eeq

In section \ref{sec:D5}, we show that all these results can be reproduced from perturbation theory on the probe D5-brane in $AdS_5\times S^5$. Using the localization formulae, we can also compute perturbative and nonperturbative $1/g$ corrections to the leading strong-coupling results computed here. It would be an interesting future problem to perform such a computation explicitly and compare them with the stringy corrections on the D-brane side.

\subsection{Symmetric loop at strong coupling\label{subsec:symmetricloop}}
\begin{figure}[t]
\centering
\begin{minipage}{0.45\hsize}
\centering
\includegraphics[clip,height=3.5cm]{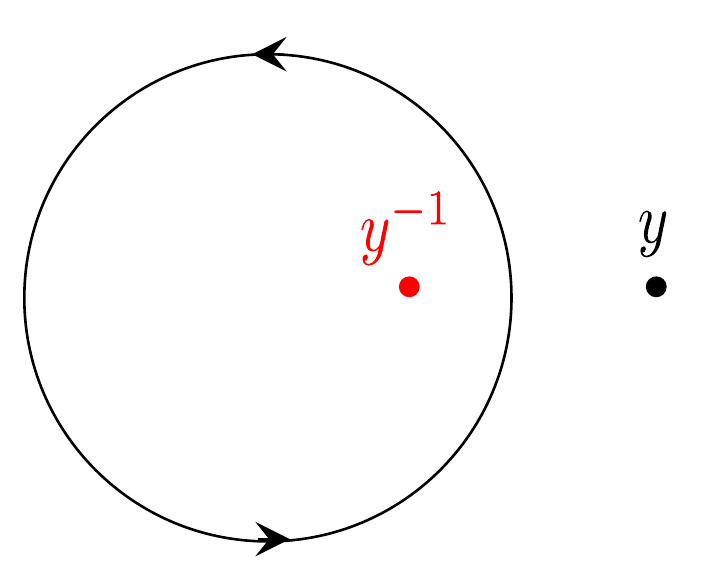}
\subcaption{$|y|>1$}
\end{minipage}
\begin{minipage}{0.45\hsize}
\centering
\includegraphics[clip,height=3.5cm]{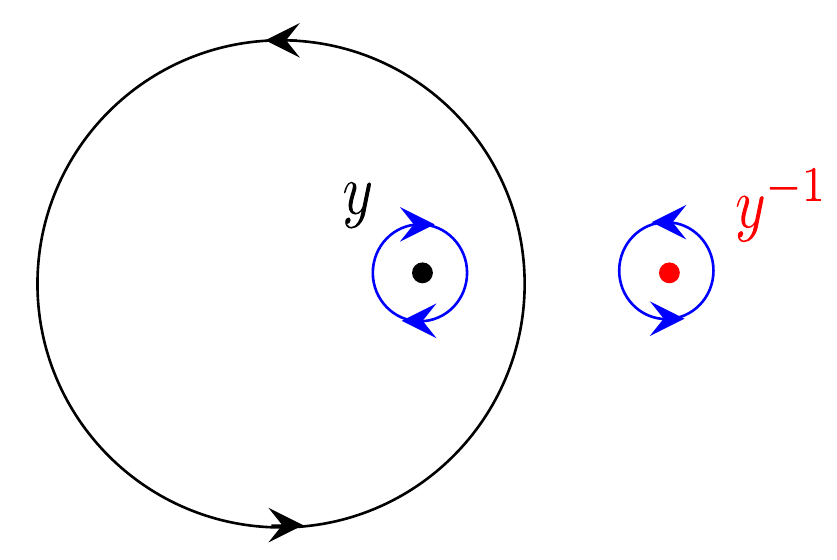}
\subcaption{$|y|<1$}
\end{minipage}
\caption{The integration contour for $x$ for (a) $|y|>1$, and (b) $|y|<1$. When $|y|>1$, the integration contour is along a unit circle. However, if we analytically continue this integral to $|y|<1$, the poles in the integrand $x=y^{\pm 1}$ cross the contour and produce extra contributions (small circles around $y^{\pm 1}$). In the strong coupling limit, such extra contributions are dominant.}\label{fig:symcontour}
\end{figure}

We now study the correlation functions on the symmetric Wilson loop in the strong coupling limit. Again we focus on the $1/2$-BPS Wilson loop and set $t_1=0$ (or equivalently $A=2\pi$).
\paragraph{Saddle point}
The saddle point equation for the symmetric Giant Wilson loop reads
\beq
\kappa=-\frac{1}{4\pi g}\oint \frac{dx (1+x^{-2})}{2\pi i}\log \left(1-e^{2\pi g (x+1/x)+t_0}\right)\comma
\eeq
which can be rewritten by integration by parts as
\beq\label{eq:rewriteintegrandsaddle}
\kappa=\oint \frac{dx}{4\pi i x}\frac{(x-x^{-1})^2}{1-e^{-2\pi g (x+1/x)-t_0}}\period
\eeq
Unlike the antisymmetric Wilson loop, we need to carefully define the right hand side of \eqref{eq:rewriteintegrandsaddle} since the integrand can be singular on the integration cycle of $x$, which is along the unit circle. For this purpose, it is convenient to parameterize $t_0$ as
\beq\label{eq:t0parameterize}
t_0 =-2\pi g\left(y+\frac{1}{y}\right)\period
\eeq 
We can then see that the integrand has poles at $x=y^{\pm 1}$. When $|y|>1$, these poles are away from the integration contour and the integral \eqref{eq:rewriteintegrandsaddle} is well-defined. However, if we analytically continue it to the $|y|<1$ region, the poles $x=y^{\pm 1}$ cross the contour and produce extra contributions to \eqref{eq:rewriteintegrandsaddle} (see also Figure \ref{fig:symcontour}). Therefore we have  
\beq\label{eq:newsaddleeq}
\kappa =\begin{cases}\oint_{U} \frac{dx}{4\pi i x}\frac{(x-x^{-1})^2}{1-e^{-2\pi g (x-y)(1-1/xy)}}\qquad &|y|>1\\ \frac{y^{-1}-y}{2\pi g}+\oint_{U} \frac{dx}{4\pi i x}\frac{(x-x^{-1})^2}{1-e^{-2\pi g (x-y)(1-1/xy)}}\qquad &|y|<1 \end{cases}\comma
\eeq
where $U$ is the contour along the unit circle and $(y^{-1}-y)/2\pi g$ is the contribution from the poles.

It turns out that the saddle point at strong coupling $g\gg 1$ is in the region $|y|<1$. This follows from the fact that the integral along the unit circle is exponentially small both for $|y|>1$ and $|y|<1$:
\beq
\oint_{U} \frac{dx}{4\pi i x}\frac{(x-x^{-1})^2}{1-e^{-2\pi g (x-y)(1-1/xy)}} \sim e^{-g \bullet}\period
\eeq
Therefore the saddle-point equation \eqref{eq:newsaddleeq} can be approximated at strong coupling as
\beq\label{eq:saddleys}
\kappa = \frac{y^{-1}-y}{2\pi g}\qquad\qquad  |y|<1 \period
\eeq
In order to make contact with the D-brane analysis in section \ref{sec:D3}, it is useful to parametrize the solution to this equation by $y=e^{-u_k}$. In terms of $u_k$, \eqref{eq:saddleys} can be rewritten as
\beq
\sinh (u_k)= 4\pi g \kappa\period
\eeq
We will later see that $u_k$ determines the size of the D3-brane in AdS.

Before proceeding, let us point out that there is a close analogy with the Bose Einstein condensation: The right hand side of \eqref{eq:rewriteintegrandsaddle} has the same structure as the distribution of free Bose gas, and the coupling $g$ can be identified with the inverse temperature $\beta$. From this point of view, the contribution from the poles in \eqref{eq:newsaddleeq} can be viewed as an analogue of the Bose-Einstein condensation. The fact that the result at $g\gg 1$ is dominated by these poles parallels the fact that, at zero temperature $(\beta\gg 1)$, all the particles in the free Bose gas are in the condensate. Below we will see that this ``Bose-Einstein condensation'' is responsible for the difference of the spectra on the antisymmetric loop and the symmetric loop at strong coupling---namely the absence of the Kaluza-Klein modes on the D3-brane dual to the symmetric loop.
\paragraph{Q-functions and the absence of Kaluza-Klein modes} Let us next analyze the $Q$-functions using the Gram-Schdmit determinant \eqref{eq:defQfunctions}. To write it down, we need to evaluate the integral
\beq\label{eq:nmnorm}
\langle n,m\rangle=\oint d\mu_2 (g(x-x^{-1}))^{n+m}\comma
\eeq  
with\fn{Here we already substituted $t_0$ with \eqref{eq:t0parameterize}.}
\beq
d\mu_2 =-\frac{1}{4\pi g}\frac{dx(1+x^{-2})}{2\pi i}\frac{1}{1-e^{-2\pi g (x-y)(1-1/xy)}}\period
\eeq
Just like the saddle-point equation \eqref{eq:rewriteintegrandsaddle}, we need to include the contribution from the poles at $x=y^{\pm 1}$ in \eqref{eq:nmnorm}. Again the contribution from the poles dominate at strong coupling and we thus have
\beq
\langle n,m\rangle=\begin{cases}\frac{g^{n+m-2}}{8\pi^2}\left(y+\frac{1}{y}\right)^3\left(y-\frac{1}{y}\right)^{n+m-1}\qquad &n+m\text{: even}\\-\frac{g^{n+m-2}}{8\pi^2}\left(y+\frac{1}{y}\right)^2\left(y-\frac{1}{y}\right)^{n+m}\qquad &n+m\text{: odd}\end{cases}\period
\eeq
Plugging these expressions into the Gram-Schmidt determinant \eqref{eq:defQfunctions}, we find that all but $Q_{0}$ and $Q_1$ are identically zero. This is because, for $Q_n$ with $n\geq 2$, there are always (at least) two different rows in the determinant which are  proportional to each other. Alternatively we can understand this as follows: The $Q$-functions define a set of orthogonal polynomials under the measure $d\mu_2$. However, at strong coupling, $d\mu_2$ has support only at two points, $x=y^{\pm 1}$. The space of functions defined at two points is two-dimensional and is spanned by $Q_0$ and $Q_1$. 

Physically this means that the higher-charge operators, $\tilde{\Phi}^{n}$ with $n\geq 2$, all decouple when $g\gg 1$, and their couplings to the modes on the D3-brane are exponentially suppressed $\sim e^{-g \bullet}$. This is consistent with the fact that the D3-brane is point-like on $S^5$ and therefore does not host Kaluza-Klein excitations coming from $S^5$. It is interesting that this is realized in the localization computation by the ``Bose-Einstein condensation'' mentioned earlier. Roughly speaking, there seems to be a qualitative correspondence between the size of the distribution of the Bose gas and the size of D3-brane on $S^5$. Note that, even though the higher-charge single-particle operators  $\tilde{\Phi}^{n}$ decouple for $n>1$, the correlation functions involving the charge-1 operator $\tilde{\Phi}$ can still be computed by taking simple area derivatives of the Wilson loop expectation value (\ref{Ws-vev}), and we will match them below with the D3-brane calculation.

We should also note that this decoupling of higher-charge operators is only true in the strict $g\to \infty$ limit. Away from the limit, there will be contribution from the integral along the unit circle \eqref{eq:nmnorm} and therefore higher-charge $Q$-functions do not vanish. In particular, at weak coupling $g\ll 1$ all these higher-charge operators exist and are visible. This explains the apparent mismatch of the spectrum of operators at weak and strong couplings discussed in section \ref{sec:setup}.
\section{Correlation functions in dCFT$_1$ from the D5-brane\label{sec:D5}}

\subsection{D5-brane solution in $AdS_5\times S^5$}
In this section, we review the D5-brane solution in the $AdS_5\times S^5$ background \cite{Camino:2001at, Yamaguchi:2006tq}. The bosonic part of the Euclidean D5-brane action takes the form
\be\label{eq:D5_DBI}
S_{D5}=T_{D5}\int d^6\xi \sqrt{\text{det}(G+F)}+i T_{D5}\int F\wedge C_4.
\ee
where $G$ is the induced metric, and we have absorbed a factor of $2\pi\alpha'$ into the worldvolume gauge field. 
The D5-brane tension $\Tf$ is given by
\be
\Tf=\dfrac{N\sqrt{\lambda}}{8\pi^4}.
\ee
To write down the D5-brane solution, we use the following parametrization of the $AdS_5\times S^5$ space:
\begin{align}
\label{eq:D5_10d_ds2_1}
&ds^2=du^2+\cosh^2 u\,ds^2_{AdS_2}+\sinh^2u\,d\Omega^2_2+d\theta^2+\sin^2\theta d\Omega^2_4\,.
\end{align}
The four-form $C_4$ which produces the the five-form flux can be written as\cite{Yamaguchi:2006tq}:
\be\label{eq:D5_CS_term}
C_4=\left(-\frac{u}{2}+\frac{1}{8}\sinh 4u\right)dH_2\wedge d\Omega_2+\left(\frac{3}{2}\theta-\sin 2\theta+\frac{1}{8}\sin 4\theta\right)d\Omega_4,
\ee
where $dH_2$ is the volume element of the $AdS_2$ space. 
The embedding of the D5-brane in the $AdS_5\times S^5$ background is parametrized by
\be
u=0, \quad \theta =\tk,
\ee
where the angle $\tk$ is related to the fundamental string charge $k$ via:
\be
k=\frac{N}{\pi}\left(\tk-\frac{1}{2}\sin 2\tk\right).
\ee
The induced worldvolume geometry of the D5-brane is then $AdS_2\times S^4$, and  the induced metric is
\be
ds^2_{D5}= ds^2_{AdS_2}+\sin^2\tk\, d\Omega^2_4\,.
\ee
For the case of the Wilson loop on an infinite straight line at the boundary, we 
can take the metric of $AdS_2$ to be that of the Poincare half-plane
\be 
ds^2_{AdS_2}=\frac{1}{r^2}(d\t^2+dr^2)\,.
\label{Poinc-plane}
\ee 
In these coordinates, the worldvolume gauge field strength of the classical solution is given by
\be
F=i\frac{\cos\tk}{r^2} d\t\wedge dr.
\ee
In addition to the bulk action, we also need to add the following boundary term to implement the correct boundary conditions \cite{Drukker:1999zq,Drukker:2005kx,Yamaguchi:2006tq}
\be\label{eq:D5_bdy}
S^A_{bdy}=-\int \! d\t\int\!  d\Omega_4\, A_\t\,\pi_A,
\ee
where $\pi_A$ is the conjugate momentum to $A_\t$
\be
\pi_A=\frac{\p\mathcal{L}_{D5}}{\p F_{\t r}}.
\ee
Adding this boundary term corresponds to choosing boundary conditions such that the momentum $\pi_A$ is fixed at the boundary (while $A_\t$ is dynamical). Indeed, the integral of $\pi_A$ over $S^4$ is related to the fundamental string charge $k$ by\footnote{The factor of $2\pi\alpha'$ is because in (\ref{eq:D5_DBI}) we have absorbed this factor into the gauge field, and the factor of $i$ is due to the Euclidean signature.}
\be 
k = -2\pi i \alpha' \int_{S^4} \frac{\p\mathcal{L}_{D5}}{\partial F_{\t r}} = \frac{N}{\pi}\left(\theta_k-\sin\theta_k \cos\theta_k\right)\,.
\ee 

Let us review how the expectation value of the circular Wilson loop in the large antisymmetric representation is obtained from the classical D5-brane action. The solution described above applies equally well to the circular loop, provided we use the Poincare disk metric of $AdS_2$ instead of (\ref{Poinc-plane}). The expectation value of the Wilson loop is obtained as
\be 
\langle W_{{\sf A}_k} \rangle = \exp\left(-S_{D5}-S^A_{bdy}\right)\,.
\ee 
Plugging in the solution above, we find
\be 
S_{D5}+S^A_{bdy} = T_{D5} {\rm vol}(AdS_2){\rm vol}(S^4) \sin^3\theta_k
\ee 
Using the well-known regularized value of the hyperbolic disk volume ${\rm vol}(AdS_2)=-2\pi$,\footnote{Equivalently, one can add to the action a boundary term corresponding to a Legendre transform in the AdS radial direction \cite{Drukker:1999zq, Drukker:2005kx}. This gives the same result as using directly the regularized volume of $AdS_2$.} as well as ${\rm vol}(S^4)=8\pi^2/3$ and the value of the D5-brane tension, one finds \cite{Yamaguchi:2006tq}
\be 
\langle W_{{\sf A}_k} \rangle = \exp\left(\frac{2N\sqrt{\lambda}}{3\pi}\sin^3\theta_k\right)\,,
\ee 
which agrees with the localization prediction, which can be obtained by evaluating (\ref{W-LargeN}) in the strong coupling limit with $k/N$ fixed \cite{Yamaguchi:2006tq, Hartnoll:2006is}. Note that the 1/2-BPS case corresponds to $a=0$ in (\ref{W-LargeN}).

If we expand the action \eqref{eq:D5_DBI} in powers of the fluctuations around the D5-brane solution and perform $KK$-reduction:
\be
S_B=\int\frac{d\t dr}{r^2} L_B,
\quad L_B=L^{(2)}+L^{(3)}+L^{(4)}+\dots,
\ee
the resulting action can be viewed as a 2d field theory on $AdS_2$ background with a manifest symmetry of $SL(2,R)\times SO(3)\times SO(5)$. The dual of this bulk $AdS_2$ theory is the defect $CFT_1$ defined by operator insertions on the straight (or circular) 1/2-BPS Wilson loop. In most of the calculations below, we will focus on the straight line geometry, but all results can be easily translated to the circle. 

\subsection{Spectrum of excitations around the D5-brane}
In this section, we expand the D-brane action around the D5-brane solution and find the spectrum of fluctuations, focusing on bosonic fields only. Since the spectrum has been computed in \cite{Harrison:2011fs,Faraggi:2011ge}, we briefly review the calculation here. For the study of the spectrum, instead of parameterizing $AdS_5\times S^5$ as \eqref{eq:D5_10d_ds2_1}, we change to $x^i$ coordinates so that the metric reads
\be\label{ds2_dis}
ds^2= \frac{(1+\frac{1}{4}x^2)^2}{(1-\frac{1}{4}x^2)^2}ds^2_{AdS_2}+\frac{dx^i dx^i}{(1-\frac{1}{4}x^2)^2}+d\theta^2+\sin^2\theta\, d\Omega^2_4,
\ee
where $i=1,\dots,3$ refers to the transverse directions. The previous $u$ coordinate is related to $x^2=x^ix^i$ by
\be
\frac{x^2}{(1-\frac{1}{4}x^2)^2}=\sinh^2u.
\ee
We use Greek letters ($\mu$,$\nu$) for $AdS_2$ coordinates and Greek letters ($\alpha$,$\beta$) for $S^4$ coordinates.

Now we consider the effective action for fluctuations $\dx^i$, $\dt$ and $f$ around the D5-brane solution, where $f$ is a 2-form in 6d spacetime representing the fluctuations of the background field strength. We expand everything to quartic order in fluctuations as we need to compute various four-point functions later. The variation of the metric in powers of fluctuations is
\be
\delta(ds^2)=\biggl(\dx^2+\frac{1}{2}\dx^4\biggr)ds^2_{AdS_2}+\biggl(1+\frac{1}{2}\dx^2\biggr) (d\dx^i)(d\dx^i)+A(\dt)d\Omega^2_4+ (d\dt)^2,
\ee
where
\be
A(\dt)=\sin 2\tk\dt+\cos 2\tk \dt^2-\frac{2}{3}\sin 2\tk\dt^3-\frac{1}{3}\cos 2\tk \dt^4.
\ee
The variation of $C_4$ in powers of fluctuations is
\bea{
\delta C_4 =& -\frac{1}{8}(12\tk-8\sin 2\tk+\sin 4\tk)-4\sin^4\tk \dt-8\cos\tk\sin^3\tk \dt^2\\
& -\frac{8}{3}(1+2\cos 2\tk)\sin^2\tk\dt^3+\frac{2}{3}(\sin 2\tk-2\sin 4\tk)\dt^4.
\eqn
}
The mass spectrum can be then obtained by expanding the action \eqref{eq:D5_DBI} to quadratic order in fluctuations around the D5-brane solution. 

\paragraph{$\dx^i$ sector}
The quadratic Euclidean action for the $\dx^i$ sector is
\be\label{S2dis}
S_{\dx}^{(2)} = \Tf\sin^3\tk\int d^6\xi \frac{\sqrt{g_4}}{r^2}\frac{1}{2}\left[\p_\m\dx^i\p^\m\dx^i+\nabla_\alpha\dx^i\nabla^\alpha\dx^i + 2\dx^i\dx^i \right],
\ee
where $g_4$ is the metric for $S^4$. To keep
the $SO(5)$ symmetry manifest, we expand the fields using the spherical harmonics defined by symmetric traceless tensor. Specifically, if we let $Y^a$ to be the five-dimensional vector specifying $S^4$:
\be
\sum_{a=1}^5 Y^a(\Omega_4)Y^a(\Omega_4)=1,
\ee
then the $\dx^i$ field is expanded as
\be
\dx^i(\t,r,\Omega_4)=\sum_{l=0}^\infty (\dx^i)_{a_1\cdots a_l}(\t,r) Y^{a_1}\cdots Y^{a_l},
\ee
where $(\dx^i)_{a_1\cdots a_l}$ is a symmetric traceless tensor field and the repeated indices are summed. In particular, we have
\be
\nabla^2_{S^4}\biggl((\dx^i)_{a_1\cdots a_l}Y^{a_1}\cdots Y^{a_l}\biggr)=-l(l+3)(\dx^i)_{a_1\cdots a_l}Y^{a_1}\cdots Y^{a_l}.
\ee 
The quadratic action for these expanded fields is
\bea{
S_{\dx}^{(2)} =& \sum_{l=0}^\infty V_l\,\Tf \sin^3 \tk\\
&\qquad \int \frac{drd\t}{r^2}\frac{1}{2}\biggl[\p_\m(\dx^i)_{a_1\cdots a_l}\p^\m(\dx^i)_{a_1\cdots a_l}+(l+2)(l+1)(\dx^i)_{a_1\cdots a_l}(\dx^i)_{a_1\cdots a_l}\biggr].
\eqn
}
The factor $V_l$ comes from the integral of spherical harmonics over $S^4$ and is defined by \be\label{eq:D5_def_Vl}
\int d\Omega_4 (\mathbf{u_1}\cdot Y)^l (\mathbf{u_2}\cdot Y)^l \equiv V_l\,(\mathbf{u_1}\cdot\mathbf{u_2})^l,
\ee
where  $\mathbf{u}^a$ denotes a five-dimensional null vector. This integral can be done analytically and we find $V_l$ to be
\be
V_l=\frac{16\pi^2\,2^l\,(l+1)!\,l!}{(2l+3)!}.
\ee

\paragraph{$\dt$ and $a_\mu$ sector}
In order to decouple $a_\mu$ from the gauge fields along the $S^4$ directions, we need to impose the gauge condition
\be\label{eq:D5_am_gauge}
\nabla^\alpha a_\alpha=0.
\ee
The quadratic Euclidean action for $\dt$ and $a_\mu$ is
\bea{\label{eq:D5_quad_OX_action}
S_{\dt, f}^{(2)} =& \Tf\int d^6\xi \frac{\sqrt{g_4}}{r^2} \biggl[\frac{\sin^3\tk}{2}\left(
\p_\m\dt\p^\m\dt+\nabla_\alpha\dt\nabla^\alpha\dt-4\dt^2\right)\\
&+\frac{\sin\tk}{2}\left(\frac{1}{2}f_{\m\n}f^{\m\n}+\nabla_\alpha a_\mu \nabla^\alpha a^\mu\right)
+2i\sin^2\tk\dt\varepsilon^{\m\n}f_{\m\n}\biggr], \eqn
}
where $\varepsilon_{\mu\nu}=\sqrt{g}\epsilon_{\mu\nu}$ is the Levi-Civita tensor\footnote{$\epsilon_{\mu\nu}$ is antisymmetric with $\epsilon_{\t r}=1$}.

We expand the fields $a_\m$ and $\dt$ in terms of the symmetric traceless tensor fields:
\be\label{eq:D5_ftheta_exp1}
a_\mu(\t,r,\Omega_4)=\sum_{l=0}^\infty (a_\mu)_{a_1\cdots a_l}(\t,r)Y^{a_1}\cdots Y^{a_l},\quad
\dt(\t,r,\Omega_4)=\sum_{l=0}^\infty \dt_{a_1\cdots a_l}(\t,r)Y^{a_1}\cdots Y^{a_l}.
\ee
Then the equations of motion for $(a_\mu)_{a_1\cdots a_l}$ and $\dt_{a_1\cdots a_l}$ derived from the action \eqref{eq:D5_quad_OX_action} are
\bea{\label{eq:D5_eom1}
-\p_\t f_{a_1\cdots a_l}+l(l+3)(a_r)_{a_1\cdots a_l}-4i\sin\tk \p_\t\dt_{a_1\cdots a_l}&=0,\\
\p_r  f_{a_1\cdots a_l}+l(l+3)(a_\t)_{a_1\cdots a_l}+4i\sin\tk \p_r\dt_{a_1\cdots a_l}&=0,\eqn\\
-\nabla_\mu\nabla^\mu\dt_{a_1\cdots a_l}+(l+4)(l-1)\dt_{a_1\cdots a_l}+\frac{4i}{\sin\tk}f_{a_1\cdots a_l}&=0,
}
where we have defined $f_{a_1\cdots a_l}\equiv\epsilon^{\mu\nu}\partial_\mu (a_\nu)_{a_1\cdots a_l}$ to simplify the notation.

Taking derivatives on both sides of the first two equations in \eqref{eq:D5_eom1}, we obtain the following set of equations:
\bea{\label{eq:D5_eom2}
\nabla_\mu\nabla^\mu f_{a_1\cdots a_l}-(l^2+3l+16)f_{a_1\cdots a_l}+4i\sin\tk(l+4)(l-1)\dt_{a_1\cdots a_l}&=0, \\
\nabla_\mu\nabla^\mu\dt_{a_1\cdots a_l}-(l+4)(l-1)\dt_{a_1\cdots a_l}-\frac{4i}{\sin\tk}f_{a_1\cdots a_l}&=0.
\eqn
}
By diagonalizing \eqref{eq:D5_eom2}, we find two types of modes with the mass spectrum 
\be\label{eq:defOX}
\begin{cases}
\displaystyle
O_{a_1\cdots a_l}=\dt_{a_1\cdots a_l}- \frac{i f_{a_1\cdots a_l}}{(4+l)\sin\tk}, &\text{with } m_l^2=l(l-1),\quad (l=1,2,\dots)\\
\\
\displaystyle
X_{a_1\cdots a_l}=(l-1)\sin\tk\dt_{a_1\cdots a_l}+i f_{a_1\cdots a_l}, & \text{with } m_l^2=(l+3)(l+4),\quad (l=0,1,\dots).
\end{cases}
\ee
The $O_{a_1\cdots a_l}$ modes start with $l=1$ because the $l=0$ mode $O_0$ is not dynamical as the equations of motion for this mode are
\be
\partial_\t O_0=\partial_r O_0=0.
\ee 
From \eqref{eq:defOX}, we can express $\dt_{a_1\cdots a_l}$ and $f_{a_1\cdots a_l}$ in terms of $O_{a_1\cdots a_l}$ and $X_{a_1\cdots a_l}$
\bea{\label{eq:dt&f_OX}
\dt_{a_1\cdots a_l}&=\frac{X_{a_1\cdots a_l}+(4+l)\sin\tk O_{a_1\cdots a_l}}{(2l+3)\sin\tk},\\
f_{a_1\cdots a_l}&=\frac{i(l+4)}{(2l+3)}\biggl[-X_{a_1\cdots a_l}+(l-1)\sin\tk O_{a_1\cdots a_l}\biggr].
\eqn
}
We will denote the $l=0$ mode of $f_{a_1\cdots a_l}$ simply as $f_0$ in the later sections.

\paragraph{$a_\alpha$ sector}
The quadratic Euclidean action for gauge fields along $S^4$ directions is
\be
S^{(2)}_{a_\alpha}=\Tf\sin\tk\int d^6\xi\frac{\sqrt{g_4}}{r^2}
\frac{1}{2}\biggl[
\p_\mu a_\alpha\p^\mu a^\alpha -a_\alpha\bigl(g_4^{\alpha\beta}\nabla^2_{S^4}-\mathcal{R}_{4}^{\alpha\beta}\bigr)a_\beta
\biggr],
\ee
where $g_4^{\alpha\beta}$ is the metric for $S^4$ and $\mathcal{R}_{4}^{\alpha\beta}=3g^{\alpha\beta}$ is the Ricci tensor for $S^4$. Since the gauge condition \eqref{eq:D5_am_gauge} is imposed to decouple $a_\alpha$ from $a_\mu$, we need to expand $a_\alpha$ in terms of the transverse vector spherical harmonics on $S^4$ as
\be
a_\alpha(\t,r,\Omega_4)=\sum_{l=1}^\infty a_l(\t,r)(\hat{Y}_\alpha)_{lm}(\Omega_4).
\ee
The transverse vector spherical harmonics $(\hat{Y}_\alpha)_{lm}$ satisfies following properties\cite{1984JMP....25.2888R,vanNieuwenhuizen:2012zk}:
\be
\nabla^2_{S^4}(\hat{Y}_\alpha)_{lm}=-(l^2+3l-1)(\hat{Y}_\alpha)_{lm},\quad
\nabla^\alpha_{S^4}(\hat{Y}_\alpha)_{lm}=0,
\quad (l=1,2,\dots).
\ee
The quadratic action for the $a_l$ modes is
\be
S_{a_l}^{(2)} = \sum_{l=1}^\infty\Tf \sin\tk\int \frac{drd\t}{r^2}\frac{1}{2}\biggl[\p_\m a_l\p^\m a_l+(l+2)(l+1)a_l a_l\biggr].
\ee

\subsection{Dual operators and two-point functions}
The holographic dictionary for the bulk fluctuation modes has been established in \cite{Harrison:2011fs}. In this section, we briefly review the dual operators for each fluctuation mode. We summarize the results in table \ref{tab:D5_spectrum}.

\paragraph{$\dx^i$ sector}
From the mass spectrum of the $(\dx^i)_{a_1\cdots a_l}$ modes, we see that the mode $(\dx^i)_{a_1\cdots a_l}$ should be dual to an operator of dimension $\Delta_l=l+2$ which transforms under $SO(3)$ as a vector. In particular, the three $l=0$ modes which we shall denote as $\dx^i_0$ are dual to the displacement operator $\mathbb{F}_{ti}$ in the ultrashort supermultiplet of $OSp(4^*|4)$. The higher $l$ modes $(l\geq 1)$ are dual to the operators in a short multiplet of $OSp(4^*|4)$ (see \cite{Harrison:2011fs} and table \ref{tab:D5_spectrum}).

\paragraph{$\dt$ and $a_\mu$ sector}
In this sector, there are two families of modes $O_{a_1\cdots a_l}$ and $X_{a_1\cdots a_l}$. From the mass spectrum, we see that the mode $O_{a_1\cdots a_l}$ should be dual to an operator of dimension $\Delta_l=l$ while the mode $X_{a_1\cdots a_l}$ should be dual to an operator of dimension $\Delta_l=l+4$. In both cases, the dual operator transforms in the symmetric representations of $SO(5)$. The modes $O_{a_1\cdots a_l}$ are dual to the protected operator $\O_l$ in the defect CFT which played the central role in the localization analysis (in particular, the $l=1$ mode $O_a$ is dual to $\O_1$ in the ultrashort multiplet of $OSp(4^*|4)$). On the other hand, the modes $X_{a_1\cdots a_l}$ are dual to supersymmetry descendants of the operator $\O_l$, i.e. they belong to the same short multiplet of $OSp(4^*|4)$.

\paragraph{$a_\alpha$ sector}
From the mass spectrum, we see that the $a_l$ mode should be dual to an operator of dimension $\Delta_l=l+2$, which is again in the short multiplet of $OSp(4^*|4)$ headed by $\O_l$.

\begin{table}
\centering
\def\arraystretch{1.5}
\begin{tabular}{||c|c| c| c| c||}
\hline
Fluctuation modes & Dual operator &$\Delta$ & $SO(3)$ & $SO(5)$\\
\hline
$O_a$ & $\O_1$& $1$ & $0$ & $(0,1)$\\
\hline
$\dx^i_0$ & ${\mathbb{F}_{t}}^i=\mathcal{Q}^2\O_1$& $2$ & $1$ & $(0,0)$\\
\hline
\hline
$O_{a_1\cdots a_l}$ $(l\geq 2)$& $\O_l$& $l$ & $0$ & $(0,l)$\\
\hline
$(\dx^i)_{a_1\cdots a_l}$ $(l\geq 1)$ & $\mathcal{Q}^2\O_{l+1}$& $l+2$ & $1$ & $(0,l)$\\
\hline
$X_{a_1\cdots a_l}$ $(l\geq 0)$&$\mathcal{Q}^4\O_{l+2}$ &$l+4$ & $0$ & $(0,l)$\\
\hline
$a_l$ $(l\geq 1)$& $\mathcal{Q}^2\O_{l+1}$& $l+2$ & $0$ & $(2,l-1)$\\
\hline
\end{tabular}
\caption{In this table we summarize the quantum numbers of the operator dual to each fluctuation mode. $\Delta$ gives the conformal dimension of the dual operator. The quantum numbers of the dual operator under $SO(3)$ and $SO(5)$ symmetry are given in terms of the Dynkin labels of the corresponding representations.}
\label{tab:D5_spectrum}
\end{table}

\paragraph{The two-point functions}
From \eqref{eq:defOX} we see that both $O_{a_1\cdots a_l}$ and $X_{a_1\cdots a_l}$ are linear combinations of $\dt$ and the 2d field strength $f_{\mu\nu}$. Therefore, the boundary value for $\dt$ and $f_{\mu\nu}$ should be fixed when varying the action. To ensure that the solutions to the equations of motion \eqref{eq:D5_eom1} are stationary under the variations satisfying these boundary conditions, we need to add the following boundary term
\be\label{eq:D5_quad_OX_bdy}
S^{(2)}_{bdy} = -\Tf\underset{r=r_0}{\int} d\t d\Omega_4 \left[4i \sin^2\tk\dt \,a_\t+ \sin\tk r^2 a_\t(\p_\t a_r-\p_r a_\t)\right],
\ee
where $r_0$ is the location of the boundary. In fact, this boundary term can be also derived from expanding the boundary term \eqref{eq:D5_bdy} to quadratic order in fluctuations. 

To compute the tree level two-point function $\langle \! \langle \O_{L_1}\!(\t_1,\vu_1)\O_{L_2}\!(\t_2,\vu_2)\rangle\!\rangle$, we need the quadratic order on-shell action for the field $O_{a_1\cdots a_l}$, which is
\bea{\label{eq:D5_quad_OX_final}
S_{on-shell}^{(2)}=&-V_l\Tf\underset{r=r_0}{\int} d\t \biggl[\frac{\sin^3\tk}{2}\dt_{a_1\cdots a_l}\p_r \dt_{a_1\cdots a_l}+\frac{8\sin^3\tk}{l(l+3)}\dt_{a_1\cdots a_l}\p_r \dt_{a_1\cdots a_l}\\
&-\frac{2i\sin^2\tk}{l(l+3)}\dt_{a_1\cdots a_l} \p_r f_{a_1\cdots a_l}- \frac{2i\sin^2\tk}{l(l+3)}f_{a_1\cdots a_l}\p_r\dt_{a_1\cdots a_l}-\frac{\sin\tk}{2l(l+3)}f_{a_1\cdots a_l}\p_r f_{a_1\cdots a_l}\biggr]\\
=&-\frac{V_l\Tf}{2}\underset{r=r_0}{\int} d\t\left[\frac{(4+l)^2\sin^3\tk}{(3+2l)l}\,O_{a_1\dots a_l}\partial_r O_{a_1\dots a_l}+\frac{\sin\tk}{(2l+3)(l+3)}\,X_{a_1\dots a_l}\partial_r X_{a_1\dots a_l}\right].
\eqn
}
We use the following normalization of the bulk-to-boundary propagator\cite{Freedman:1998tz}
\be
K_\D(r,\t;\t')=\mathcal{C}_\D\left[\frac{r}{r^2+(\t-\t')^2}\right]^\D, \quad
\C_\D=\frac{\Gamma(\D)}{\sqrt{\pi}\Gamma(\D-\frac{1}{2})}. 
\ee
With this normalization, we find that the tree level two-point function of the dual boundary operator $\O_{L}$ is
\bea{\label{eq:D5_def_c2pt}
\langle\!\langle \O_{L_1}(\t_1,\vu_1) \O_{L_2}(\t_2,\vu_2)\rangle\!\rangle
=&\delta_{L_1 L_2}\,\frac{\Tf\sin^3\tk\,\pi^2(L_1+4)^2(2L_1-1)\Gamma^2(L_1)}{2^{L_1-2}(2L_1+3)^2\Gamma(L_1-\frac{1}{2})\Gamma(L_1+\frac{3}{2})}\frac{(\mathbf{u_1}\cdot\mathbf{u_2})^{L_1}}{(\t_{12})^{2L_1}}\\
\equiv&\delta_{L_1 L_2}\, c_{L_1} \frac{(\mathbf{u_1}\cdot\mathbf{u_2})^{L_1}}{(\t_{12})^{2L_1}}.
\eqn
}
As we will also need the tree level two-point function of $\mathbb{F}_{ti}$ later, we provide the result here:
\be
\langle\!\langle {\mathbb{F}_t}^i(\t_1){\mathbb{F}_t}^j(\t_2)\rangle\!\rangle=\langle \dx^i_0(\t_1)\dx^j_0(\t_2)\rangle_{AdS_2}=\delta^{ij}\, \frac{16\pi\Tf\sin^3\tk}{\t_{12}^{4}}.
\label{xx-D5}
\ee

\subsection{Three-point functions of $S^5$ fluctuations}
\label{sec:D5_3pt}
In this section, we compute the three-point function
\be
\langle\!\langle \O_{L_1}(\t_1,\vu_1) \O_{L_2}(\t_2,\vu_2)
\O_{L_3}(\t_3,\vu_3)\rangle\!\rangle,
\ee
from the expanded D-brane action. This requires the knowledge of the cubic interaction vertices of $\dt$ and $a_\mu$, which we find to be
\bea{\label{eq:D5_S3_1}
L^{(3)}_{\dt,f}=&\Tf\int d\Omega_4
\biggl[
\cos\tk\biggl(\nabla_\alpha a_\mu\nabla^\alpha a^\mu+f_{\mu\nu}f^{\mu\nu}\biggr)\dt
-\frac{i\cot\tk}{4}\varepsilon^{\mu\nu}f_{\mu\nu}\biggl(
\nabla_\alpha a_\mu\nabla^\alpha a^\mu+\frac{1}{2}f_{\mu\nu}f^{\mu\nu}\biggr)\\
&+\sin\tk\sin 2\tk\biggl(
\partial_\mu\dt\partial^\mu\dt+\frac{1}{2}\nabla_\alpha\dt\nabla^\alpha\dt-2\dt^2
\biggr)\dt
+\frac{i\sin 2\tk}{2}\varepsilon^{\mu\nu}\nabla_\alpha a_\mu\nabla^\alpha\dt\partial_\nu\dt\\
&-\frac{i\sin 2\tk}{8}\varepsilon^{\mu\nu}f_{\mu\nu}
\biggl(
\partial_\mu\dt\partial^\mu\dt-\nabla_\alpha\dt\nabla^\alpha\dt-12\dt^2
\biggr)
\biggr].
\eqn
}
The relevant cubic coupling for $\langle\!\langle\O_{L_1}\O_{L_2}\O_{L_3} \rangle\!\rangle$ can be then extracted from \eqref{eq:D5_S3_1} after we substitute the expressions \eqref{eq:dt&f_OX} into \eqref{eq:D5_S3_1}. 

Using the $SO(5)$ symmetry, the general three-point function of three $\O_{L}$ operators can be written as 
\bea{\label{eq:D5_OkOlOm_1}
&\langle\!\langle \O_{L_1}(\t_1,\vu_1)\O_{L_2}(\t_2,\vu_2)\O_{L_3}(\t_3,\vu_3)\rangle\!\rangle\\
=&f_{L_1L_2L_3}(\t_1,\t_2,\t_3)\times (\mathbf{u}_1\cdot\mathbf{u}_2)^{L_{12|3}}(\mathbf{u}_2\cdot\mathbf{u}_3)^{L_{23|1}}(\mathbf{u}_1\cdot\mathbf{u}_3)^{L_{13|2}}
\eqn,
}
where $L_{ij|k}\equiv(L_i+L_j-L_k)/2$. The $f_{L_1L_2L_3}$ can be computed from the bulk cubic coupling \eqref{eq:D5_S3_1} and we find it to be
\bea{\label{eq:D5_cklm_1}
f_{L_1L_2L_3}=&\frac{2(4+L_1)(4+L_2)(4+L_3)L_{12|3}L_{23|1}L_{13|2}(\Sigma^2-1)(3+\Sigma)}{L_1L_2L_3(3+2L_1)(3+2L_2)(3+2L_3)}\\
&\times\Tf\sin^2\tk\cos\tk\, V_{L_1,L_2, L_3}\,
\times\int\frac{drd\t}{r^2}K_{L_1}(r,\t;\t_1)K_{L_2}(r,\t;\t_2)K_{L_3}(r,\t;\t_3),
\eqn
}
where $\Sigma\equiv L_1+L_2+L_3$. We have defined $V_{L_1,L_2,L_3}$ to be
\be
\int d\Omega_4 (\vu_1\cdot Y)^{L_1} (\vu_2\cdot Y)^{L_2}(\vu_3\cdot Y)^{L_3} \equiv V_{L_1,L_2,L_3} \left[(\mathbf{u}_1\cdot\mathbf{u}_2)^{L_{12|3}}(\mathbf{u}_2\cdot\mathbf{u}_3)^{L_{23|1}}(\mathbf{u}_1\cdot\mathbf{u}_3)^{L_{13|2}}\right],
\ee
which can be computed as shown in the Appendix, and we find that
\be
V_{L_1,L_2,L_3}=\frac{(1+(-1)^{L_1+L_2+L_3})}{2}\frac{8\pi^2\,(\sqrt{2})^\Sigma\,(\Sigma+2)L_1!\,L_2!\,L_3!\,\left(\frac{\Sigma}{2}\right)!}{(\Sigma+3)!\,L_{12|3}!\,L_{23|1}!\,L_{13|2}!}.
\ee
The result for the bulk integral in \eqref{eq:D5_cklm_1} is \cite{Freedman:1998tz}
\bea{\label{eq:D5_3pt_bulkint}
&\int\frac{drd\t}{r^2}K_{L_1}(r,\t;\t_1)K_{L_2}(r,\t;\t_2)K_{L_3}(r,\t;\t_3)\\
=&\frac{\Gamma(\frac{\Sigma}{2}-\frac{1}{2})\,\Gamma(L_{12|3})\,\Gamma(L_{23|1})\,\Gamma(L_{13|2})}{2\pi\,\Gamma(L_1-\frac{1}{2})\Gamma(L_2-\frac{1}{2})\Gamma(L_3-\frac{1}{2}) (\t_{12})^{2L_{12|3}}(\t_{23})^{2L_{23|1}}(\t_{13})^{2L_{13|2}}}.
\eqn
}
Putting everything together, the 3-point function is given by
\begin{equation}
\label{3pt-final}
\langle\!\langle\O_{L_1}(\t_1,\vu_1)\O_{L_2}(\t_2,\vu_2)\O_{L_3}(\t_3,\vu_3)\rangle\!\rangle\\
={\cal C}_{L_1,L_2,L_3}\frac{(\mathbf{u}_1\cdot\mathbf{u}_2)^{L_{12|3}}(\mathbf{u}_2\cdot\mathbf{u}_3)^{L_{23|1}}(\mathbf{u}_1\cdot\mathbf{u}_3)^{L_{13|2}}}
{(\t_{12})^{2L_{12|3}}(\t_{23})^{2L_{23|1}}(\t_{13})^{2L_{13|2}}}
\end{equation}
with the 3-point structure constants taking the simple factorized form
\begin{equation}
\label{3pt-final2}
{\cal C}_{L_1,L_2,L_3} =
8(1+(-1)^{L_1+L_2+L_3}) \pi^{\frac{3}{2}} T_{D5}\sin^2\theta_k \cos\theta_k \prod_{i=1}^3 
\frac{\Gamma(L_i)(4+L_i)}{2^{\frac{L_i}{2}}\Gamma\left(L_i-\frac{1}{2}\right)(2L_i+3)}\,.
\end{equation}

Note that, although the bulk 3-point integral \eqref{eq:D5_3pt_bulkint} has a pole when one of the $L_{ij|k}$ is zero, the pole is canceled by the $L_{ij|k}$ factor in \eqref{eq:D5_cklm_1}. As a result, the three-point function is always finite, when computed by analytic continuation in the charges. 

To compare with the prediction of localization, we can do a conformal transformation to the circular Wilson loop and set the polarizations to
\be\label{eq:D5_topo_config_pol}
\vu_i=(\cos\t_i,\sin\t_i,0,i,0).
\ee
The normalized three-point function with the topological configuration is then
\bea{\label{eq:D5_VLLL_expressionf}
&\frac{\langle \!\langle  \tO_{L_1}\tO_{L_2}\tO_{L_3}\rangle\!\rangle}{\sqrt{\langle \!\langle \tO_{L_1}\tO_{L_1}\rangle\!\rangle 
\langle \!\langle \tO_{L_2}\tO_{L_2}\rangle\!\rangle 
\langle \!\langle \tO_{L_3}\tO_{L_3}\rangle\!\rangle }}\\
=&\frac{(1+(-1)^{L_1+L_2+L_3})}{2}\sqrt{\frac{\left(L_1+\frac{1}{2}\right)\left(L_2+\frac{1}{2}\right)\left(L_3+\frac{1}{2}\right)}{2\pi^3 \Tf}}\frac{\cos\tk}{(\sin\tk)^{5/2}}.
\eqn
}
Using the relation $2\pi^3 \Tf=N\sqrt{\lambda}/(4\pi) = Ng$, we find the result agrees with the prediction of localization.

As we have pointed out previously, for $L_{ij|k}=0$,
the bulk integral \eqref{eq:D5_3pt_bulkint} is divergent while the prefactor in the first line of \eqref{eq:D5_cklm_1} has a zero. The zero actually results from the vanishing of the bulk cubic coupling. In fact, the three-point function is called extremal in this case and one expects the corresponding bulk cubic coupling to vanish \cite{DHoker:1999jke}. For the case that $L_1=L_2=1$ and $L_3=2$ which is relevant to the calculation in section \ref{sec:D5_o2o2o1o1}, by expanding the bulk action explicitly we find that the corresponding cubic coupling from \eqref{eq:D5_S3_1} is
\be\label{eq:D5_3pt_o2o1o1_bulk}
\frac{\Tf 8\pi^2\sin\tk\sin 2\tk}{245}\int \frac{drd\t}{r^2}
\left(
18\p_\m O_a\p^\m O_b O_{ab}+\p_\m O_a O_b\p^\m O_{ab}-17 O_a O_b O_{ab}    
\right),
\ee
which indeed vanishes on-shell. To avoid this subtlety and reproduce the result \eqref{eq:D5_VLLL_expressionf} from the bulk calculation, we shall use the following approach. At the boundary the single particle operator $\O_{2}$ can be mixed with the two-particle operator $:\!\!\mathcal{O}_1\mathcal{O}_1\!\!:$. Therefore, from the bulk point of view, it is reasonable to consider the bulk dual for the boundary operator $\O_2$ to be the linear combination
\be\label{eq:D5_def_O'2}
O'_{ab}\equiv O_{ab}+\frac{c}{\Tf}O_a O_b.
\ee
The coefficient $c$ is fixed by demanding that the direct bulk computation of the three-point function $\langle\!\langle O_{ab}^{\prime}O_c O_d\rangle\!\rangle$ reproduces the result \eqref{3pt-final}. As we have shown in \eqref{eq:D5_3pt_o2o1o1_bulk}, the bulk coupling between $O_{ab}$ and $O_a$ vanishes on-shell. Therefore, the bulk calculation of the three-point function $\langle\!\langle O_{ab}^{\prime}O_c O_d\rangle\!\rangle$ only receives contribution from the part $\langle\!\langle :\!\! O_aO_b\!\!:O_cO_d\rangle\!\rangle$. Evaluating this by simple Wick contractions, we obtain
\begin{equation}
\frac{c}{T_{D5}}\langle \!\langle :\!\!O_1(\t_1,\mathbf{u}_1)^2\!\!: O_1(\t_2,\mathbf{u}_2) O_1(\t_3,\mathbf{u}_3)  \rangle \!\rangle
= \frac{c}{T_{D5}}\cdot 2 c_1^2 \frac{(\mathbf{u}_1\cdot\mathbf{u}_2)(\mathbf{u}_1\cdot\mathbf{u}_3)}{\t_{12}^2 \t_{13}^2}
\end{equation}
where $c_1$ is the 2-point function coefficient defined in \eqref{eq:D5_def_c2pt}, for $L=1$. 
Requiring that this matches (\ref{3pt-final}) for $L_1=2$, $L_2=L_3=1$, we find 
\be
c=\frac{27}{56\pi^2}\frac{\cos\tk}{\sin^4\tk}.
\ee
A similar analysis can be carried out for extremal three-point functions involving higher charge operators. We will see in section \ref{subsec:higherKK} that the contribution of the two-particle state in $O'_{ab}$ is necessary in order to obtain the correct result for the 4-point function $\langle \O_2\O_2\O_1\O_1\rangle$. 

\subsection{Four-point function of $AdS_5$ fluctuations}
\begin{figure}
\begin{center}
\includegraphics[width=0.7\textwidth]{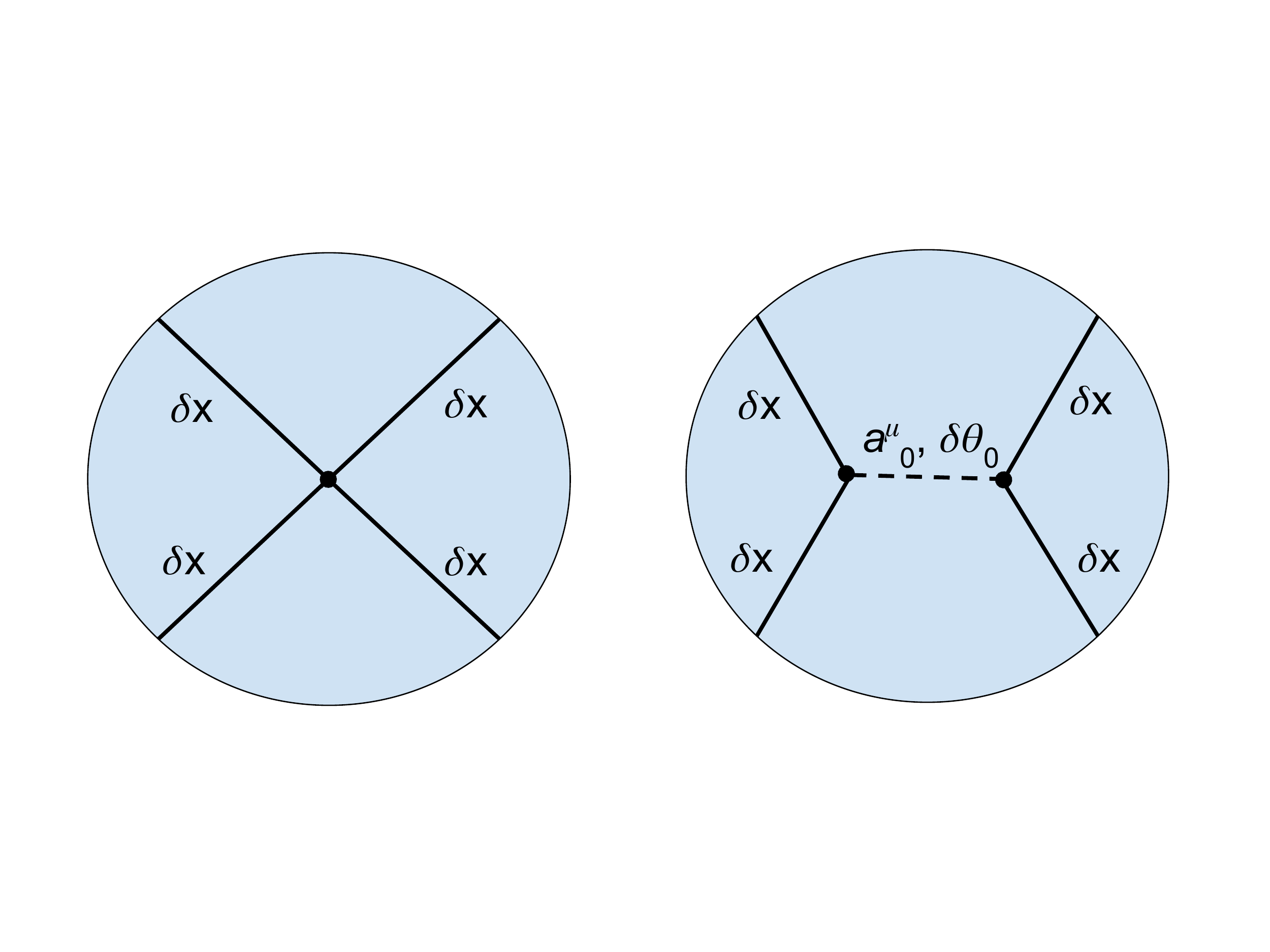}
\end{center}
\vskip -2cm
\caption{Witten diagrams for computing the connected part of the four-point function $\langle\dx^{i_1}_0\dx^{i_2}_0\dx^{i_3}_0\dx^{i_4}_0\rangle$. The $l=0$ modes of $\dt$ and $a_\mu$ fields are exchanged in the exchange diagram.}
\label{fig:D5-4pt-4x}
\end{figure}

In this section, we compute the connected part of the four-point function
\be
\langle\!\langle {\mathbb{F}_t}^{i_1}(\t_1){\mathbb{F}_t}^{i_2}(\t_2){\mathbb{F}_t}^{i_3}(\t_3){\mathbb{F}_t}^{i_4}(\t_4)\rangle\!\rangle=
\langle\dx_0^{i_1}(\t_1)\dx_0^{i_2}(\t_2)\dx_0^{i_3}(\t_3)\dx_0^{i_4}(\t_4)\rangle_{AdS_2}.
\ee
The relevant quartic vertices from expanding the D5-brane action are
\bea{\label{eq:D5_4x_1}
L^{(4)}_{xxxx}=&\frac{\pi^2\Tf \sin^3 \tk}{3}\biggl[(\p_\m\dx_0^i\p^\m\dx_0^i)^2-2(\p_\m\dx_0^i\p_\n\dx_0^i)(\p^\m\dx_0^j\p^\n\dx_0^j)\\
&+2(\p_\m\dx_0^i\p^\m\dx_0^i)\dx_0^2 +4\dx_0^2\dx_0^2
-\cot^2\tk \left(\p_\m\dx_0^i\p^\m\dx_0^i+2\dx_0^2\right)^2
\biggr],
\eqn
}
which leads to the contact diagram in figure \ref{fig:D5-4pt-4x}. The contribution from the exchange diagram in figure \ref{fig:D5-4pt-4x} results from the following cubic vertices:
\bea{
L_{xxf_0}&=-\frac{4i\,\pi^2 \Tf\sin\tk\cos\tk}{3}\left(\p_\m\dx_0^i\p^\m\dx_0^i+ 2\dx_0^2\right)f_0,\\
L_{xx\dt_0}&=\frac{16\pi^2 \Tf\sin^2\tk\cos\tk}{3}\left(\p_\m\dx_0^i\p^\m\dx_0^i+ 2\dx_0^2\right)\dt_0.
\eqn
}

When computing the contribution from the exchange diagram, the field $\dx^i_0$ is put on-shell. Therefore, we can use the equations of motion for the external fields $\dx_0^i$ and simplify the cubic vertices to
\be\label{eq:D5_3pt_cubic}
-\frac{ 4i\,\pi^2 \Tf\sin\tk\cos\tk}{3}\dx_0^i\,\p^\m\dx_0^i\biggl[-\p_\m f_0-4i\sin\tk \p_\m\dt_0\biggr].
\ee
We emphasize here that this cubic coupling is only correct when $\dx^i_0$ is on-shell. To compute the exchange diagram, we need to use the bulk propagator $G_{pq}(\t,r;\t',r')$ with $p,q\in\{\t,r,\theta\}$ defined by the bulk two-point functions:
\bea{\label{eq:D5_Gprop_def}
G_{\mu\nu}(\t,r;\t',r')&=\bigl\langle (a_\mu)_0(\t,r)(a_\nu)_0(\t',r')\bigr\rangle,\\
G_{\mu\theta}(\t,r;\t',r')&=\bigl\langle (a_\mu)_0(\t,r)\dt_0(\t',r')\bigr\rangle,\eqn\\
G_{\theta\theta}(\t,r;\t',r')&=\bigl\langle \dt_0(\t,r)\dt_0(\t',r')\bigr\rangle.
}
Since the quadratic action is not diagonal in $a_\mu$ and $\dt$, the bulk propagator $G_{pq}$ satisfies the following equations derived from \eqref{eq:D5_eom1}:
\bea{\label{eq:D5_G_tf_0}
-\nabla^\mu (\varepsilon^{\alpha\beta}\partial_\alpha G_{\beta\gamma'})-4i\sin\tk \nabla^\mu G_{\theta\gamma'}
&=\frac{3r^2}{8\pi^2\Tf\sin\tk}{\varepsilon^\mu}_{\gamma'}\delta^2(\t,r;\t',r'),\\
-\nabla^\mu (\varepsilon^{\alpha\beta}\partial_\alpha G_{\beta\theta})-4i\sin\tk \nabla^\mu G_{\theta\theta}
&=0,\eqn\\
-\nabla_\m\nabla^\m G_{\theta p}-4 G_{\theta p}+ \frac{4i}{\sin\tk} \varepsilon^{\m\n}\p_\m G_{\n p}
&=\frac{3r^2}{8\pi^2\Tf\sin^3\tk}\delta_{\theta p}\delta^2(\t,r;\t',r'),
}
where we have suppressed the dependence of $G_{pq}$ on the coordinates to simplify the notation. Due to the structure of \eqref{eq:D5_3pt_cubic}, we find that the exchange diagram can be reduced to a contact diagram with the following effective quartic coupling:
\be\label{eq:D5_4x_intout}
L_{exchange}=\frac{\pi^2\Tf \sin\tk\cos^2\tk}{3}\left(\p_\m\dx_0^i\p^\m\dx_0^i+2\dx_0^2\right)^2.
\ee
It follows that the connected part of the four-point function can be computed effectively from a single contact diagram with the quartic coupling:
\bea{\label{eq:D5_4x_final}
L^{eff}_{xxxx}=&\frac{8\pi^2\Tf \sin^3 \tk}{3}
\biggl[\frac{1}{8}(\p_\m\dx_0^i\p^\m\dx_0^i)^2-\frac{1}{4}(\p_\m\dx_0^i\p_\n\dx_0^i)(\p^\m\dx_0^j\p^\n\dx_0^j)\biggr.\\
&+\biggl.\frac{1}{4}(\p_\m\dx_0^i\p^\m\dx_0^i)\dx_0^2 +\frac{1}{2}\dx_0^2\dx_0^2\biggr]. 
\eqn
}
This effective quartic coupling in fact takes the identical form as the one appeared in the fundamental string case \cite{Giombi:2017cqn}, but with a different prefactor. Using the result in \cite{Giombi:2017cqn}, we find that the connected part of the normalized four-point function is (to get the normalized 
correlation function, we divide by the two-point function normalization factor in (\ref{xx-D5})):
\be\label{eq:D5_4x_f1}
\langle\dx_0^{i_1}(\t_1)\dx_0^{i_2}(\t_2)\dx_0^{i_3}(\t_3)\dx_0^{i_4}(\t_4)\rangle = 
\frac{3}{8\pi^2 \Tf \sin^3\tk} \frac{G^{i_1 i_2 i_3 i_4}_{4x}(\chi)}{\t_{12}^4\t_{34}^4},
\ee
where the expression of $G^{i_1 i_2 i_3 i_4}_{4x}(\chi)$ is given in Appendix \ref{sec:app_funcs}. 

Note that if we take the string limit defined by
\be 
\frac{k}{N}\to 0,\quad (\tk)^3\to \frac{3\pi k}{2N},
\ee 
the four-point function \eqref{eq:D5_4x_f1} then becomes
\be\label{eq:D5_4x_f2}
\langle\dx_0^{i_1}(\t_1)\dx_0^{i_2}(\t_2)\dx_0^{i_3}(\t_3)\dx_0^{i_4}(\t_4)\rangle \to
\frac{2\pi}{k\sqrt{\lambda}} \frac{G^{i_1 i_2 i_3 i_4}_{4x}(\chi)}{\t_{12}^4\t_{34}^4}.
\ee
Comparing with the result in \cite{Giombi:2017cqn}, we see that the D-brane result reduces to the result calculated from $k$ weakly coupled coincident strings.

\subsection{Two $AdS_5$ and two $S^5$ fluctuations of D5-brane}
In this section, we compute the connected part of the four-point function
\be
\langle\!\langle {\mathbb{F}_t}^{i_1}(\t_1){\mathbb{F}_t}^{i_2}(\t_2)\Phi_{a_1}(\t_3)\Phi_{a_2}(\t_4)\rangle\!\rangle
=\langle \dx_0^{i_1}(\t_1) \dx_0^{i_2}(\t_2) O_{a_1}(\t_3) O_{a_2}(\t_4)\rangle_{AdS_2}.
\ee
In previous section, we have shown that the four-point function $\langle\dx_0^{i_1}\dx_0^{i_2}\dx_0^{i_3}\dx_0^{i_4}\rangle$ has the same form as in the fundamental string case. Then the supersymmetry uniquely fixes the four-point function $\langle \dx_0^{i_1} \dx_0^{i_2} O_{a_1} O_{a_2}\rangle$. In fact, we expect it to have the same form as the correlator $\langle \dx^{i_1}\dx^{i_2}\dy_{a_1}\dy_{a_2}\rangle$ computed in the fundamental string case \cite{Giombi:2017cqn} but with the same prefactor as in \eqref{eq:D5_4x_final}. We verify this by explicitly calculating the four-point function using the effective action for the fluctuations.

\begin{figure}
\begin{center}
\includegraphics[width=0.7\textwidth]{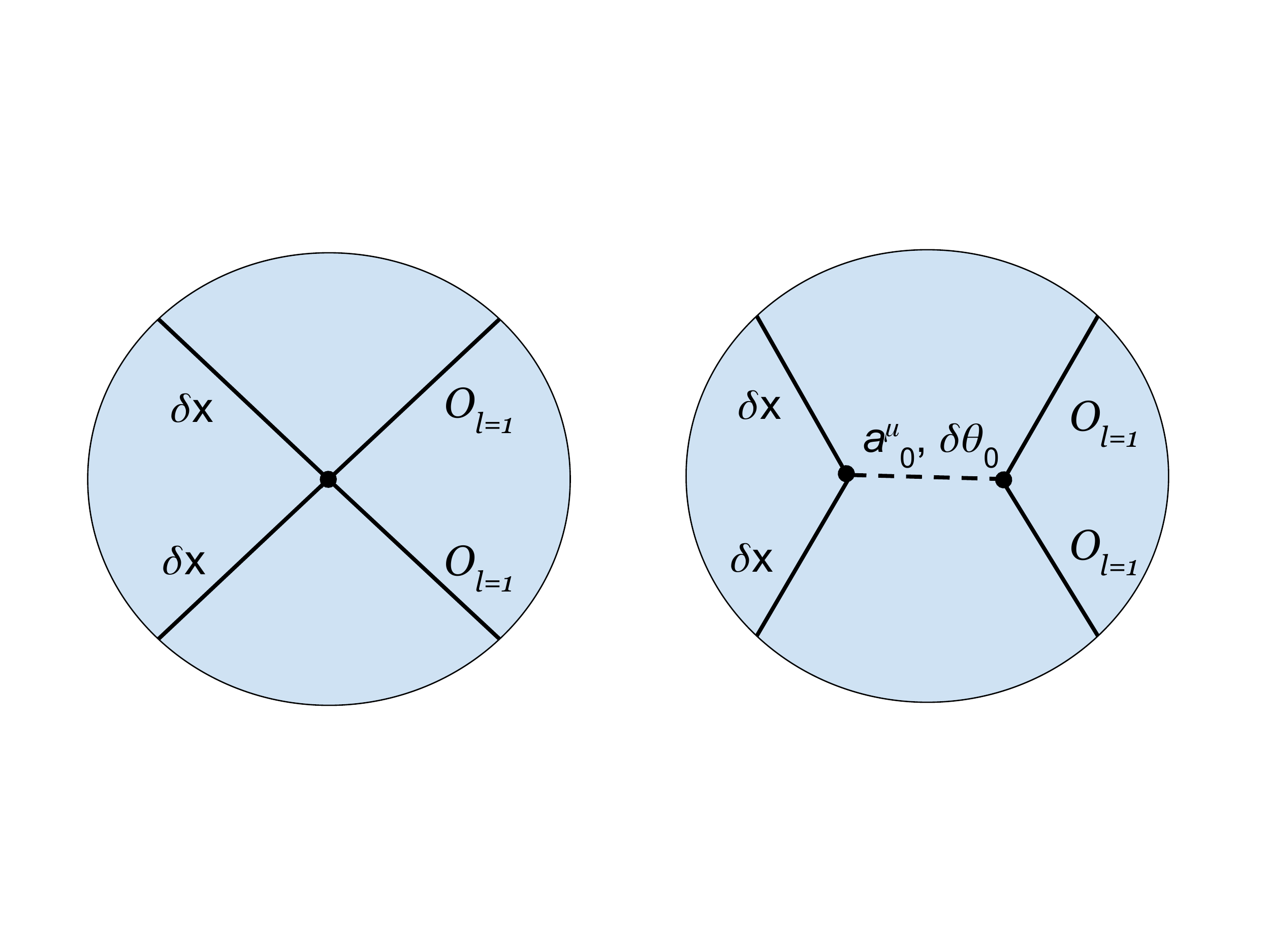}
\end{center}
\vskip -2cm
\caption{Witten diagrams for computing the connected part of the four-point function $\langle\dx^{i_1}_0\dx^{i_2}_0 O_{a_1} O_{a_2}\rangle$. The $l=0$ modes of $\dt$ and $a_\mu$ fields are exchanged in the exchange diagram.}
\label{fig:D5-4pt-2x2O_1}
\end{figure}

There is a quartic coupling from the expanded D-brane action
\bea{\label{eq:D5_2O2X_1}
L^{(4)}_{xxoo}=&\frac{2\pi^2\Tf\sin^3\tk}{15}
\biggl[
\frac{(4-\cos 2\tk)}{\sin^2\tk}\p_\m O_a\p^\m O_a\p_\n\dx_0^i\p^\n\dx_0^i\\
&-10\p_\m O_a\p_\n O_a\p^\m\dx_0^i\p^\n\dx_0^i
+6\cot^2\tk\p_\m O_a\p^\m O_a\dx_0^i\dx_0^i\\
&+16\cot^2\tk O_a O_a\p_\m\dx_0^i\p^\m\dx_0^i
+32\cot^2\tk O_a O_a\dx_0^i\dx_0^i
\biggr],
\eqn
}
which leads to the contact diagram in figure \ref{fig:D5-4pt-2x2O_1}. The exchange diagram in figure \ref{fig:D5-4pt-2x2O_1} results from the cubic couplings \eqref{eq:D5_3pt_cubic} and\footnote{There are also cubic couplings between $(\dx^i)_a$, $\dx^i_0$ and $O_a$, which leads to the Witten diagram with bulk $(\dx^i)_a$ fields being exchanged. However, if we put $\dx^i_0$ and $O_a$ on-shell, then this cubic coupling vanishes.}
\begin{align}
L_{oof_0}&=\frac{2i\pi^2\Tf\sin 2\tk}{15}\left(3\p_\m O_a\p^\m O_a+16 O_a O_a\right)f_0,
\label{eq:D5_o1o1f0}
\\
L_{oo\dt_0}&=-\frac{8\pi^2\Tf\sin 2\tk\sin\tk}{15}\left(
\p_\m O_a\p^\m O_a\dt_0-4O_a\p_\m O_a\p^\m\dt_0+4O_a O_a\dt_0\right).
\label{eq:D5_o1o1dt0}
\end{align}

Due to the special form of the coupling \eqref{eq:D5_3pt_cubic}, the exchange diagram in figure \ref{fig:D5-4pt-2x2O_1} can be again reduced to a contact diagram with the effective quartic coupling:
\be\label{eq:D5_2O2X_2}
L_{exchange}=-\frac{2\pi^2\Tf\cos^2\tk\sin\tk}{15}
\left(3\p_\m O_a\p^\m O_a+16 O_aO_a\right)
\left(\p_\n\dx_0^i\p^\n\dx_0^i+2\dx^i_0\dx^i_0\right).
\ee
Combining \eqref{eq:D5_2O2X_1} and \eqref{eq:D5_2O2X_2}, we see that the four-point function can be computed from a contact diagram with the effective quartic coupling
\be\label{eq:D5_xxoo_eff}
L^{eff}_{xxoo}=\frac{8\pi^2\Tf\sin^3\tk}{3}
\biggl(
\frac{1}{4}\p_\m O_a\p^\m O_a\p_\n\dx_0^i\p^\n\dx_0^i
-\frac{1}{2}\p_\m O_a\p_\n O_a\p^\m\dx_0^i\p^\n\dx_0^i
\biggr).
\ee
The form is exactly what we expect from the fundamental string case and the prefactor agrees with \eqref{eq:D5_4x_final}. Using the result in \cite{Giombi:2017cqn}, we find that the connected part of the normalized four-point function takes the form:
\be
\langle \dx_0^{i_1}(\t_1) \dx_0^{i_2}(\t_2) O_{a_1}(\t_3) O_{a_2}(\t_4)\rangle = \delta^{i_1i_2}\delta_{a_1a_2}\,
\frac{3}{8\pi^2 \Tf \sin^3\tk}\frac{G_{2x2y}(\chi)}{\t_{12}^4\t_{34}^2} ,
\ee
where the expression for $G_{2x2y}(\chi)$ is given in Appendix \ref{sec:app_funcs}.

\subsection{Four $S^5$ fluctuations of D5-brane}
\begin{figure}
\begin{center}
\includegraphics[width=0.7\textwidth]{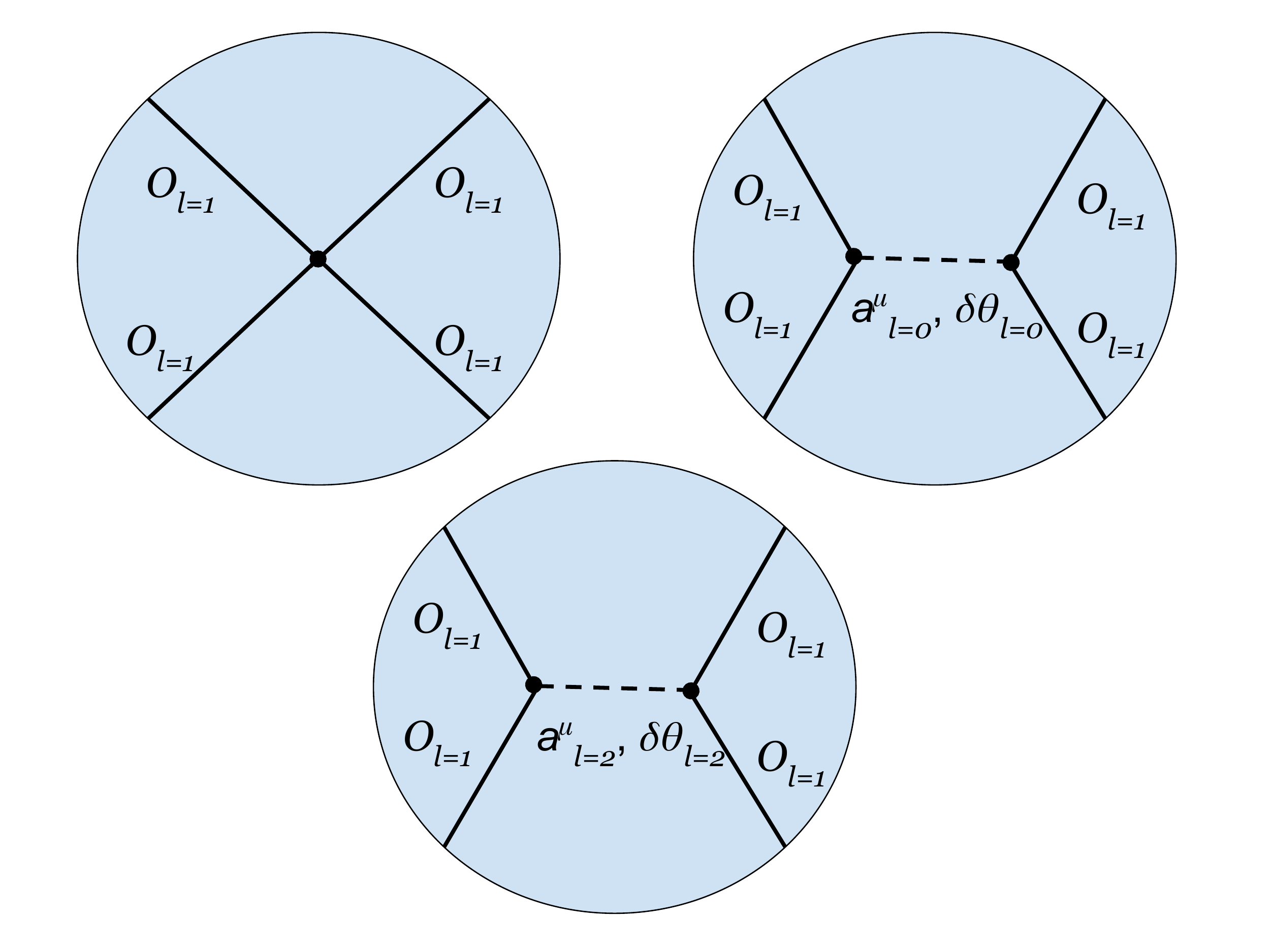}
\end{center}
\vskip -1cm
\caption{Witten diagrams for computing the connected part of the four-point function $\langle O_{a_1}O_{a_2}O_{a_3}O_{a_4}\rangle$. Both $l=0$ and $l=2$ modes of $\dt$ and $a_\mu$ fields are exchanged in the exchange diagrams.}
\label{fig:D5-4pt-4O_1}
\end{figure}

In this section, we compute the four-point function 
\be
\langle\!\langle \Phi_{a_1}(\t_1) \Phi_{a_2}(\t_2) \Phi_{a_3}(\t_3) \Phi_{a_4}(\t_4)
\rangle\!\rangle=
\langle O_{a_1}(\t_1) O_{a_2}(\t_2) O_{a_3}(\t_3) O_{a_4}(\t_4)\rangle_{AdS_2}.
\ee
The supersymmetry fixes $\langle O_{a_1}O_{a_2}O_{a_3}O_{a_4}\rangle$ to take the same form as $\langle y_{a_1}y_{a_2}y_{a_3}y_{a_4}\rangle$ in \cite{Giombi:2017cqn}. The Witten diagrams for the D-brane calculation are shown in figure \ref{fig:D5-4pt-4O_1}. The contact diagram in figure \ref{fig:D3-4pt-2y2x} results from the quartic coupling from the expanded action:
\bea{\label{eq:D5_4O_1}
L^{(4)}_{oooo}=&\frac{\pi^2\Tf\sin\tk}{105}
\biggl[
-80\cos^2\tk O_aO_aO_bO_b-14(13-8\sin^2\tk)\p_\m O_a\p^\m O_a O_bO_b
\\
&
-(11-46\sin^2\tk)\p_\m O_a\p^\m O_a\p_\n O_b\p^\n O_b
-2(4+31\sin^2\tk) \p_\m O_a \p_\n O_a\p^\m O_b \p^\n O_b
\biggr].
\eqn
}
The other two diagrams in figure \ref{fig:D5-4pt-4O_1} involve the exchange of $l=0$ and $l=2$ modes of $\dt$ and $a_\mu$ fields. 

\paragraph{Exchange of $l=0$ modes}
In this case, the cubic couplings involved are \eqref{eq:D5_o1o1f0} and \eqref{eq:D5_o1o1dt0}. Using the fact that $O_a$ is put on-shell in the calculation of the Witten diagram, we can wirte the cubic couplings as
\bea{\label{eq:D5_o1o1l0_cubic}
\frac{2i\pi^2\Tf\sin 2\tk}{15}&\biggl[
3O_a\p_\m O_a(-\p^\m f_0-4i\sin\tk \p^\m\dt_0)\\
&-4i\sin\tk\,O_aO_a\biggl(-\nabla^2\dt_0-4\dt_0+\frac{4i}{\sin\tk}f_0\biggr)
\biggr].
\eqn
}
In this form, we see that the exchange diagram can be reduced to a contact diagram with the effective quartic coupling:
\bea{\label{eq:D5_2O2O_l=0}
L_{exc,l=0}=&
\frac{\pi^2\Tf\cos^2\tk\sin\tk}{75}
\biggl(
64O_aO_aO_bO_b+128\p_\m O_a\p^\m O_a O_bO_b\\
&+9\p_\m O_a\p^\m O_a\p_\n O_b\p^\n O_b
\biggr).
\eqn
}

\paragraph{Exchange of $l=2$ modes}
In this case, the cubic couplings involved are
\begin{align}
L_{oof_2}=&\frac{i 8\pi^2 \Tf \sin 2\tk}{105}\left[-\p_\m O_a\p^\m O_b f_{ab}+5\varepsilon^{\m\n}O_a\p_\m O_b (a_\n)_{ab}
+\frac{11}{2}O_a O_b f_{ab}\right],\\
L_{oo\dt_2}=&\frac{8\pi^2\Tf \sin 2\tk\sin\tk}{105}\biggl(8\p_\m O_a\p^\m O_b\dt_{ab}
+3O_a\p_\m O_b \p^\m \dt_{ab}-3O_a O_b\dt_{ab}\biggr).
\end{align}
Using integration by parts and the on-shellness of the external $O_a$, the cubic couplings can be brought to the form
\bea{\label{eq:D5_cubic_o1o1ex}
\frac{8\pi^2\Tf\sin 2\tk}{105}&\biggl\{-i O_a\p_\m O_b\bigl[
-\p^\m f_{ab}+10\varepsilon^{\m\n}(a_\n)_{ab}-4i\sin\tk\p^\m\dt_{ab}
\bigr]\\
&-\frac{\sin\tk}{2}O_aO_b\biggl(-\nabla^2\dt_{ab}+6\dt_{ab}+\frac{4i}{\sin\tk} f_{ab}\biggr)\biggr\}. 
\eqn
}
To compute the exchange diagram, we need to use the bulk propagator $G^{ab\,a'b'}_{pq}(\t,r;\t',r')$ with $p,q\in\{\t,r,\theta\}$ defined similarly as in \eqref{eq:D5_Gprop_def}. The bulk propagator $G^{ab\,a'b'}_{pq}$ satisfies the following equations derived from \eqref{eq:D5_eom1}:
\bea{\label{eq:D5_G_tf(ij)}
-\nabla^\mu (\varepsilon^{\alpha\beta}\partial_\alpha G_{\beta\gamma'}^{ab\, a'b'})+10\varepsilon^{\mu\nu}G_{\nu\gamma'}^{ab\,a'b'}-4i\sin\tk \nabla^\mu G_{\theta\gamma'}^{ab\,a'b'}
&=\frac{105 M^{ab\,a'b'}r^2}{16\pi^2\Tf\sin\tk}{\varepsilon^\mu}_{\gamma'}\delta^2(\t,r;\t',r'), \\
-\nabla^\mu (\varepsilon^{\alpha\beta}\partial_\alpha G_{\beta\theta}^{ab\,a'b'})+10\varepsilon^{\mu\nu}G_{\nu\theta}^{ab\,a'b'}-4i\sin\tk \nabla^\mu G_{\theta\theta}^{ab\,a'b'}
&=0, \eqn\\
-\nabla_\m\nabla^\m G_{\theta p}^{ab\,a'b'}+6 G_{\theta p}^{ab\,a'b'}+ \frac{4i}{\sin\tk} \varepsilon^{\m\n}\p_\m G_{\n p}^{ab\,a'b'}
&=\frac{105M^{ab\,a'b'}r^2}{16\pi^2\Tf\sin^3\tk}\delta_{\theta p}\delta^2(\t,r;\t',r'),
}
where $M^{ab\,a'b'}$ is defined as
\be
M^{ab\, a'b'}\equiv\frac{1}{2}\left(\delta^{aa'}\delta^{bb'}+\delta^{ab'}\delta^{ba'}-\frac{2}{5}\delta^{ab}\delta^{a'b'}\right).
\ee
By examining the structure of the cubic coupling \eqref{eq:D5_cubic_o1o1ex}, we see that the exchange diagram can be again reduced to a contact diagram with the effective quartic coupling:
\bea{\label{eq:D5_2O2O_l=2}
L_{exc,l=2}=
&\frac{2\pi^2\Tf\cos^2\tk\sin\tk}{525}
\biggl(
20\p_\m O_a\p_\n O_a \p^\m O_b \p^\n O_b
-4\p_\m O_a\p^\m O_a \p_\n O_b \p^\n O_b\\
&
+7\p_\m O_a\p^\m O_a O_bO_b
-24 O_aO_aO_bO_b
\biggr).
\eqn
}

\paragraph{The four-point function}
Combining \eqref{eq:D5_2O2O_l=0} and \eqref{eq:D5_2O2O_l=2} with \eqref{eq:D5_4O_1}, we find that the four-point function can be computed from a contact diagram with the effective quartic coupling:
\bea{\label{eq:D5_4O_eff}
L^{eff}_{oooo}=&
\frac{8\pi^2\Tf\sin^3\tk}{3}
\biggl(
-\frac{1}{4}\p_\m O_a\p^\m O_a O_bO_b
+\frac{1}{8}\p_\m O_a\p^\m O_a\p_\n O_b\p^\n O_b\\
&-\frac{1}{4}\p_\m O_a \p_\n O_a\p^\m O_b \p^\n O_b
\biggr).
\eqn
}
The structure of the vertices is the same as what we expect from the fundamental string case \cite{Giombi:2017cqn}. Computing the Witten diagram with quartic coupling \eqref{eq:D5_4O_eff}, we find that the connected part of the normalized four-point function is
\be\label{eq:D5_o1o1o1o1_index}
\langle O^{a_1}(\t_1) O^{a_2}(\t_2) O^{a_3}(\t_3) O^{a_4}(\t_4)\rangle =
\frac{3}{8\pi^2\Tf\sin^3\tk}
\frac{G^{a_1 a_2 a_3 a_4}_{4y}(\chi)}{\t_{12}^2\t_{34}^2} ,
\ee
where the expression of $G^{a_1 a_2 a_3 a_4}_{4y}(\chi)$ is given in Appendix \ref{sec:app_funcs}.

To compare with the prediction of localization, we again transform to the circular Wilson loop and setting the polarizations to \eqref{eq:D5_topo_config_pol}. The normalized four-point function then becomes
\be\label{eq:D5_o1o1o1o1_2}
\frac{\langle\!\langle \tO_{1} \tO_{1} \tO_{1} \tO_{1}\rangle\!\rangle}{\langle\!\langle \tO_{1} \tO_{1}\rangle\!\rangle^2}
=-\frac{9}{16\pi^3\Tf\sin^3\tk}=-\frac{9}{8Ng\sin^3\tk},
\ee
which agrees with the prediction of localization.

\subsection{Four $S^5$ fluctuations including higher KK modes\label{subsec:higherKK}}
\label{sec:D5_o2o2o1o1}
In this section, we compute the four-point function which includes the $l=2$ KK modes: 
\be
\langle\!\langle \Phi^{a_1b_1}(\t_1) \Phi^{a_2b_2}(\t_2) \Phi^{a_3}(\t_3) \Phi^{a_4}(\t_4) 
\rangle\!\rangle=
\langle  O'^{a_1b_1}(\t_1) O'^{a_2b_2}(\t_2)O^{a_3}(\t_3) O^{a_4}(\t_4)\rangle_{AdS_2},
\ee
where $O'^{ab}$ is defined in \eqref{eq:D5_def_O'2}. The four-point function can be written as the sum of two pieces
\be\label{eq:D5_4pt_O2O2O1O1_def}
\langle O'^{a_1b_1} O'^{a_2b_2}O^{a_3} O^{a_4} \rangle=
\langle O^{a_1b_1} O^{a_2b_2}O^{a_3} O^{a_4}\rangle+
\frac{c^2}{\Tf^2}\langle :\!\!O^{a_1}O^{b_1}\!\!: :\!\!O^{a_2}O^{b_2}\!\!:O^{a_3}O^{a_4}\rangle.
\ee

\begin{figure}
\begin{center}
\includegraphics[width=0.7\textwidth]{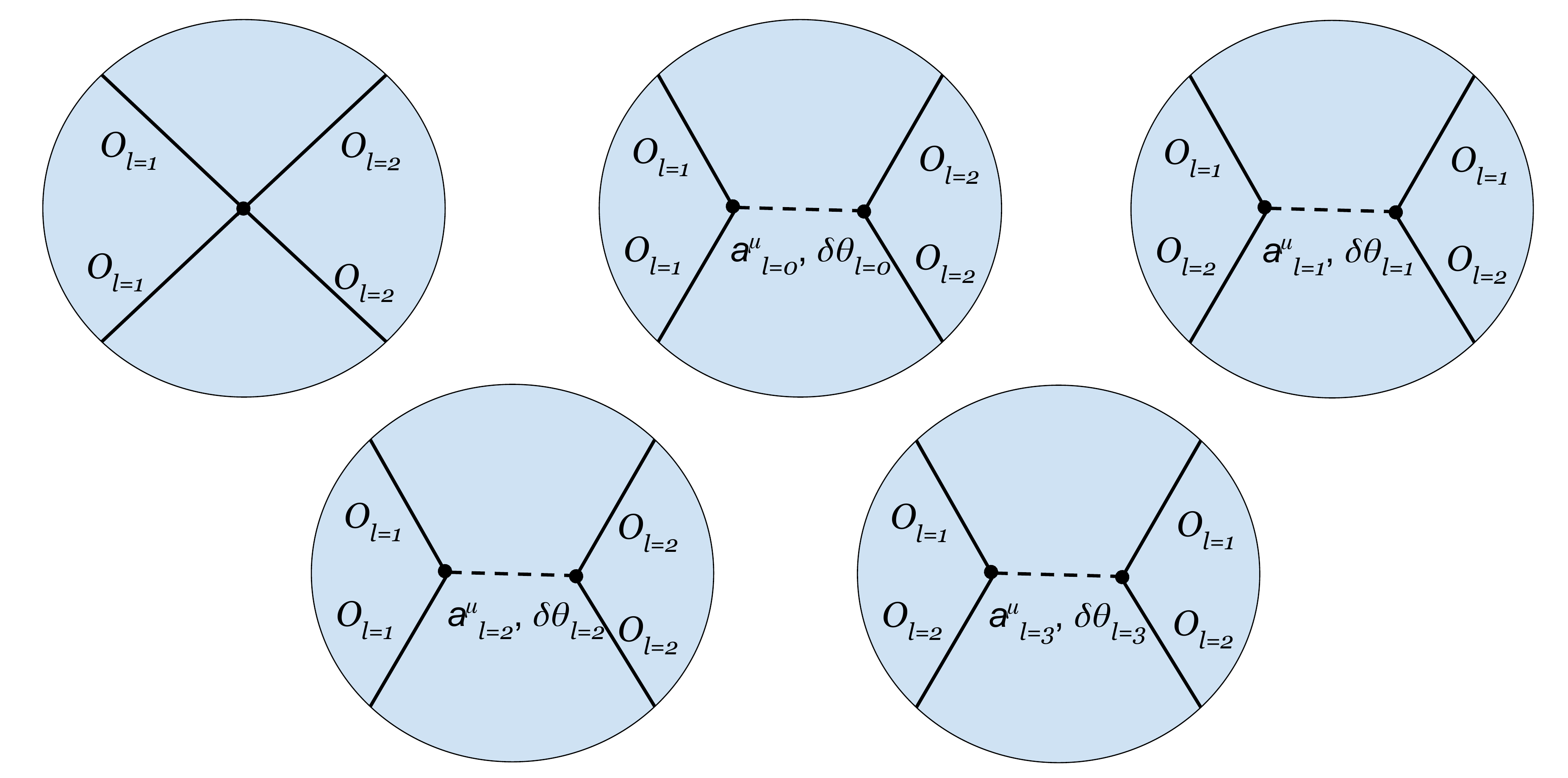}
\end{center}
\vskip -1cm
\caption{Witten diagrams for computing the connected part of the four-point function $\langle O_{a_1}O_{a_2}O_{a_3b_3}O_{a_4b_4}\rangle$. The $l=0,1,2,3$ modes of $\dt$ and $a_\mu$ fields are exchanged in the exchange diagrams.}
\label{fig:D5-4pt-2O_12O_2}
\end{figure}

We shall first compute the piece $\langle O^{a_1b_1} O^{a_2b_2}O^{a_3} O^{a_4}\rangle$.
The Witten diagrams involved are given in figure \ref{fig:D5-4pt-2O_12O_2}. The contact diagram comes from the quartic couplings in the expansion of the D-brane action:
\bea{\label{eq:D5_o1o1o2o2_quartic}
L^{(4)}_{o_1o_1o_2o_2}=
&\frac{16\pi^2\Tf\sin\tk}{5145}\biggl[
\frac{(-71+55\cos 2\tk)}{2}\p_\m O_a\p_\n O_a\p^\m O_{bc}\p^\n O_{bc}\\
&+\frac{(23-40\cos 2\tk)}{2}\p_\m O_a\p^\m O_a\p_\n O_{bc}\p^\n O_{bc}\\
&+(28-35\cos 2\tk)\p_\m O_a\p_\n O_b\p^\m O_{ac}\p^\n O_{bc}\\
&+\frac{(-79+47\cos 2\tk)}{2}\p_\m O_a\p_\n O_b\p^\n O_{ac}\p^\m O_{bc}\\
&+\frac{(-61+65\cos 2\tk)}{2}\p_\m O_a\p^\m O_b\p_\n O_{ac}\p^\n O_{bc}\\
&-(277+88\cos 2\tk)\p_\m O_a\p^\m O_a O_{bc}O_{bc}\\
&+(80-46\cos 2\tk)\p_\m O_a\p^\m O_b O_{ac}O_{bc}\\
&-36(13+6\cos 2\tk)\p_\m O_a O_b \p^\m O_{ac}O_{bc}
-156\cos^2 \tk O_a O_a O_{bc}O_{bc}\\
&\biggl.-840\cos^2 \tk O_a O_b O_{ac}O_{bc}
\biggr].
\eqn	
}
We note here that in deriving \eqref{eq:D5_o1o1o2o2_quartic} we have used the fact that both $O_a$ and $O_{ab}$ are on-shell in the computation of the Witten diagram so that the equations of motion can be applied. 
The other diagrams in figure \ref{fig:D5-4pt-2O_12O_2} involve the exchange of higher KK modes of $\dt$ and $a_\mu$ fields. 

\paragraph{Exchange of $l=0$ modes}
The cubic vertices appear in the exchange diagram are \eqref{eq:D5_o1o1l0_cubic} and
\begin{align}
L_{o_2o_2f_0}&=\frac{72i\pi^2\Tf\sin 2\tk}{1715}\left(\p_\m O_{ab}\p^\m O_{ab}+22O_{ab}O_{ab}\right)f_0,\\
L_{o_2o_2\dt_0}&=-\frac{48\pi^2\Tf\sin 2\tk\sin \tk}{1715}\left(\p_\m O_{ab}\p^\m O_{ab}\dt_0
-20O_{ab}\p_\m O_{ab}\p^\m\dt_0+32O_{ab}O_{ab}\dt_0\right).
\end{align}
As in the previous cases, the exchange diagram can be reduced to a contact diagram with the effective quartic coupling:
\bea{
L_{exc, l=0}=&\frac{216\pi^2\Tf\cos^2\tk\sin\tk}{8575}\biggl(
\p_\m O_a\p^\m O_a\p_\n O_{bc}\p^\n O_{bc}
+\frac{122}{3}\p_\m O_a\p^\m O_a O_{bc}O_{bc}\\
&+\frac{80}{3}O_a O_a O_{bc} O_{bc}
\biggr).
\eqn
}

\paragraph{Exchange of $l=1$ modes}
The cubic vertices appear in the exchange diagram are
\bea{
L_{o_1o_2f_1}=&\frac{8i\pi^2\Tf\sin 2\tk}{245}\biggl[3\p_\m O_a\p^\m O_{ab} f_b-16\varepsilon^{\m\n}\p_\m O_a O_{ab}(a_\n)_b\\
&+12\varepsilon^{\m\n}O_a\p_\m O_{ab}(a_\n)_b+42O_aO_{ab}f_b
\biggr],
\eqn
\\
L_{o_1o_2\dt_1}=&\frac{8\pi^2\Tf\sin 2\tk\sin\tk}{245}\biggl(
\p_\m O_a\p^\m O_{ab}\dt_b+20\p_\m O_a O_{ab}\p^\m\dt_b\\&+13O_a\p_\m O_{ab}\p^\m\dt_b-34O_aO_{ab}\dt_b
\biggr).
\eqn
}
By using the on-shellness of the external $O_a$ and $O_{ab}$ when computing the diagram, we can express the coupling in the form:
\bea{\label{eq:D5_o1o2l1_cubic}
\frac{8i\pi^2\Tf\sin 2\tk}{245}&\biggl\{
\bigl(-2\p_\m O_a O_{ab}+5O_a\p_\m O_{ab}\bigr)
\bigl[-\p_\m f_b+4\varepsilon^{\m\n}(a_\n)_b-4i\sin\tk\p_\m\dt_b\bigr]\\
&-10i\sin\tk\, O_aO_{ab}\biggl(-\nabla^2\dt_b+\frac{4i}{\sin\tk}f_b\biggr)
\biggr\}.
\eqn
}
We also need the bulk propagator $G^{a\, a'}_{pq}(\t,r;\t',r')$, which satisfies the equations:
\begin{align}\label{eq:D5_G_tf(i)}
-\nabla^\mu (\varepsilon^{\alpha\beta}\partial_\alpha G_{\beta\gamma'}^{a\, a'})+4\varepsilon^{\mu\nu}G_{\nu\gamma'}^{a\,a'}-4i\sin\tk \nabla^\mu G_{\theta\gamma'}^{a\,a'}
&=\frac{15 \delta^{a\,a'}r^2}{8\pi^2\Tf\sin\tk}{\varepsilon^\mu}_{\gamma'}\delta^2(\t,r;\t',r'), \\
-\nabla^\mu (\varepsilon^{\alpha\beta}\partial_\alpha G_{\beta\theta}^{a\,a'})+4\varepsilon^{\mu\nu}G_{\nu\theta}^{a\,a'}-4i\sin\tk \nabla^\mu G_{\theta\theta}^{a\,a'}
&=0,\\
-\nabla_\m\nabla^\m G_{\theta p}^{a\,a'}+ \frac{4i}{\sin\tk} \varepsilon^{\m\n}\p_\m G_{\n p}^{a\,a'}
&=\frac{15\delta^{a\,a'}r^2}{8\pi^2\Tf\sin^3\tk}\delta_{\theta p}\delta^2(\t,r;\t',r').
\end{align}
From the form of the coupling \eqref{eq:D5_o1o2l1_cubic}, it follows that the exchange diagram can be reduced to a contact diagram with the effective quartic coupling:
\bea{
L_{exc, l=1}=&
\frac{432\pi^2\Tf\cos^2\tk\sin\tk}{12005}\biggl(
\p_\m O_a\p_\n O_b\p^\m O_{ac}\p^\n O_{bc}
+\frac{196}{9} \p_\m O_a\p^\m O_b O_{ac}O_{bc}\\
&+\frac{580}{9} \p_\m O_a O_b \p^\m O_{ac}O_{bc}
+100 O_a O_b O_{ac}O_{bc}
\biggr).
\eqn
}

\paragraph{Exchange of $l=2$ modes}
The cubic couplings in the exchange diagram are \eqref{eq:D5_cubic_o1o1ex} and
\begin{align}
L_{o_2o_2f_2}=&\frac{16i\pi^2\Tf\sin 2\tk}{5145}\biggl[
\p_\m O_{ab}\p^\m O_{ac}f_{bc}+20\varepsilon^{\m\n}O_{ab}\p_\m O_{ac} (a_\n)_{bc}+112 O_{ab}O_{ac}f_{bc}
\biggr],\\
L_{o_2o_2\dt_2}=&\frac{32\pi^2\Tf\sin 2\tk\sin\tk}{5145}\biggl(
13\p_\m O_{ab}\p^\m O_{ac}\dt_{bc}+30O_{ab}\p_\m O_{ac}\p^\m\dt_{bc}-64O_{ab}O_{ac}\dt_{bc}
\biggr).
\end{align}
As before, the exchange diagram can be reduced to a contact diagram with the effective quartic coupling:
\bea{
L_{exc,l=2}=
&-\frac{32\pi^2\Tf\cos^2\tk\sin\tk}{5145}\biggl(
\p_\m O_a\p^\m O_b\p_\n O_{ac}\p^\n O_{bc}
-\frac{1}{5}\p_\m O_a\p^\m O_a\p_\n O_{bc}\p^\n O_{bc}
\\
&+119\p_\m O_a\p^\m O_b O_{ac}O_{bc}-\frac{119}{5}\p_\m O_a\p^\m O_a O_{bc}O_{bc}
+90 O_a O_b O_{ac}O_{bc}\\
&-18 O_a O_a O_{bc} O_{bc}
\biggr).
\eqn
}

\paragraph{Exchange of $l=3$ modes}
The cubic couplings appear in the exchange diagram are
\begin{align}
L_{oof_3}=&-\frac{32i \pi^2\Tf\sin 2\tk}{735}\left[\p_\m O_a\p^\m O_{bc}f_{abc}-3\varepsilon^{\m\n}\p_\m(O_a O_{bc})(a_\n)_{bc}-7O_a O_{bc}f_{abc}\right],\\
L_{oo\dt_3}=&\frac{32\pi^2\Tf\sin 2\tk\sin\tk}{735}\left[9\p_\m O_a\p^\m O_{bc}\dt_{abc}+\frac{3}{2}\p_\m(O_a O_{bc})\p^\m\dt_{abc}-5O_aO_{bc}\dt_{abc}\right].
\end{align} 
Using the on-shellness of the external $O_a$ and $O_{ab}$ when computing the diagram, we can expressed the cubic vertices as
\bea{\label{eq:D5_o1o2l3_cubic}
-\frac{16i\pi^2\Tf\sin 2\tk}{735}&\biggl\{\p_\m(O_a O_{bc})\bigl[
-\p^\m f_{abc}+18\varepsilon^{\m\n}(a_\n)_{abc}-4i\sin\tk\p_\m\dt_{abc}
\bigr]\\
&-2i\sin\tk\,O_aO_{bc}\left(-\nabla^2\dt_{abc}+14\dt_{abc}+\frac{4i}{\sin\tk}f_{abc}\right)
\biggr\}. 
\eqn
}
The bulk propagator $G^{abc\,a'b'c'}_{pq}(\t,r;\t',r')$ needed in the computation satisfies the following equations:
\begin{align}\label{eq:D5_G_tf(abc)}
-\nabla^\mu (\varepsilon^{\alpha\beta}\partial_\alpha G_{\beta\gamma'}^{abc\, a'b'c'})+18\varepsilon^{\mu\nu}G_{\nu\gamma'}^{abc\,a'b'c'}-4i\sin\tk \nabla^\mu G_{\theta\gamma'}^{abc\,a'b'c'}
&=\frac{315 M^{abc\,a'b'c'}r^2}{16\pi^2\Tf\sin\tk}{\varepsilon^\mu}_{\gamma'}\delta^2(\t,r;\t',r'), \\
-\nabla^\mu (\varepsilon^{\alpha\beta}\partial_\alpha G_{\beta\theta}^{ab\,a'b'})+18\varepsilon^{\mu\nu}G_{\nu\theta}^{abc\,a'b'c'}-4i\sin\tk \nabla^\mu G_{\theta\theta}^{abc\,a'b'c'}
&=0,\\
-\nabla_\m\nabla^\m G_{\theta p}^{abc\,a'b'c'}+14 G_{\theta p}^{abc\,a'b'c'}+ \frac{4i}{\sin\tk} \varepsilon^{\m\n}\p_\m G_{\n p}^{abc\,a'b'c'}
&=\frac{315M^{abc\,a'b'c'}r^2}{16\pi^2\Tf\sin^3\tk}\delta_{\theta p}\delta^2(\t,r;\t',r'),
\end{align}
where $M^{abc\,a'b'c'}$ is defined as
\bea{
M^{abc\, a'b'c'}=&\frac{1}{6}\biggl[\delta^{aa'}\delta^{bb'}\delta^{cc'}+\delta^{aa'}\delta^{bc'}\delta^{cb'}
+\delta^{ab'}\delta^{bc'}\delta^{ca'}+\delta^{ab'}\delta^{ba'}\delta^{cc'}+\delta^{ac'}\delta^{bb'}\delta^{ca'}\\
&+\delta^{ac'}\delta^{ba'}\delta^{ca'}
-\frac{2}{7}\bigl(\delta^{ab}\delta^{ca'}\delta^{b'c'}+\delta^{ab}\delta^{cb'}\delta^{a'c'}+\delta^{ab}\delta^{cc'}\delta^{a'b'}
+\delta^{ac}\delta^{ba'}\delta^{b'c'}\\
&+\delta^{ac}\delta^{bb'}\delta^{a'c'}
+\delta^{ac}\delta^{bc'}\delta^{a'b'}
+\delta^{bc}\delta^{aa'}\delta^{b'c'}+\delta^{bc}\delta^{ab'}\delta^{a'c'}+\delta^{bc}\delta^{ac'}\delta^{a'b'}
\bigr)
\biggr].
\eqn
}
From the form of \eqref{eq:D5_o1o2l3_cubic}, we see that the exchange diagram can be reduced to a contact diagram with the effective quartic coupling:
\bea{
L_{exc,l=3}=
&\frac{128\pi^2\Tf\cos^2\tk\sin\tk}{5145}\biggl(
\p_\m O_a\p_\n O_a\p^\m O_{bc}\p^\n O_{bc}-\frac{4}{7}\p_\m O_a\p_\n O_b\p^\m O_{ac}\p^\n O_{bc}\\
&+2\p_\m O_a\p_\n O_b\p^\n O_{ac}\p^\m O_{bc}
-\frac{3}{2}\p_\m O_a\p^\m O_a O_{bc}O_{bc}-6\p_\m O_a\p^\m O_b O_{ac}O_{bc}\\
&-\frac{54}{7}\p_\m O_a O_b\p^\m O_{ac} O_{bc}
-12 O_a O_a O_{bc} O_{bc} -\frac{120}{7}O_a O_b O_{ac}O_{bc}\biggr). \eqn
}

\paragraph{The four-point function}
Summing up all the diagrams, we find that the connected part of the four-point function can be computed from a single contact diagram with the effective quartic coupling:
\bea{\label{eq:D5_o2o2o1o1_feffv}
L^{eff}_{o_1o_1o_2o_2}=
&\frac{24\pi^2\Tf\sin^3\tk}{245}\biggl(
\p_\m O_a\p^\m O_a\p_\n O_{bc}\p^\n O_{bc}-2\p_\m O_a\p_\n O_a\p^\m O_{bc}\p^\n O_{bc}
\\
&-2\p_\m O_a\p^\m O_b\p_\n O_{ac}\p^\n O_{bc}+2\p_\m O_a\p_\n O_b\p^\m O_{ac}\p^\n O_{bc}
-2\p_\m O_a\p_\n O_b\p^\n O_{ac}\p^\m O_{bc}\\
&-6\p_\m O_a\p^\m O_a O_{bc} O_{bc}+4\p_\m O_a\p^\m O_b O_{ac} O_{bc}
-8\p_\m O_a O_b \p^\m O_{ac} O_{bc}\biggr).
\eqn
}
This leads to the following result for the connected part of the unnormalized four-point function:
\be
\langle O^{a_1 b_1}(\t_1)O^{a_2 b_2}(\t_2)O^{a_3}(\t_3)O^{a_4}(\t_4)\rangle
=-\frac{24\pi^2\Tf\sin^3\tk}{245}(\C_{\D=2}\C_{\D=1})^2 Q_{2O_22O_1}^{a_1b_1a_2b_2a_3a_4},
\ee
where
\bea{
Q_{2O_22O_1}^{a_1b_1a_2b_2a_3a_4}=&
8\biggl[-5D_{2211}-4\t_{12}^2 D_{3311}+4\t_{13}^2 D_{3221}+4\t_{14}^2 D_{3212}+4\t_{23}^2 D_{2321}+4\t_{24}^2 D_{2312}
\\
&\quad+2\t_{34}^2 D_{2222}
+8\left(\t_{12}^2\t_{34}^2-\t_{13}^2\t_{24}^2-\t_{14}^2\t_{23}^2\right)D_{3322}
\biggr]\delta^{a_1a_2}\delta^{b_1b_2}\delta^{a_3 a_4}\\
&+8\biggl[-5D_{2211}+4\t_{12}^2 D_{3311}+4\t_{14}^2 D_{3212}+4\t_{23}^2 D_{2321}+2\t_{34}^2 D_{2222}
\\
&\quad
-8\left(\t_{12}^2\t_{34}^2-\t_{13}^2\t_{24}^2+\t_{14}^2\t_{23}^2\right)D_{3322}
\biggr]\delta^{a_1a_3}\delta^{b_1b_2}\delta^{a_2 a_4}\\
&+8\biggl[-5D_{2211}+4\t_{12}^2 D_{3311}+4\t_{13}^2 D_{3221}+4\t_{24}^2 D_{2312}+2\t_{34}^2 D_{2222}
\\
&\quad
-8\left(\t_{12}^2\t_{34}^2+\t_{13}^2\t_{24}^2-\t_{14}^2\t_{23}^2\right)D_{3322}
\biggr]\delta^{a_1a_4}\delta^{b_1b_2}\delta^{a_2 a_3},
\eqn
}
where the function $D_{\D_1\D_2\D_3\D_4}$ is defined in Appendix \ref{sec:app_funcs}.
In terms of the confromal cross-ratios, the four-point function can be expressed as
\be\label{eq:D5_o2o2o1o1_index}
\langle O^{a_1 b_1}(\t_1)O^{a_2 b_2}(\t_2)O^{a_3}(\t_3)O^{a_4}(\t_4)\rangle
=-\frac{24\pi^3\Tf\sin^3\tk}{245}\frac{(\C_{\D=2}\C_{\D=1})^2}{\t_{12}^4\t_{34}^2} G^{a_1b_1a_2b_2a_3a_4}(\chi).
\ee
The function $G^{a_1b_1a_2b_2a_3a_4}(\chi)$ is defined as
\be
G^{a_1b_1a_2b_2a_3a_4}(\chi)=\biggl[
G_1(\chi)\delta^{a_1a_2}\delta^{b_1b_2}\delta^{a_3 a_4}
+G_2(\chi)\delta^{a_1a_3}\delta^{b_1b_2}\delta^{a_2 a_4}
+G_3(\chi)\delta^{a_1a_4}\delta^{b_1b_2}\delta^{a_2 a_3}
\biggr],
\ee
where
\bea{
G_1(\chi)=&\frac{3}{(\chi-1)^3}\biggl[
-4+12\chi-9\chi^2-2\chi^3+5\chi^4-2\chi^5\\
&+\left(
-4+14\chi-18\chi^2+10\chi^3-6\chi^5+6\chi^6-2\chi^7
\right)\frac{\log|1-\chi|}{\chi}\\
&+(6-6\chi+2\chi^2)\chi^4\log|\chi|
\biggr],	\eqn \\
G_2(\chi)=&\frac{3}{2(\chi-1)^3}\biggl[
4\chi-15\chi^2+11\chi^3+9\chi^4-15\chi^5+6\chi^6\\
&+\left(
4-12\chi+14\chi^2-10\chi^3+16\chi^5-18\chi^6+6\chi^7
\right)\log|1-\chi|\\
&+(-16+18\chi-6\chi^2)\chi^5\log|\chi|
\biggr],	\eqn \\
G_3(\chi)=&\frac{3}{2(\chi-1)^3}\biggl[
-4\chi+5\chi^2+9\chi^3-8\chi^4+4\chi^5\\
&+\left(
-4+12\chi-14\chi^2+6\chi^3+6\chi^4-10\chi^5+4\chi^6
\right)\log|1-\chi|\\
&+(-16+14\chi-4\chi^2)\frac{\chi^5}{\chi-1}\log|\chi|
\biggr].	\eqn 
}

The second piece in \eqref{eq:D5_4pt_O2O2O1O1_def} can be computed easily and we find
\bea{\label{eq:D5_o2o2o1o1_extra1}
&\frac{c^2}{\Tf^2}\langle :\!\!O^{a_1}O^{b_1}\!\!:\!(\t_1):\!\!O^{a_2}O^{b_2}\!\!:\!(\t_2)O^{a_3}(\t_3)O^{a_4}(\t_4)\rangle \\
=& \frac{c^2}{\Tf^2}\frac{c_1^3}{\t_{12}^4\t_{34}^2}
\left[
2\delta^{a_1a_2}\delta^{b_1b_2}\delta^{a_3 a_4}
+4\chi^2\delta^{a_1a_3}\delta^{b_1b_2}\delta^{a_2 a_4}
+4\frac{\chi^2}{(1-\chi)^2}\delta^{a_1a_4}\delta^{b_1b_2}\delta^{a_2 a_3}
\right].
\eqn
}
The first term in the bracket of \eqref{eq:D5_o2o2o1o1_extra1} does not contribute to the connected part of the four-point function as it is proportional to $\langle \!\langle  :\!\!\mathcal{O}_1\mathcal{O}_1\!\!::\!\!\mathcal{O}_1\mathcal{O}_1\!\!:\rangle\!\rangle\langle\!\langle:\!\!\mathcal{O}_1\!\!::\!\!\mathcal{O}_1\!\!:\rangle\!\rangle$.

Summing up the contribution from the two pieces in \eqref{eq:D5_4pt_O2O2O1O1_def}, we find the connected part of the normalized four-point function is
\be
\langle\!\langle \O_{2}(\t_1,\vu_1) \O_{2}(\t_2,\vu_2) \O_{1}(\t_3,\vu_3) \O_{1}(\t_4,\vu_4) \rangle\!\rangle
=\frac{(\mathbf{u_1}\cdot\mathbf{u_2})^2(\mathbf{u_3}\cdot\mathbf{u_4})}{\t_{12}^4\t_{34}^2}\mathcal{G}(\chi,\xi,\zeta),
\ee
where 
\be\label{eq:D5_o2o2o1o1_G}
\mathcal{G}(\chi,\xi,\zeta)=-\frac{1}{16\pi^3\Tf\sin^3\tk}
\biggl[
G_1(\chi)+\xi G_2(\chi)+\zeta G_3(\chi)
-45\cot^2\tk\chi^2 \biggl(\xi +\frac{\zeta}{(1-\chi)^2}\biggr)
\biggr].
\ee
One can check explicitly that the function $\mathcal{G}(\chi,\xi,\zeta)$ indeed satisfies the superconformal identities \eqref{eq:superconformal_id}. By transforming to the circular Wilson loop and setting the polarizations to \eqref{eq:D5_topo_config_pol}, we find the normalized four-point function becomes
\be
\frac{\langle\!\langle \tO_{2} \tO_{2} \tO_{1}\tO_{1}\rangle\!\rangle}{\langle\!\langle \tO_{2}\tO_{2}\rangle\!\rangle \langle\!\langle \tO_{1} \tO_{1}\rangle\!\rangle}\\
=-\frac{15(1-6\cot^2\tk)}{16\pi^3\Tf\sin^3\tk}
=-\frac{15(1-6\cot^2\tk)}{8Ng\sin^3\tk}.
\ee
Remarkably, this is in precise agreement with the prediction of localization.

\section{Correlation functions in dCFT$_1$ from the D3-brane\label{sec:D3}}
\subsection{D3-brane solution in $AdS_5\times S^5$}
In this section we review the D3-brane solution in $AdS_5\times S^5$ background \cite{Drukker:2005kx}. The bosonic part of the Euclidean action for the D3-brane is given by
\be\label{eq:D3_action}
S_{D3}=T_{D3}\int d^4\sigma\sqrt{\text{det}(G+F)}-T_{D3}\int C_4\,.
\ee
The D3-brane tension is
\be
T_{D3}=\frac{N}{2\pi^2}.
\ee

To write down the D3-brane solution, it is convenient to parametrize the $AdS_5\times S^5$ space as
\be\label{eq:D3_metric}
ds^2_{AdS_5\times S^5}=\cosh^2 u\,ds^2_{AdS_2}+\sinh^2 u\, d\Omega^2_2+du^2+\frac{dy^ady^a}{(1+\tfrac{1}{4}y^2)^2}.
\ee
The four-form potential $C_4$ is
\be
C_4=\left(-\frac{u}{2}+\frac{\sinh 4u}{8}\right)\frac{\sin\theta}{r^2}d\t \wedge dr \wedge d\theta\wedge d\phi,
\ee
where $(\t,r)$ are the Poincare coordinates for the Euclidean $AdS_2$ (suitable in the case of straight Wilson line at the boundary), and $(\theta,\phi)$ are the coordinates for $S^2$.
The embedding of the D3-brane solution in $AdS_5\times S^5$ is given by the $AdS_2\times S^2$ hyper-surface parametrized by $u=u_k$ in $AdS_5$ and an arbitrary point on $S^5$. For simplicity, we can choose $y_0^a=0$. The value of $u_k$ is related to the fundamental string charge $k$ dissolved on the brane via \cite{Drukker:2005kx}
\be\label{def_uk}
\sinh u_k=\frac{k\sqrt{\lambda}}{4N}.
\ee
The background gauge field strength is
\be
F=i\frac{\cosh u_k}{r^2}d\t\wedge dr.
\ee
As in the D5-brane case, we need to add the following boundary term to the action to implement the correct boundary conditions \cite{Drukker:2005kx,Faraggi:2011bb}
\be\label{eq:D3_bdy}
S^A_{bdy}=-\int\! d\t\int\! d\Omega_2\, A_\t\,\pi_A,
\ee
with $\pi_A$ being the conjugate momentum to $A_\t$:
\be
\pi_A=\frac{\p\mathcal{L}_{D3}}{\p F_{\t r}}.
\ee
As explained in the D5 brane case above, the boundary term ensures that the 
momentum conjugate to $A$ is held fixed at the boundary. This is related to fundamental string charge as 
\be 
k = -2\pi i \alpha' \int_{S^2} \frac{\p\mathcal{L}_{D3}}{\partial F_{\t r}} = \frac{4N}{\sqrt{\lambda}}\sinh u_k\,,
\ee 
and fixing $k$ means fixing the rank of the symmetric representation of the Wilson loop operator. 

The expectation value of the circular Wilson loop at strong coupling can be obtained by using the hyperbolic disk coordinates on $AdS_2$ and evaluating the D3 brane classical action supplemented by the boundary term (\ref{eq:D3_bdy})
\be 
\langle W_{{\sf S}_k} \rangle = \exp\left(-S_{D3}-S^A_{bdy}\right)\,.
\ee 
Using the solution above, we find
\be 
S_{D3}+S^A_{bdy} = \frac{1}{2} T_{D3} {\rm vol}(AdS_2){\rm vol}(S^2) \left(u_k+\sinh u_k \cosh u_k \right)\,.
\ee 
This yields (as for the D5 brane, we use ${\rm vol}(AdS_2)=-2\pi$ instead of adding a boundary term for the AdS radial coordinate):
\be 
\langle W_{{\sf S}_k} \rangle = \exp\left(2N (u_k+\sinh u_k \cosh u_k)\right)\,.
\label{WSk}
\ee 
This agrees with the localization prediction \cite{Drukker:2005kx, Hartnoll:2006is}, which can be obtained by evaluating (\ref{W-LargeN}) at strong coupling with $k/N$ fixed (note that, as already discussed above, the result in (\ref{W-LargeN})  applies to the more general 1/8-BPS Wilson loops, which are just related to the 1/2-BPS one by a rescaling of the coupling $\lambda\rightarrow \lambda(1-a^2)$).

\subsection{Spectrum of excitations around the D3-brane}
To obtain the mass spectrum, we need to consider the quadratic action for the fluctuations $\dy^a$, $\du$ and $f$ around the D3-brane solution, where $f$ is a 2-form representing the fluctuations of the background field strength. Since the spectrum has been computed in \cite{Faraggi:2011bb}, we briefly review the calculation here. The variation of the metric up to quartic order in fluctuations is
\bea{\label{eq:D3_varmetric}
\delta(ds^2)=&\left(\sinh 2u_k\du+\cosh 2u_k\du^2+\tfrac{2}{3}\sinh 2u_k\du^3+\tfrac{1}{3}\cosh 2u_k\du^4\right)(ds^2_{AdS_2}+d\Omega^2_2)\\
&+(d\du)^2+(1-\frac{1}{2}\dy^2)(d\dy)^2. 
\eqn
}
The variation of the four-form $C_4$ up to quartic order in fluctuations is
\be\label{eq:D3_varC4}
\delta C_4=\frac{\sin\theta}{r^2}\left(\sinh^2 2u_k\du+\sinh 4u_k\du^2+\tfrac{4}{3}\cosh 4u_k\du^3+\tfrac{4}{3}\sinh 4u_k\du^4\right)d\t \wedge dr \wedge d\theta\wedge d\phi.
\ee
We use Greek letters $(\mu,\nu)$ for the coordinates of $AdS_2$ and Geek letters $(\alpha,\beta)$ for the coordinates of $S^2$. The mass spectrum can be obtained by expanding the action \eqref{eq:D3_action} to quadratic order in fluctuations around the D3-brane solution. 

\paragraph{$\dy^a$ sector}
The quadratic Euclidean action for $\dy^a$ sector is
\be
S^{(2)}_{\dy}=\frac{\Tt\sinh 2u_k}{2}\int d^4\xi \frac{\sqrt{g_2}}{r^2}\frac{1}{2}
\biggl(\partial_\mu\dy^a\partial^\mu\dy^a+\nabla_\alpha\dy^a\nabla^\alpha\dy^a\biggr),
\ee
where $g_2$ is the metric for $S^2$. Similar to the D5-brane case, we expand the field $\dy^a$ in terms of symmetric traceless tensor fields:
\be
\dy^a(\t,r,\Omega_2)=\sum_{l=0}^\infty(\dy^a)_{i_1\cdots i_l}(\t,r)Y^{i_1}\cdots Y^{i_l},
\ee
where $Y^i$ is the three-dimensional vector specifying $S^2$:
\be
\sum_{i=1}^3 Y^i(\Omega_2)Y^i(\Omega_2)=1.
\ee
We have
\be
\nabla^2_{S^2}\biggl((\dy^a)_{i_1\cdots i_l}Y^{i_1}\cdots Y^{i_l}\biggr)=-l(l+1)(\dy^a)_{i_1\cdots i_l}Y^{i_1}\cdots Y^{i_l}.
\ee
The quadratic action for the $(\dy^a)_{i_1\cdots i_l}$ modes is 
\be
S^{(2)}_{\dy}=\sum_{l=0}^\infty\frac{V_l\Tt\sinh 2u_k}{2}\int \frac{drd\t}{r^2}\frac{1}{2}
\biggl[\partial_\mu(\dy^a)_{i_1\cdots i_l}\partial^\mu(\dy^a)_{i_1\cdots i_l}+l(l+1)(\dy^a)_{i_1\cdots i_l}(\dy^a)_{i_1\cdots i_l}\biggr].
\ee
The factor $V_l$ comes from the integral of spherical harmonics over $S^2$ and is defined by 
\be\label{eq:D3_Vl_def}
\int d\Omega_2(\mathbf{u}_1\cdot Y)^l(\mathbf{u}_2\cdot Y)^l\equiv V_l(\mathbf{u}_1\cdot\mathbf{u}_2)^l,
\ee
where $\mathbf{u}$ is a three-dimensional null vector. Using the same method as in the D5-brane case, we find that 
\be
V_l=\frac{4\pi(l!)^2\,2^l}{(2l+1)!}.
\ee

\paragraph{$\du$ sector}
The quadratic Euclidean action in $\du$ sector is
\be
S^{(2)}_{\du}=\frac{\Tt\sinh 2u_k}{2}\int d^4\xi \frac{\sqrt{g_2}}{r^2}\frac{1}{2}
\biggl(\partial_\mu\du\partial^\mu\du+\nabla_\alpha\du\nabla^\alpha\du\biggr).
\ee
Expanding the $\du$ field in terms of the symmetric traceless tensor fields
\be
\du(\t,r,\Omega_2)=\sum_{l=0}^\infty \du_{i_1\cdots i_l}(\t,r)Y^{i_1}\cdots Y^{i_l},
\ee
we find that the quadratic action for these modes is
\be
S^{(2)}_{\dy}=\sum_{l=0}^\infty\frac{V_l\Tt\sinh 2u_k}{2}\int \frac{drd\t}{r^2}\frac{1}{2}
\biggl[\partial_\mu \du_{i_1\cdots i_l}\partial^\mu\du_{i_1\cdots i_l}+l(l+1)\du_{i_1\cdots i_l}\du_{i_1\cdots i_l}\biggr].
\ee

\paragraph{Gauge field sector}
In order to decouple $a_\mu$ from the gauge fields along $S^2$ direction, we impose the gauge condition:
\be
\nabla^\alpha a_\alpha=0.
\ee
The field $a_\mu$ can be expanded in terms of symmetric traceless tensor fields expand while the field $a_\alpha$ needs to be expanded using transverse vector spherical harmonics $(\hat{Y}_\alpha)_{lm}$:
\be
a_\mu(\t,r,\Omega_2)=\sum_{l=0}^\infty(a_\mu)_{i_1\cdots i_l}(\t,r)Y^{i_1}\cdots Y^{i_l},\quad
a_\alpha(\t,r,\Omega_2)=\sum_{l=1}^\infty
a_l(\t,r)(\hat{Y}_{\alpha})_{lm}(\Omega_2).
\ee
The transverse vector spherical harmonics satisfy the following properties \cite{1984JMP....25.2888R,vanNieuwenhuizen:2012zk}
\be
\nabla^2_{S^2}(\hat{Y}_\alpha)_{lm}=-(l^2+l-1)(\hat{Y}_\alpha)_{lm},\quad
\nabla^\alpha_{S^2}(\hat{Y}_\alpha)_{lm}=0,
\quad (l=1,2,\dots).
\ee
The quadratic action for the $(a_\mu)_{i_1\cdots i_l}$ modes is
\be
S^{(2)}_{a_\mu}=\sum_{l=1}^\infty \frac{V_l\Tt\coth u_k}{2}\int\frac{d\t dr}{r^2}
\biggl[\frac{1}{2}(f_{\mu\nu})_{i_1\cdots i_l}(f^{\mu\nu})_{i_1\cdots i_l}+l(l+1)(a_\mu)_{i_1\cdots i_l}(a^\mu)_{i_1\cdots i_l}\biggr],
\ee
where we have omitted the $l=0$ mode of $a_\mu$ because it is not dynamical. The quadratic action for the $a_l$ modes is
\be
S^{(2)}_{a_\alpha}=\sum_{l=1}^\infty \pi\Tt\coth{u_k}\int\frac{d\t dr}{r^2}
\frac{1}{2}\biggl[\partial_\mu a_l\partial^\mu a_l+l(l+1)a_l^2\biggr].
\ee

\subsection{Dual operators and two-point functions}
In this section, we discuss the dual operators for the bulk fluctuation modes. Unlike in the D5-brane case, although there have been discussions on the holographic dictionary in \cite{Faraggi:2011bb}, we think there remain some questions on the identification of the dual operators. In table \ref{tab:D3_spectrum}, we summarize the quantum numbers of the dual operators. 

\paragraph{$\dy^a$ sector}
From the mass spectrum, we see that the mode $(\dy^a)_{i_1\cdots i_l}$ should be dual to an operator of dimension $\Delta_l=l+1$ which transforms as a $SO(5)$ vector. In particular, the five $l=0$ modes which we shall denote as $\dy^a_0$ are dual to the five scalars $\Phi^a$ in the ultrashort supermultiplet of $OSp(4^*|4)$. Note that the for $l>0$, the protected operators $\O_l$ do not appear as single-particle states in the D3 brane spectrum, unlike the D5 brane case discussed above. This agrees with the localization analysis in section \ref{subsec:symmetricloop}. Note that, as in the fundamental string case \cite{Giombi:2017cqn, Giombi:2018qox}, one still has protected ``multi-particle" operators with $\Delta=l$ in the totally symmetric representation of $SO(5)$ built from symmetrized products of $\dy^a$. 

\paragraph{$\du$ and $a_\mu$ sector}
From the mass spectrum, we see that both $\du_{i_1\cdots i_l}$ and $(a_\mu)_{i_1\cdots i_l}$ should be dual to the operators of dimension $\Delta_l=l+1$ which transform in the spin-$l$ representation of $SO(3)$. In particular, $\du_0$ should be dual to an operator of dimension $\Delta=1$ which is a singlet under both $SO(3)$ and $SO(5)$. There is no natural candidate for a protected operator with these quantum numbers on the gauge theory side. A possible resolution to this puzzle is that $\du_0$ belongs to a semi-short multiplet of $OSp(4^*|4)$ (this can be thought of as a long multiplet at the unitarity bound, see \cite{Liendo:2018ukf}), and as soon as we move away from the strict strong coupling limit, this operator may acquire anomalous dimension and become part of a long multiplet. It would be interesting to clarify this further. On the other hand, both $\du_i$ and $(a_\mu)_i$ have the correct quantum numbers to be dual to the displacement operator $\mathbb{F}_{ti}$ in the ultrashort multiplet of $OSp(4^*|4)$. By computing the various four-point functions, we find that the dual of $\mathbb{F}_{ti}$ should be a linear combination of $\du_i$ and $(a_\mu)_i$ fields. Specifically, we will find that if we decompose $\du_i$ and $(a_\mu)_i$ as
\be\label{eq:D3_def_chi}
\sqrt{2}\,\du_i=\sqrt{\frac{1}{3}}\chi_i+\sqrt{\frac{2}{3}}\psi_i,\quad
\frac{i}{\sinh u_k}f_i=-\sqrt{\frac{2}{3}}\chi_i+\sqrt{\frac{1}{3}}\psi_i,
\ee
where $f_i\equiv\varepsilon^{\mu\nu}\partial_\mu (a_\nu)_i$, then the mode $\chi_i$ appears to be the bulk mode dual to the displacement operator $\mathbb{F}_{ti}$, at least to the order we are working. Given the four-point function $\langle\!\langle \Phi^{a_1}\Phi^{a_2}\Phi^{a_3}\Phi^{a_4}\rangle\!\rangle$, the supersymmetry uniquely fix the
four-point function $\langle\!\langle \mathbb{F}_{ti_1}\mathbb{F}_{ti_2}\Phi^{a_1}\Phi^{a_2}\rangle\!\rangle$ and $\langle\!\langle \mathbb{F}_{ti_1}\mathbb{F}_{ti_2}\mathbb{F}_{ti_3}\mathbb{F}_{ti_4}\rangle\!\rangle$. As a test of our identification of $\chi_i$ as dual to the displacement operator, we will verify that it has the correct four-point functions by computing them in section \ref{sec:D3_2x2y} and \ref{sec:D3_4x}. 

\paragraph{$a_\alpha$ sector}
From the mass spectrum, we see that the mode $a_l$ should be dual to an operator of dimension $\Delta_l=l+1$ which transforms in the spin-$l$ representation of $SO(3)$. It would be interesting to clarify to which supermultiplet this mode belongs, and its gauge theory interpretation. 

\begin{table}
\centering
\def\arraystretch{1.5}
\begin{tabular}{||c| c| c| c||}
\hline
Fluctuation modes & $\Delta$ & $SO(3)$ & $SO(5)$\\
\hline
$(\dy^a)_{i_1\cdots i_l}$ $(l\geq 0)$ & $l+1$ & $l$ & $(0,1)$\\
\hline
$\du_{i_1\cdots i_l}$ $(l\geq 0)$& $l+1$ & $l$ & $(0,0)$\\
\hline
$(a_\mu)_{i_1\cdots i_l}$ $(l\geq 1)$& $l+1$ & $l$ & $(0,0)$\\
\hline
$a_l$ $(l\geq 1)$& $l+1$ & $l$ & $(0,0)$\\
\hline
\end{tabular}
\caption{In this table we summarize the quantum numbers of the operator dual to each fluctuation mode. $\Delta$ gives the conformal dimension of the dual operator. The quantum numbers of the dual operator under $SO(3)$ and $SO(5)$ symmetry are given in terms of the Dynkin labels of the corresponding representations.}
\label{tab:D3_spectrum}
\end{table}

\paragraph{The two-point functions}
As in the D5-brane case, we need to include the following boundary term to ensure the correct boundary conditions for the gauge fields
\be
-\int\!d\t \int\!d\Omega_4\,\coth u_kr^2\,a_\t(\p_\t a_r-\p_r a_\t).
\ee
Including this boundary term, we find that the tree level two-point function of the operator $\mathbb{F}_{ti}$ is 
\be
\langle\!\langle \mathbb{F}_{ti}(\t_1)\mathbb{F}_{tj}(\t_2)\rangle\!\rangle=
\langle\chi_i(\t_1)\,\chi_j(\t_2)\rangle_{AdS_2} =\delta_{ij}\, \frac{2\Tt\sinh 2u_k}{\t_{12}^{4}}.
\ee
The tree level two-point function for the operator $\Phi^a$ is
\be
\langle\!\langle \Phi^a(\t_1)\Phi^b(\t_2)\rangle\!\rangle=
\langle\dy^a_0(\t_1)\,\dy^b_0(\t_2)\rangle_{AdS_2} =\delta^{ab}\, \frac{2\Tt\sinh 2u_k}{\t_{12}^{2}}.
\ee

\subsection{Four-point function of $S^5$ fluctuations of D3-brane}
\begin{figure}
\begin{center}
\includegraphics[width=0.7\textwidth]{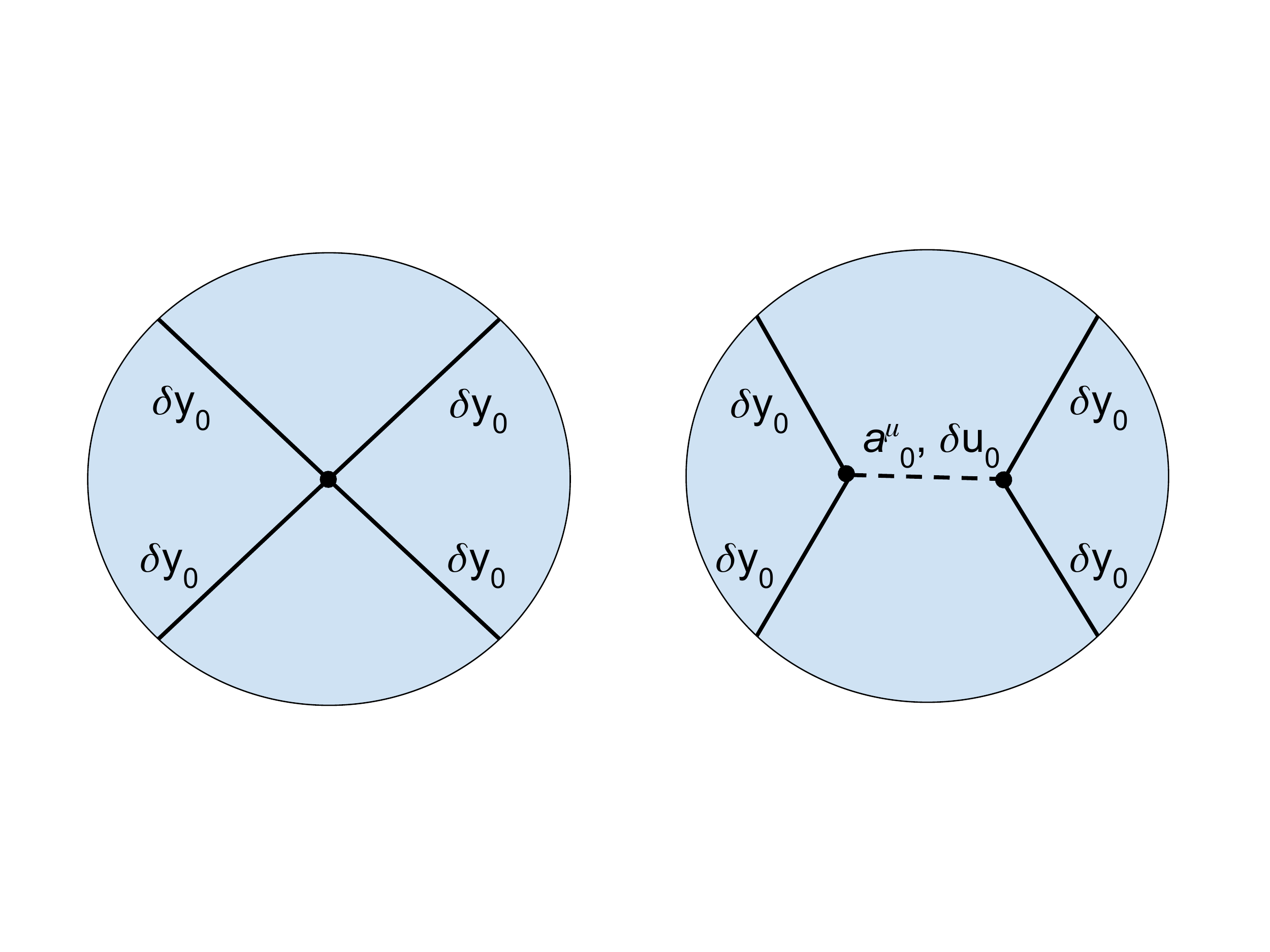}
\end{center}
\vskip -2cm
\caption{Witten diagrams for computing the connected part of the four-point function $\langle \dy^{a_1}_0\dy^{a_2}_0\dy^{a_3}_0\dy^{a_4}_0\rangle$. The $l=0$ modes of $\du$ and $a_\mu$ fields are exchanged in the exchange diagram.}
\label{fig:D3-4pt-4y}
\end{figure}

In this section, we compute the connected part of the tree level four-point function 
\be
\langle\!\langle \Phi^{a_1}(\t_1)\Phi^{a_2}(\t_2)\Phi^{a_3}(\t_3)\Phi^{a_4}(\t_4)\rangle\!\rangle
=
\langle\dy_0^{a_1}(\t_1)\dy_0^{a_2}(\t_2)\dy_0^{a_3}(\t_3)\dy_0^{a_4}(\t_4)\rangle_{AdS_2}.
\ee
The Witten diagrams involved are shown in the figure \ref{fig:D3-4pt-4y}. The quartic coupling in the contact diagram is obtained from expanding the D3-brane action and we find
\bea{\label{eq:D3_4y_exp}
L^{(4)}_{yyyy}=&
\frac{\pi \Tt}{2} \biggl[\biggl(\coth u_k-\frac{4}{\sinh 2u_k}\biggr)\partial_\mu\dy_0^a\partial^\mu\dy_0^a\partial_\nu\dy_0^b\partial^\nu\dy_0^b\biggr.\\
&-2\tanh u_k\partial_\mu\dy_0^a\partial_\nu\dy_0^a\partial^\mu\dy_0^b\partial^\nu\dy_0^b-\sinh 2u_k\partial_\mu\dy_0^a\partial^\mu\dy_0^a\dy_0^b\dy_0^b\biggr].
\eqn
}
The cubic couplings in the exchange diagrams are
\begin{align}
\label{eq:D3_yyu0_cubic}
L_{yyu_0}&=4\pi \Tt\sinh^2 u_k\partial_\mu\dy_0^a\partial^\mu\dy_0^a\du_0,\\
\label{eq:D3_yyf0_cubic}
L_{yyf_0}&=-2i\pi\Tt\csch u_k\partial_\mu\dy_0^a\partial^\mu\dy_0^a\,f_0,
\end{align}
where $f_0=\varepsilon^{\m\n}\p_\m(a_\n)_0$.

To compute the exchange-diagrams, we need the bulk propagator $G_{uu}(\t,r;\t',r')$ for $\du_0$ and $G_{\mu\nu'}(\t,r;\t',r')$ for $(a_\mu)_0$. These propagators satisfy the following equations:
\begin{align}
-\nabla^\mu(\varepsilon^{\alpha\beta}\partial_\alpha G_{\beta\gamma'})&=\frac{\tanh u_k\, r^2}{4\pi\Tt}{\varepsilon^\mu}_{\gamma'}\delta(\t-\t')\delta(r-r'), \\
-\nabla^\mu\nabla_\mu G_{uu}&=\frac{ r^2}{2\pi\Tt\sinh 2u_k}\delta(\t-\t')\delta(r-r'),
\end{align}
where we have suppressed the dependence of the propagators on the coordinates. As in D5-brane case, due to the special form of the cubic coupling, we find that the exchange diagrams can be reduced to a single contact diagram with the effective coupling 
\be\label{eq:D3_4y_effexc}
L^{eff}_{exc}=\pi\Tt\biggl(\frac{1}{\sinh 2u_k}\partial_\mu\dy_0^a\partial^\mu\dy_0^a\partial_\nu\dy_0^b\partial^\nu\dy_0^b+
\frac{\sinh^3 u_k}{\cosh u_k}\partial_\mu\dy_0^a\partial^\mu\dy_0^a\dy_0^b\dy_0^b
\biggr),
\ee
by using integration by parts and the on-shellness of the external $\dy^a_0$.
Combing \eqref{eq:D3_4y_effexc} and \eqref{eq:D3_4y_exp}, we see that the connected part of the four-point function can be computed from a single contact diagram with the effective quartic coupling
\bea{\label{eq:D3_4y_eff_3}
L^{eff}_{yyyy}=&
4\pi\Tt\tanh u_k\biggl(
\frac{1}{8}\partial_\mu\dy_0^a\partial^\mu\dy_0^a\partial_\nu\dy_0^b\partial^\nu\dy_0^b
-\frac{1}{4}\partial_\mu\dy_0^a\partial_\nu\dy_0^a\partial^\mu\dy_0^b\partial^\nu\dy_0^b\biggr.\\
&-\frac{1}{4}\partial_\mu\dy_0^a\partial^\mu\dy_0^a\dy_0^b\dy_0^b
\biggr). \eqn
}
This effective coupling has the identical form as \eqref{eq:D5_4O_eff} in the D5-brane case except the prefactor. Using the same normalization for the bulk-to-boundary propagator as in the D5-brane case, we find that the connected part of the normalized four-point function is
\be\label{eq:D3_4y_final}
\langle\dy_0^{a_1}(\t_1)\dy_0^{a_2}(\t_2)\dy_0^{a_3}(\t_3)\dy_0^{a_4}(\t_4)\rangle =\frac{1}{4\pi\Tt\sinh u_k\cosh^3 u_k}\frac{G_{4y}^{a_1 a_2 a_3 a_4}(\chi)}{\t_{12}^2\t_{34}^2} .
\ee
We can again compare this result to the localization analysis by transforming to the circle and choosing the ``topological" configuration of the polarization vectors. The result is
\be\label{eq:D3_yyyy_top}
\frac{\langle\!\langle \tO_{1} \tO_{1} \tO_{1} \tO_{1}\rangle\!\rangle}{\langle\!\langle \tO_{1} \tO_{1}\rangle\!\rangle^2}
=-\frac{3}{8\pi^2\Tt\sinh u_k\cosh^3 u_k}=-\frac{3}{4N\sinh u_k\cosh^3 u_k}.
\ee
This again precisely agrees with the localization prediction for the 4-point function, which in this case just reduces to taking simple area derivatives of the Wilson loop expectation value (given by (\ref{WSk}) with the replacement $\lambda \rightarrow \lambda A(4\pi-A)/(4\pi^2)$).  

Note that if we take the string limit defined by
\be
\frac{k}{N}\to 0,\quad u_k\to\frac{k\sqrt{\lambda}}{4N},
\ee
then the normalized four-point function \eqref{eq:D3_4y_final} becomes
\be
\langle\dy_0^{a_1}(\t_1)\dy_0^{a_2}(\t_2)\dy_0^{a_3}(\t_3)\dy_0^{a_4}(\t_4)\rangle \to \frac{2\pi}{k\sqrt{\lambda}}
\frac{G_{4y}^{a_1 a_2 a_3 a_4}(\chi)}{\t_{12}^2\t_{34}^2} .
\ee
As in the D5-brane case, the D3-brane result reduces to the result calculated from $k$ weakly coupled string.

\subsection{Two $AdS_5$ and two $S^5$ fluctuations of D3-brane}
\label{sec:D3_2x2y}
\begin{figure}
\begin{center}
\includegraphics[width=0.7\textwidth]{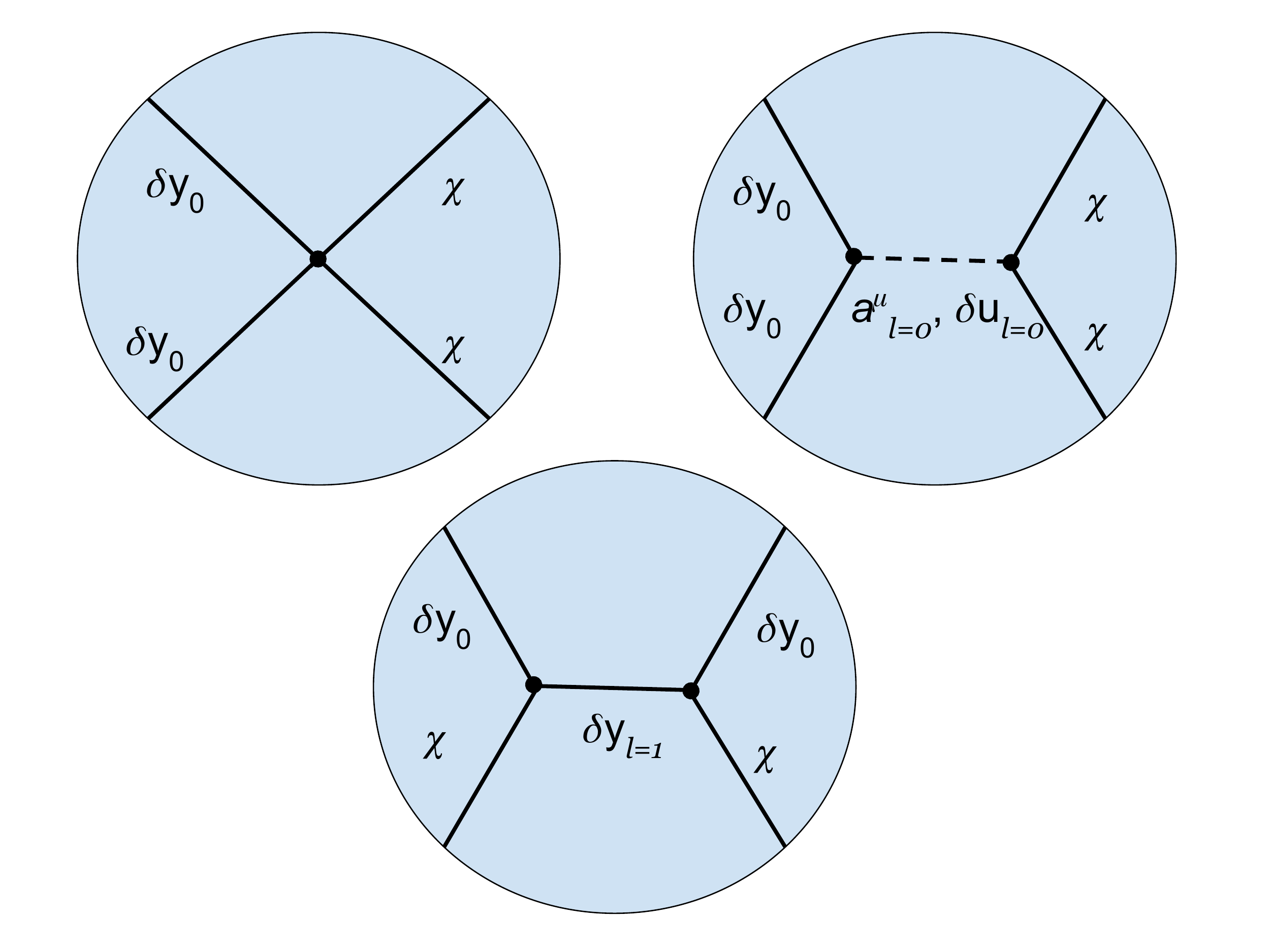}
\end{center}
\vskip -1cm
\caption{Witten diagrams for computing the connected part of the four-point function $\langle \chi_{i_1}\chi_{i_2}\dy^{a_1}_0\dy^{a_2}_0\rangle$. The $l=0$ modes of $\du$ and $a_\mu$ fields as well as the $l=1$ modes of $\dy^a$ field are exchanged in the exchange diagrams.}
\label{fig:D3-4pt-2y2x}
\end{figure}

In this section, we compute the connected part of the tree level four-point function 
\be
\langle\!\langle \mathbb{F}_{t i_1}(\t_1)\mathbb{F}_{t i_2}(\t_2)
\Phi^{a_1}(\t_3)\Phi^{a_2}(\t_4)\rangle\!\rangle=
\langle\chi_{i_1}(\t_1)\chi_{i_2}(\t_2)\dy^{a_1}_0(\t_3)\dy^{a_2}_0(\t_4)\rangle_{AdS_2}.
\ee
Since the four-point function $\langle\dy^{a_1}_0\dy^{a_2}_0\dy^{a_3}_0\dy^{a_4}_0\rangle$ has the same form as in the fundamental string case, the supersymmetry then uniquely fixes the result for $\langle\chi_{i_1}\chi_{i_2}\dy^{a_1}_0\dy^{a_2}_0\rangle$ if $\chi_i$ is dual to the displacement operator $\mathbb{F}_{ti}$. We shall show that it is indeed the case below. The diagrams involved in the calculation are shown in figure \ref{fig:D3-4pt-2y2x}. The contact diagram in figure \ref{fig:D3-4pt-2y2x} results from the quartic coupling in the expanded D3-brane action
\bea{\label{eq:D3_xxyy_quartic}
L^{(4)}_{\chi\chi yy}=&
\frac{2\pi\Tt}{36}
\biggl[\frac{\bigl(3\cosh  2u_k-1\bigr)}{\sinh  2u_k}\p_\m\chi_i\p^\m\chi_i\p_\n\dy^a_0\p^\n\dy^a_0\\
&-6\tanh u_k\p_\m\chi_i\p_\n\chi_i\p^\m\dy^a_0\p^\n\dy^a_0
+\frac{(3+\cosh 4u_k)}{\sinh 2u_k}\chi_i\chi_i\p_\m\dy^a_0\p^\m\dy^a_0\biggr].
\eqn
}
The other two exchange diagrams in figure \ref{fig:D3-4pt-2y2x} involve the exchange of $l=0$ and $l=1$ modes of the bulk fields.

\paragraph{Exchange of $l=0$ modes}
In this case, the cubic couplings involved are \eqref{eq:D3_yyu0_cubic}, \eqref{eq:D3_yyf0_cubic} and
\bea{
\label{eq:D3_xxu0_cubic}
L_{\chi\chi u_0}=&\frac{\pi\Tt}{9}(\cosh 2u_k+3)\left(2\chi_i\p_\m\chi_i\p^\m\du_0
+\p_\m\chi_i\p^\m\chi_i\du_0+2\chi_i\chi_i\du_0\right),
\eqn\\
\label{eq:D3_xxf0_cubic}
L_{\chi\chi f_0}=&\frac{i\pi\Tt}{9\sinh u_k}\left(\p_\m\chi_i\p^\m\chi_i+2\chi_i\chi_i\right)f_0.
\eqn \\
}
As before, these exchange diagrams can be reduced to contact diagrams after using the on-shellness of the external $\chi_i$ and $\dy^a_0$ and performing integration by parts. The end result can be summarized as a single contact diagram with the effective quartic coupling 
\bea{\label{eq:D3_xxyyl0_eff}
L_{exc,l=0}=
&-\frac{\pi \Tt}{18}\biggl[(\cosh 2u_k+3)\tanh u_k\p_\m\dy^a_0\p^\m\dy^a_0\chi_i\chi_i\\
&-\frac{2}{\sinh 2u_k}\bigl(\p_\m\dy^a_0\p^\m\dy^a_0\p_\n\chi_i\p^\n\chi_i 
+2\p_\m\dy^a_0\p^\m\dy^a_0\chi_i\chi_i\bigr)\biggr].
\eqn
}

\paragraph{Exchange of $l=1$ modes}
In this case, the cubic coupling involved is
\be
L_{yy_1\chi}=\frac{4\pi\Tt}{3}\sqrt{\frac{2}{3}}\left(\cosh^2 u_k\,\chi_i\p_\m\dy^a_0\p^\m\dy^a_i
+\dy^a_i\p_\m\dy^a_0\p^\m\chi_i\right).
\ee
Using the fact that $\dy^a_0$ and $\chi_i$ are put on-shell in the calculation of the Witten diagram, we can rewrite the cubic coupling as
\be\label{eq:D3_xxy1_eff}
\frac{2\pi\Tt}{3}\sqrt{\frac{2}{3}}\chi_i\dy^a_0(-\nabla^2+2)(\dy^a)_i. 
\ee
To compute the diagram, we need the bulk propagator $G_{ij'}^{ab'}(\t,r;\t'r')$ for $\dy^a_i$, which satisfies the equation
\be
\left(-\nabla^\mu\nabla_\mu + 2\right)G_{ij'}^{ab'}=\delta_{ij'}\delta^{ab'}\frac{3r^2}{2\pi\Tt\sinh 2u_k}\delta(\t-\t')\delta(r-r').
\ee
Due to the form of the cubic coupling \eqref{eq:D3_xxy1_eff}, we see that the exchange diagram can be reduced to a contact diagram with the effective coupling
\be\label{eq:D3_xxyyl1_eff}
L_{exc,l=1}=
-\frac{\pi\Tt}{9}\sinh^2 u_k\tanh u_k \p_\m\dy^a_0\p^\m\dy^a_0\chi_i\chi_i.
\ee

\paragraph{The four-point function}
Combining \eqref{eq:D3_xxyyl0_eff} and \eqref{eq:D3_xxyyl1_eff} with \eqref{eq:D3_xxyy_quartic}, we find that the four-point function can be computed from a single contact diagram with the effective quartic coupling
\be
L^{eff}_{\chi\chi yy}=
\frac{2\pi\Tt\tanh u_k}{3}
\biggl(\frac{1}{4}\,\p_\m\chi_i\p^\m\chi_i\p_\n\dy^a_0\p^\n\dy^a_0
-\frac{1}{2}\, \p_\m\chi_i\p_\n\chi_i\p^\m\dy^a_0\p^\n\dy^a_0
\biggr).
\ee
The effective coupling has the same form as \eqref{eq:D5_xxoo_eff}, which we have expected from the supersymmetry. It follows that the connected part of the normalized tree level four-point function is
\be\label{eq:D3_2x2y_4ptf}
\langle\chi_{i_1}(\t_1)\chi_{i_2}(\t_2)\dy^{a_1}_0(\t_3)\dy^{a_2}_0(\t_4)\rangle
=\delta_{i_1 i_2}\delta^{a_1 a_2}\frac{1}{4\pi\Tt\sinh u_k\cosh^3 u_k}\frac{G_{2x2y}(\chi)}{\t_{12}^4\t_{34}^2}.
\ee
We note that the prefactor in \eqref{eq:D3_2x2y_4ptf} also agrees with \eqref{eq:D3_4y_final}.

\subsection{Four $AdS_5$ fluctuations of D3-brane}
\label{sec:D3_4x}
\begin{figure}
\begin{center}
\includegraphics[width=0.7\textwidth]{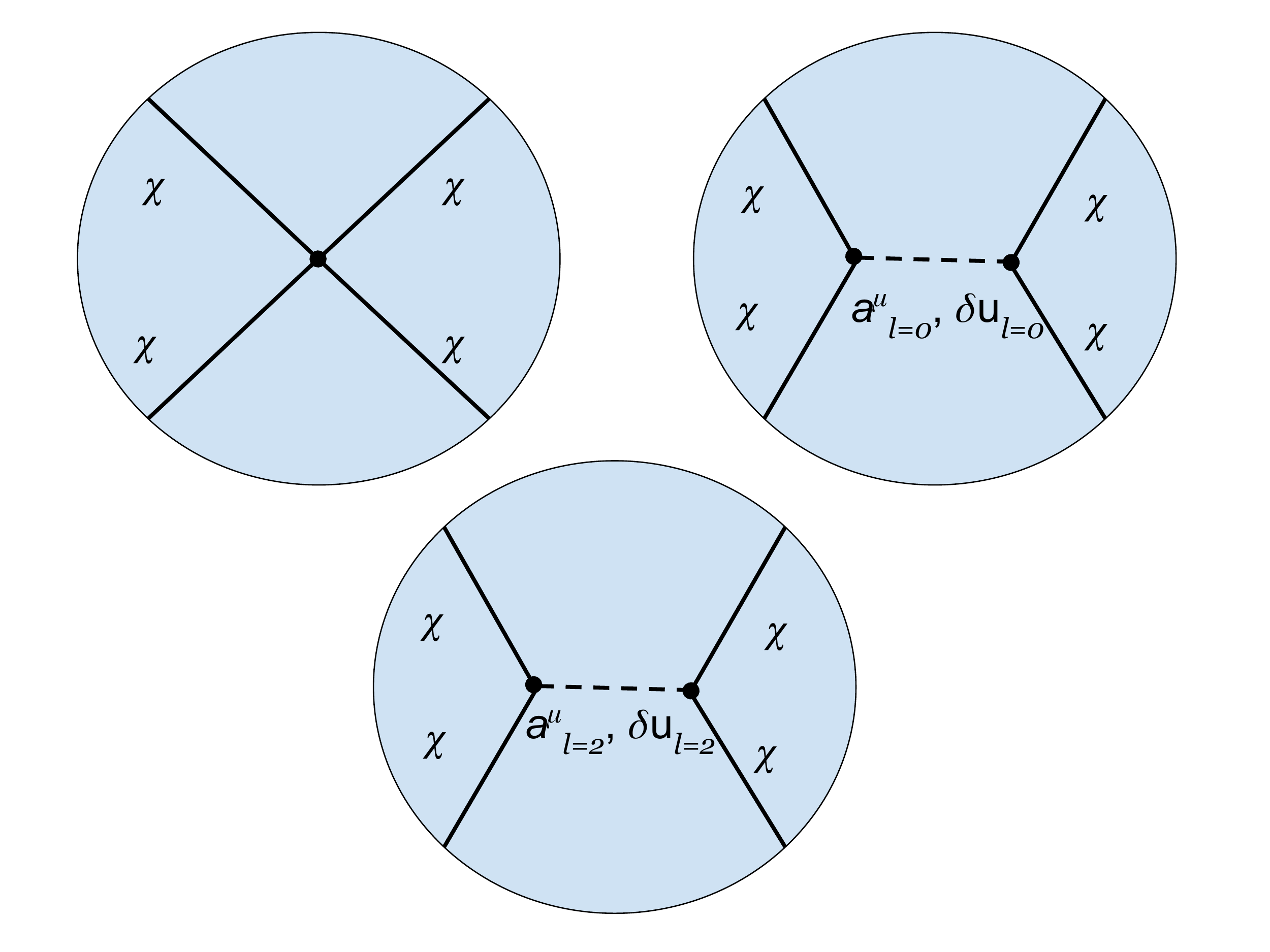}
\end{center}
\vskip -1cm
\caption{Witten diagrams for computing the connected part of the four-point function $\langle \chi_{i_1}\chi_{i_2}\chi_{i_3}\chi_{i_4}\rangle$. Both $l=0$ and $l=2$ modes of $\du$ and $a_\mu$ fields are exchanged in the exchange diagrams.}
\label{fig:D3-4pt-4x}
\end{figure}

In this section, we compute the connected part of the tree level four-point function
\be 
\langle\!\langle{\mathbb{F}_{t}}^{i_1}(\t_1){\mathbb{F}_{t}}^{i_2}(\t_2){\mathbb{F}_{t}}^{i_3}(\t_3){\mathbb{F}_{t}}^{i_4}(\t_4)\rangle\!\rangle=
\langle\chi^{i_1}(\t_1)\chi^{i_2}(\t_2)\chi^{i_3}(\t_3)\chi^{i_4}(\t_4)\rangle_{AdS_2}.
\ee
The relevant quartic vertices from expanding the D3-brane action are:
\bea{\label{eq:D3_xxxx_quartic}
L^{(4)}_{\chi\chi\chi\chi}=&
\frac{\pi\Tt}{1080}\tanh u_k\biggl[(14+\coth^2 u_k)\p_\m\chi_i\p^\m\chi_i\p_\n\chi_j\p^\n\chi_j\\
&-(22+8\coth^2 u_k)\p_\m\chi_i\p_\n\chi_i\p^\m\chi_j\p^\n\chi_j
+\frac{(-15+5\cosh 2u_k)}{\sinh^2u_k}\chi_i\chi_i\p_\m\chi_j\p^\m\chi_j\\
&+\frac{(-18+10\cosh 2u_k)}{\sinh^2 u_k}\chi_i\chi_i\chi_j\chi_j\biggr],
\eqn
}
which leads to the contact diagram in figure \ref{fig:D3-4pt-4x}. The exchange diagrams in figure \ref{fig:D3-4pt-4x} involves the exchange of $l=0$ and $l=2$ particles.

\paragraph{Exchange of $l=0$ modes}
The cubic couplings involved are \eqref{eq:D3_xxu0_cubic} and \eqref{eq:D3_xxf0_cubic}. As before, since the external $\chi_i$ in the calculation of the exchange diagram is put on-shell, we can use the equations of motion for $\chi_i$ and integration by parts to rewrite the cubic couplings as
\be
\frac{\pi\Tt}{18}\biggl[(\cosh 2u_k+3)\,\chi_i^2\,(-\nabla^2\du_0)
+\frac{2i}{\sinh u_k}\left(\p_\m\chi_i\p^\m\chi_i+2\chi^2_i\right)f_0\biggr].
\ee
It follows that the exchange diagrams can be reduced to a contact diagram with the effective coupling
\bea{\label{eq:D3_xxxxl0_eff}
L_{exc,l=0}=
&\frac{\pi \Tt}{648\sinh 2u_k}\biggl[(3+\cosh 2u_k)^2(\chi_i\chi_i\,\p_\m\chi_j\p^\m\chi_j+2\chi_i\chi_i\,\chi_j\chi_j)\\
&+2\left(\p_\m\chi_i\p^\m\chi_i\p_\n\chi_j\p^\n\chi_j+4\chi_i\chi_i\, \p_\m\chi_j\p^\m\chi_j+4\chi_i\chi_i\,\chi_j\chi_j\right)\biggr].
\eqn
}

\paragraph{Exchange of $l=2$ modes}
In this case, the cubic couplings involved are
\bea{
L_{\chi\chi u_2}&=\frac{2\pi\Tt}{45}\biggl[2\cosh 2u_k(\chi_i\p_\m\chi_j +\p_\m\chi_i\,\chi_j)\p^\m\du_{ij}+(5\cosh 2u_k-3)\chi_i\chi_j \du_{ij}\\
&+(\cosh 2u_k+3)\p_\m\chi_i\p^\m\chi_j\du_{ij}\biggr],\\
L_{\chi\chi f_2}&=-\frac{2i\pi\Tt}{45\sinh u_k}\biggl[6\varepsilon^{\mu\nu}\chi_i\p_\m\chi_j(a_{\nu})_{ij}+(\chi_i\chi_j+2\p_\m\chi_i\p^\m\chi_j)f_{ij}\biggr].
\eqn
}
Using integration by parts and the on-shellness of the external $\chi_i$ in the calculation of the diagram, the cubic couplings can be expressed as
\be\label{eq:D3_xxl2_cubic}
\frac{\pi\Tt}{45}(\cosh 2u_k-3)\biggl[\chi_i\chi_j(-\nabla^2 +6)\du_{ij}
-\frac{4i}{\sinh u_k} \chi_i\p_\m\chi_j\bigl(-\nabla^\m f_{ij}+6\varepsilon^{\m\n}(a_\n)_{ij}\bigr)\biggr].
\ee
To compute the exchange diagram, we also need the bulk propagator $G_{uu}^{iji'j'}(\t,r;\t',r')$ for $\du_{ij}$ and $G_{\mu\nu'}^{iji'j'}(\t,r;\t',r')$ for $(a_\mu)_{ij}$, which satisfy the equations
\begin{align}
(-\nabla^\mu\nabla_\mu+6)G_{uu}^{ij\,i'j'}&=\frac{15 r^2M^{ij\,i'j'}}{4\pi\Tt  \sinh 2u_k }\delta^2(\t,r;\t',r'),\\
-\nabla^\mu (\varepsilon^{\alpha\beta}\partial_\alpha G_{\beta\gamma'}^{ij\, i'j'})+6\varepsilon^{\mu\beta}G_{\beta\gamma'}^{ij\, i'j'}
&=\frac{15\tanh u_k\, r^2}{8\pi\Tt}{\varepsilon^\mu}_{\gamma'}M^{ij\,i'j'}\delta^2(\t,r;\t',r'), 
\end{align}
where $M^{ij\,i'j'}$ is defined as
\be
M^{ij\, i'j'}=\frac{1}{2}\left(\delta^{ii'}\delta^{jj'}+\delta^{ij'}\delta^{ji'}-\frac{2}{3}\delta^{ij}\delta^{i'j'}\right).
\ee
From the form of the cubic coupling \eqref{eq:D3_xxl2_cubic}, we see that the exchange diagram can be again reduced to a contact diagram with the effective quartic coupling
\bea{\label{eq:D3_xxxxl2_eff}
L_{exc,l=2}=
&\frac{\pi \Tt}{648\sinh 2u_k}\biggl[-(\cosh 2u_k-3)^2\left(\chi_i\chi_i\,\p_\m\chi_j\p^\m\chi_j+2\chi_i\chi_i\,\chi_j\chi_j\right)\\
&+\frac{48}{5}\p_\m\chi_i\p_\n\chi_i\p^\m\chi_j\p^\n\chi_j
-\frac{16}{5}\p_\m\chi_i\p^\m\chi_i\p_\n\chi_j\p^\n\chi_j\\
&-8\chi_i\chi_i\,\p_\m\chi_j\p^\m\chi_j-\frac{112}{5}\chi_i\chi_i\,\chi_j\chi_j\biggr].
\eqn
}

\paragraph{The four-point function}
Combining \eqref{eq:D3_xxxxl0_eff} and \eqref{eq:D3_xxxxl2_eff} with \eqref{eq:D3_xxxx_quartic}, we find that the four-point function can be computed from a single contact diagram with the effective quartic coupling
\bea{
L^{eff}_{\chi\chi\chi\chi}=&
\frac{\pi\Tt\tanh u_k}{9}\biggl(\frac{1}{8}\p_\m\chi_i\p^\m\chi_i\p_\n\chi_j\p^\n\chi_j
-\frac{1}{4}\p_\m\chi_i\p_\n\chi_i\p^\m\chi_j\p^\n\chi_j\\
&+\frac{1}{4}\chi_i\chi_i\,\p_\m\chi_j\p^\m\chi_j+\frac{1}{2}\chi_i\chi_i\,\chi_j\chi_j\biggr),
\eqn
}
which has the same form as \eqref{eq:D5_4x_final}. This agrees again with our expectation from the supersymmetry. It follows that the connected part of the normalized four-point function is
\be
\langle\chi^{i_1}(\t_1)\chi^{i_2}(\t_2)\chi^{i_3}(\t_3)\chi^{i_4}(\t_4)\rangle=\frac{1}{4\pi\Tt\sinh u_k\cosh^3 u_k}\frac{G^{i_1i_2i_3i_4}_{4x}(\chi)}{\t_{12}^4\t_{34}^4},
\ee
with the same prefactor as in \eqref{eq:D3_4y_final} and \eqref{eq:D3_2x2y_4ptf}.

\section{Conclusion}\label{sec:conclusion}
In this paper, we studied the correlation functions of insertions on the $1/2$-BPS Wilson loop in $\mathcal{N}=4$ SYM. In particular we focused on the Giant Wilson loops---the Wilson loops in large-rank symmetric or antisymmetric representations whose sizes are of order $N$. On the gauge-theory side, we computed the correlation functions of protected scalar insertions using a combination of various techniques developed earlier; supersymmetric localization \cite{Pestun:2009nn}, the loop equation \cite{Giombi:2020pdd}, the Gram-Schmidt orthogonalization \cite{Giombi:2018qox}, the Fermi Gas formalism \cite{Marino:2011eh} and the Clustering method \cite{Jiang:2016ulr}.  We next performed an analysis on the AdS side using the dual description in terms of D-branes. Both for the antisymmetric and the symmetric representations, we computed the four-point functions of elementary fluctuations on the D-brane, which are dual to either the displacement operators or the single scalar insertions on the Wilson loop. For the Wilson loops in the antisymmetric representations that are dual to the D3-branes in $AdS_2\times S^4$, we also computed a set of correlation functions involving the Kaluza-Klein modes coming from the reduction of the $S^4$ worldvolume. In a special supersymmetric configuration, these correlators reproduce the results of supersymmetric localization, providing nontrivial evidence for the holographic duality.

There are several interesting future directions to pursue: One obvious generalization of our analysis is to include the single-trace operators in $\mathcal{N}=4$ SYM and compute the bulk-defect correlation functions. Such correlators, which are crucial inputs for formulating the defect crossing equation \cite{Liendo:2018ukf}, were analyzed in \cite{Giombi:2018hsx} for the Wilson loop in the fundamental representation. By combining the techniques in this paper and the ones in \cite{Giombi:2018hsx}, it should be possible to perform the computation.

Another generalization would be to consider the Wilson loops in even larger representations; namely the representations whose sizes are of order $N^2$. Such Wilson loops are known to be dual to so-called bubbling geometries \cite{Yamaguchi:2006te,Lunin:2006xr,DHoker:2007mci}. In this case, the insertions on the Wilson loop are expected to be described by supergravity states propagating in such geometries. It would be interesting to make this statement precise by computing the defect CFT correlators both in the gauge theory and in supergravity.

It would also be interesting to analyze the insertions on the Giant Wilson loops from integrability. For the Wilson loops in the antisymmetric representations, some attempts were made in \cite{Correa:2013em} to compute a reflection matrix corresponding to the Giant Wilson loop, but a complete answer has not been obtained yet. The correlation functions computed by the localization in this paper admit simple integral representations involving the $Q$-function-like polynomials, suggesting a possibility of formulating the Quantum Spectral Curve for the Giant Wilson loops. Once the Quantum Spectral Curve is obtained, it would be extremely interesting to see how the operator spectrum interpolates between the spectrum of insertions in $\mathcal{N}=4$ SYM at weak coupling and the spectrum of fluctuations on the D-brane. In particular, this may help to demystify the puzzle for the D5-brane discussed in section \ref{subsec:giant}; namely the absence of the AdS Kaluza-Klein modes at strong coupling.

Yet another direction would be to understand the relation to the twisted holography discussed in \cite{Ishtiaque:2018str} and make contact with $\mathfrak{gl}(M)$ Yangian discussed there. For this purpose, one needs to consider a product of $M$ Wilson loops in the antisymmetric representations, and compute the correlators of insertions. This is certainly more complicated than what we did in this paper, but the methods developed in this paper are likely generalizable to such cases.
\label{sec:conclusion}

\subsection*{Acknowledgement}
SG and SK are grateful to CERN for hospitality during completion of this work. The work of SG and JJ is supported in part by the US  NSF under Grants No.~PHY-1620542 and PHY-1914860.  The work of SK is supported by DOE grant number DE-SC0009988. 

\appendix
\section{Relevant functions in the holographic calculation}\label{sec:app_funcs}
In this appendix, we give the definitions and the expressions for the various functions that appear in the D-brane computation of the correlation functions. The $D$-function appears in the computation of tree level four-point functions that only involve contact diagrams \cite{DHoker:1999kzh,Liu:1998ty,Dolan:2000ut}. In the general case of $AdS_{d+1}$, the $D$-function is defined as 
\be
D_{\D_1\D_2\D_3\D_4}(\vec{\bf{x}}_1,\vec{\bf{x}}_2,\vec{\bf{x}}_3,\vec{\bf{x}}_4)\equiv
\int\frac{dr\, d^d\vec{\bf{x}}}{r^{d+1}}\prod_{i=1}^4 \biggl(\frac{r}{r^2+(\vec{\bf{x}}-\vec{\bf{x}}_i)^2}\biggr)^{\D_i}.
\ee
The various functions appear in the result of the four-point functions have been first computed in \cite{Giombi:2017cqn} and we simply quote the results below.

The function $G_{2x2y}(\chi)$ is given by
\be
G_{2x2y}(\chi)=-\frac{2}{\pi}\biggl[1-\biggl(\frac{1}{2}-\frac{1}{\chi}\biggr)\log|1-\chi|\biggr].
\ee
Both $G^{i_1 i_2 i_3 i_4}_{4x}(\chi)$ and $G^{a_1 a_2 a_3 a_4}_{4y}(\chi)$ can be decomposed into singlet $(S)$, symmetric traceless $(T)$ and antisymmetric $(A)$ parts as 
\bea{
G_{4x}^{i_1 i_2 i_3 i_4}(\chi)=&
G^{(S)}_{4x}(\chi)\delta^{i_1i_2}\delta^{i_3i_4}
+G^{(T)}_{4x}(\chi)\left(\delta^{i_1i_3}\delta^{i_2i_4}
+\delta^{i_1i_4}\delta^{i_2i_3}
-\frac{2}{3}\delta^{i_1i_2}\delta^{i_3i_4}\right)
\\
&+
G^{(A)}_{4x}(\chi)\left(\delta^{i_1i_3}\delta^{i_2i_4}-\delta^{i_2i_3}\delta^{i_1i_4}
\right), 
\eqn
}
with
\bea{
G^{(S)}_{4x}(\chi)=&\frac{1}{6\pi}\biggl[
-\frac{(24\chi^8-90\chi^7+125\chi^6-76\chi^5+125\chi^4-306\chi^3+438\chi^2-288\chi+72)}{3(\chi-1)^4}\\
&-\frac{2(4\chi^6-\chi^5-6\chi+12)}{\chi}\log|1-\chi|\\
&+\frac{2\chi^4(4\chi^6-21\chi^5+45\chi^4-50\chi^3+30\chi^2-6\chi+2)}{(\chi-1)^5}\log|\chi|
\biggr],\eqn\\
G^{(T)}_{4x}(\chi)=&\frac{1}{4\pi}\biggl[
-\frac{(48\chi^4-198\chi^3+313\chi^2-230\chi+115)\chi^4}{6(\chi-1)^4}
-(8\chi-5)\chi^4\log|1-\chi|\\
&+\frac{(8\chi^6-45\chi^5+105\chi^4-130\chi^3+90\chi^2-30\chi+10)\chi^4}{(\chi-1)^5}\log|\chi|
\biggr],\eqn\\
G^{(A)}_{4x}(\chi)=&\frac{1}{4\pi}\biggl[
-\frac{(\chi-2)(48\chi^6-90\chi^5+91\chi^4+4\chi^3-17\chi^2+18\chi-6)\chi}{6(\chi-1)^4}\\
&-(8\chi^5-3\chi^4+2)\log|1-\chi|
+\frac{(\chi-2)(8\chi^4-27\chi^3+41\chi^2-28\chi+14)\chi^5}{(\chi-1)^5}\log|\chi|
\biggr],
\eqn
}
and
\bea{
G_{4y}^{a_1 a_2 a_3 a_4}(\chi)=&
G^{(S)}_{4y}(\chi)\delta^{a_1a_2}\delta^{a_3a_4}
+G^{(T)}_{4y}(\chi)\left(\delta^{a_1a_3}\delta^{a_2a_4}
+\delta^{a_2a_3}\delta^{a_1a_4}
-\frac{2}{5}\delta^{a_1a_2}\delta^{a_3a_4}\right)
\\
&+
G^{(A)}_{4y}(\chi)\left(\delta^{a_1a_3}\delta^{a_2a_4}-\delta^{a_2a_3}\delta^{a_1a_4}
\right), 
\eqn
}
with
\bea{
G^{(S)}_{4y}(\chi)=&\frac{1}{10\pi}\biggl[
-\frac{2(\chi^4-4\chi^3+9\chi^2-10\chi+5)}{(\chi-1)^2}
+\frac{\chi^2(2\chi^4-11\chi^3+21\chi^2-20\chi+10)}{(\chi-1)^3}\log|\chi|
\\
&-\frac{(2\chi^4-5\chi^3-5\chi+10)}{\chi}\log|1-\chi|
\biggr],\eqn \\
G^{(T)}_{4y}(\chi)=&\frac{1}{2\pi}\biggl[
-\frac{\chi^2(2\chi^2-3\chi+3)}{2(\chi-1)^2}
+\frac{\chi^4(\chi^2-3\chi+3)}{(\chi-1)^3}\log|\chi|-\chi^3\log|1-\chi|
\biggr],\eqn \\
G^{(A)}_{4y}(\chi)=&\frac{1}{2\pi}\biggl[
\frac{\chi(-2\chi^3+5\chi^2-3\chi+2)}{2(\chi-1)^2}
+\frac{\chi^3(\chi^3-4\chi^2+6\chi-4)}{(\chi-1)^3}\log|\chi|\\
&-(\chi^3-\chi^2-1)\log|1-\chi|
\biggr]\eqn.
}

\section{Calculation of $V_{L}$ and $V_{L_1,L_2,L_3}$}
In this appendix, we derive the expressions for $V_L$ and $V_{L_1,L_2,L_3}$ appear in the D5-brane calculation. We consider the following generating function
\be\label{eq:D5_VL_generating}
I[\mathbf{J}]=\int d\Omega_4\, e^{\mathbf{J}\cdot Y}=V_{S^3}\int_0^\pi d\theta\,\sin^3\theta\, e^{|\mathbf{J}|\cos\theta}=\frac{8\pi^2}{\mathbf{J}^2}\left(\cosh|\mathbf{J}|-\frac{\sinh|\mathbf{J}|}{|\mathbf{J}|}\right)
\ee
where $\mathbf{J}$ is a five-dimensional vector and $Y$ is the unit five-dimensional vector specifying $S^4$. We can express \eqref{eq:D5_VL_generating} as a series in power of $\mathbf{J}^2$:
\be\label{eq:D5_VL_series}
I[\mathbf{J}]=16\pi^2\sum^\infty_{n=0} \frac{(n+1)}{(2n+3)!}\left(\mathbf{J}^2\right)^{n}
\ee

To compute $V_L$, we set $\mathbf{J}=\mathbf{u}_1+\mathbf{u}_2$ so that $\mathbf{J}^2=2\, \mathbf{u}_1\cdot\mathbf{u}_2$. One can then compute $V_L$ by extracting the coefficient of the $(\mathbf{u}_1\cdot\mathbf{u}_2)^L$ term in \eqref{eq:D5_VL_series} multiplied by $(L!)^2$ from expanding the exponential in \eqref{eq:D5_VL_generating}:
\be
V_L=\frac{16\,\pi^2\, 2^L\, (L!)^2\,(L+1)}{(2L+3)!}.
\ee
The $V_L$ defined in \eqref{eq:D3_Vl_def} in the D3-brane calculation can be computed analogously.

To compute $V_{L_1,L_2,L_3}$, we set $\mathbf{J}=\mathbf{u}_1+\mathbf{u}_2+\mathbf{u}_3$ so that $\mathbf{J}^2=2(\mathbf{u}_1\cdot\mathbf{u}_2+\mathbf{u}_2\cdot\mathbf{u}_3+\mathbf{u}_1\cdot\mathbf{u}_3)$. Now we need to extract the coefficient of the term $(\mathbf{u}_1\cdot\mathbf{u}_2)^{L_{12|3}}(\mathbf{u}_2\cdot\mathbf{u}_3)^{L_{23|1}}(\mathbf{u}_1\cdot\mathbf{u}_3)^{L_{13|2}}$ in \eqref{eq:D5_VL_series} multiplied by $L_1!\, L_2!\, L_3!$ from the expansion of the exponential:
\bea{\label{eq:D5_VLLL_expression}
V_{L_1,L_2,L_3}&=\frac{(1+(-1)^{L_1+L_2+L_3})}{2}\frac{8\pi^2\,(\sqrt{2})^\Sigma\,(\Sigma+2)L_1!\,L_2!\,L_3!}{(\Sigma+3)!}\binom{\frac{\Sigma}{2}}{L_{12|3}}\binom{L_3}{L_{23|1}}\\
&=\frac{(1+(-1)^{L_1+L_2+L_3})}{2}\frac{8\pi^2\,(\sqrt{2})^\Sigma\,(\Sigma+2)L_1!\,L_2!\,L_3!\,\left(\frac{\Sigma}{2}\right)!}{(\Sigma+3)!\,L_{12|3}!\,L_{23|1}!\,L_{13|2}!},
\eqn
}
where the last two terms of the first line in \eqref{eq:D5_VLLL_expression} stand for the binomial coefficients.

\bibliographystyle{JHEP}
\bibliography{GiantRefs}
\end{document}